%% file: main.tex
\documentclass[conference]{IEEEtran}

\usepackage[hyphens]{url}
\usepackage[pdftex]{graphicx}
\usepackage[breaklinks,colorlinks]{hyperref}
\usepackage[usenames,dvipsnames, table]{xcolor}
\hypersetup{citecolor=blue,linkcolor=blue}
\usepackage{listings}
\usepackage{multirow}
\usepackage{pifont}
\usepackage{threeparttable}
\usepackage{pbox}
\usepackage[square,sort&compress,numbers]{natbib}

\usepackage{amssymb}
\usepackage{color}
\usepackage{xspace}
\usepackage{amsmath}
\usepackage{graphicx}
\usepackage{subcaption}
\usepackage{caption}
\usepackage{algorithm}
\usepackage{algpseudocode}
\usepackage{enumitem} 
\usepackage{adjustbox}
\usepackage{tikz}
\usepackage{soul}
\usepackage{mdwlist}
\usepackage{booktabs}
\usepackage{upquote}
\usepackage{multicol}
\usepackage{tabularx}
\usepackage{makecell}
\usepackage{wrapfig}

\newcommand{\ie}{\mbox{\emph{i. e.,\ }}}
\newcommand{\eg}{\mbox{\emph{e.g.,\ }}}
\newcommand{\point}[1]{\par\vspace{0.1in}\noindent{\bf #1:}}
\newcommand{\rom}[1]{\textit{\expandafter\romannumeral #1}}

\newcommand{\MOREBUGDRILLER}{{\bf 43.4\% }}
\newcommand{\MOREBUGQSYM}{{\bf 44.3\% }}

\newcommand{\BUGDIFFDRILLER}{{\bf 88 }}
\newcommand{\BUGDIFFQSYM}{{\bf 76 }}

\newcommand{\MORECOVDRILLER}{{\bf 19.68\% }}
\newcommand{\MORECOVQSYM}{{\bf 15.18\% }}
\newcommand{\OOBNUM}{{\bf 102 }}
\newcommand{\LOGICNUM}{{\bf 141 }}
\newcommand{\REALBUGNUM}{{\bf 243 }}
\newcommand{\ALLNUM}{{\bf 481 }}

\newcommand{\CONFIRMNUM}{{\bf 16} }
\newcommand{\EXPNUM}{{\bf 25 }}

\newcommand{\TOTOALBUGNUM}{{\bf 481 }}
\newcommand{\BENCHNUM}{{\tt 11 }}
\newcommand{\PROGNUM}{{\tt 8 }}

\newcommand{\comment}[1]

\algnewcommand\algorithmicinput{\textbf{INPUT:}}
\algnewcommand\algorithmicoutput{\textbf{OUTPUT:}}
\algnewcommand\INPUT{\item[\algorithmicinput]}
\algnewcommand\OUTPUT{\item[\algorithmicoutput]}

\newcommand{\afl}{{\sc AFL}\xspace}
\newcommand{\aflgo}{{\sc AFLGo}\xspace}
\newcommand{\tfuzz}{{\sc TFuzz}\xspace}
\newcommand{\angora}{{\sc Angora}\xspace}
\newcommand{\driller}{{\sc Driller}\xspace}
\newcommand{\qsym}{{\sc QSYM}\xspace}
\newcommand{\savior}{{\sc SAVIOR}\xspace}

\newcommand{\bugguidedsearch}{{bug-guided verification}\xspace}
\newcommand{\capbugguidedsearch}{{Bug-guided verification}\xspace}

\newcommand{\rulesep}{\unskip\ \vrule\ }

\floatstyle{ruled}
\newfloat{code fragment}{thp}{lop}
\floatname{code fragment}{Code Fragment}

\definecolor{dkgreen}{rgb}{0,0.6,0}
\definecolor{gray}{rgb}{0.5,0.5,0.5}
\definecolor{mauve}{rgb}{0.58,0,0.82}

\title{\savior: Towards Bug-Driven Hybrid Testing}
\begin{document}

\author{
        \IEEEauthorblockN{
                Yaohui Chen,\IEEEauthorrefmark{1}
                Peng Li,\IEEEauthorrefmark{2}
                Jun Xu,\IEEEauthorrefmark{3}
                Shengjian Guo, \IEEEauthorrefmark{2}
                Rundong Zhou, \IEEEauthorrefmark{2}
                Yulong Zhang, \IEEEauthorrefmark{2}
                Tao Wei, \IEEEauthorrefmark{2}
                Long Lu,\IEEEauthorrefmark{1}
        }
        \\
        \IEEEauthorblockA{
        \begin{tabular}{c c c}
                \IEEEauthorrefmark{1}Northeastern University    &
                \IEEEauthorrefmark{2}Baidu USA    &
                \IEEEauthorrefmark{3}Stevens Institute of Technology  \\
        \end{tabular}
        }
}

\maketitle

\input{abstract}

\input{intro}
\input{background}

\input{design}

\input{design2}

\input{impl}
\input{eval}

\input{related}

\input{conclude}
\input{ack}

\bibliographystyle{IEEEtranS}
\bibliography{ref}
\input{appendix}

\end{document}

%% file: abstract.tex
\begin{abstract}
Hybrid testing combines fuzz testing and concolic execution. It leverages fuzz
testing to test easy-to-reach code regions and uses concolic execution to
explore code blocks guarded by complex branch conditions. As a result, hybrid
testing is able to reach deeper into program state space than fuzz testing or
concolic execution alone.
Recently, hybrid testing has seen significant advancement. However, 
its code coverage-centric design is inefficient in vulnerability detection. 
First, it blindly selects seeds for concolic execution and aims to 
explore new code continuously. 
However, as statistics show, a large portion of the explored code is often
bug-free. Therefore, giving equal attention to every part of the code during
hybrid testing is a non-optimal strategy.  It slows down the detection of
real vulnerabilities by over {\bf  43\%}. 
Second, classic hybrid testing quickly moves on after reaching a chunk of code,
rather than examining the hidden defects inside. It may frequently miss subtle
 vulnerabilities despite that it has already explored the
vulnerable code paths.

We propose \savior, a new hybrid testing framework pioneering a
bug-driven principle. Unlike the existing hybrid testing tools, \savior
prioritizes the concolic execution of the seeds that are likely to uncover more
vulnerabilities. 
Moreover, \savior verifies all vulnerable program locations along the executing
program path. By modeling faulty situations using SMT constraints, \savior
reasons the feasibility of vulnerabilities and generates concrete test cases as
proofs.
Our evaluation shows that the bug-driven approach outperforms mainstream
automated testing techniques, including state-of-the-art hybrid testing systems driven
by code coverage. On average, \savior detects vulnerabilities \MOREBUGDRILLER
faster than \driller and \MOREBUGQSYM faster than \qsym, leading to 
the discovery of \BUGDIFFDRILLER  and \BUGDIFFQSYM more unique bugs, respectively. According to the evaluation on \BENCHNUM well fuzzed benchmark programs, within the first 24 hours, \savior triggers \TOTOALBUGNUM UBSAN violations, among which \REALBUGNUM are real bugs.

\end{abstract}

%% file: intro.tex
\section{Introduction} \label{sec:intro}

Software inevitably contains
defects~\cite{smullyan1992godel,TheMytho90:online}. A large amount of these
defects are security vulnerabilities that can be exploited for malicious
purposes~\cite{mu2018understanding}. This type of vulnerable code has become 
a fundamental threat against software security.  
Contributed from both academia and industry,
automated software testing techniques have gained remarkable advances in 
finding software vulnerabilities. In particular, people have widely used 
fuzz testing~\cite{takanen2008fuzzing, tool-afl} and concolic 
execution~\cite{Sen07a,majumdar2007hybrid} to disclose a great amount of 
vulnerabilities every year. Nevertheless, the inherent limitations of these
two techniques impede their further applications. On one hand, 
fuzz testing quickly tests a program, but it hardly explores code regions 
guarded by complex conditions. On the other hand, concolic execution 
excels at solving path conditions but it frequently directs the execution 
into code branches containing a large number of execution paths (\eg loop).
Due to these shortcomings, using fuzz testing or concolic execution alone 
often ends with large amounts of untested code after exhausting the time budget. 
To increase code coverage, recent works have experimented the idea of 
hybrid testing, which combines both fuzz testing and concolic execution
~\cite{qsyminsu,driller,jfuzz}.

The goal of hybrid testing is to utilize fuzzing in path exploration and 
leverage concolic execution to solve hard-to-resolve conditions. A hybrid 
approach typically lets fuzz testing run as much as possible. When 
the fuzzer barely makes any progress, the hybrid controller switches to the 
concolic executor which re-runs the generated seeds from fuzzing. During the 
run, the concolic executor checks each conditional branch to see whether its 
sibling branches remain untouched. If so, the concolic executor solves the 
constraints of the new branch and contributes a new seed for fuzzing. 
In general, this hybrid approach guides the fuzzer to new regions for deeper 
program space exploration.

As shown in recent works~\cite{driller,qsyminsu}, hybrid testing creates new 
opportunities for higher code coverage. However, its coverage-driven 
principle unfortunately results in inefficiency when the end goal is vulnerability detection. Two key issues cause such inefficiency. First, existing approaches value all
the seeds from fuzzing equally. However, the code regions reachable by a number 
of seeds might lack vulnerabilities and testing them is expensive 
(\eg constraint solving and extra fuzzing). Consequently, hybrid testing often 
exhausts the assigned time budget way before it finds any vulnerability. 
Second, hybrid testing could fail to identify a vulnerability even if it reaches
the vulnerable code via the correct path. This is because hybrid testing primarily 
concentrates on covering the encountered code blocks in the manner of random exercise. 
This strategy oftentimes has low chances to satisfy the subtle conditions to 
reveal a vulnerability.

\begin{figure*}[t]
    \begin{subfigure}[t]{0.4\textwidth}
        \centering
        \includegraphics[scale=0.69]{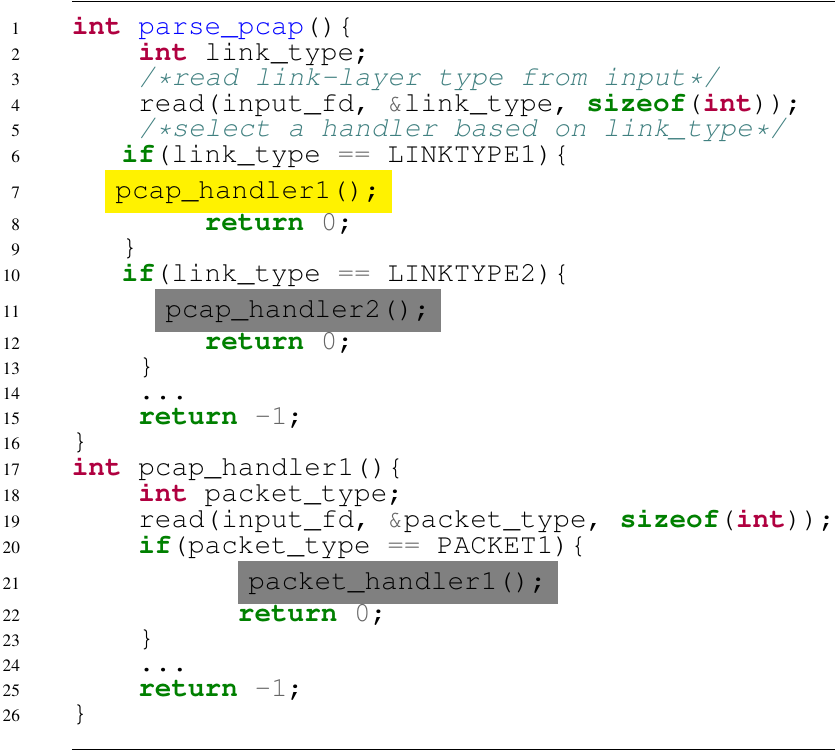}
        \caption{A simplified version of the packet-parsing code in {\tt tcpdump-4.9.2}, in which {\tt pcap\_handler2} contains vulnerabilities.} 
	\label{fuzz:code}
    \end{subfigure}~
    \begin{subfigure}[t]{0.28\textwidth}
        \centering
        \includegraphics[scale=0.3]{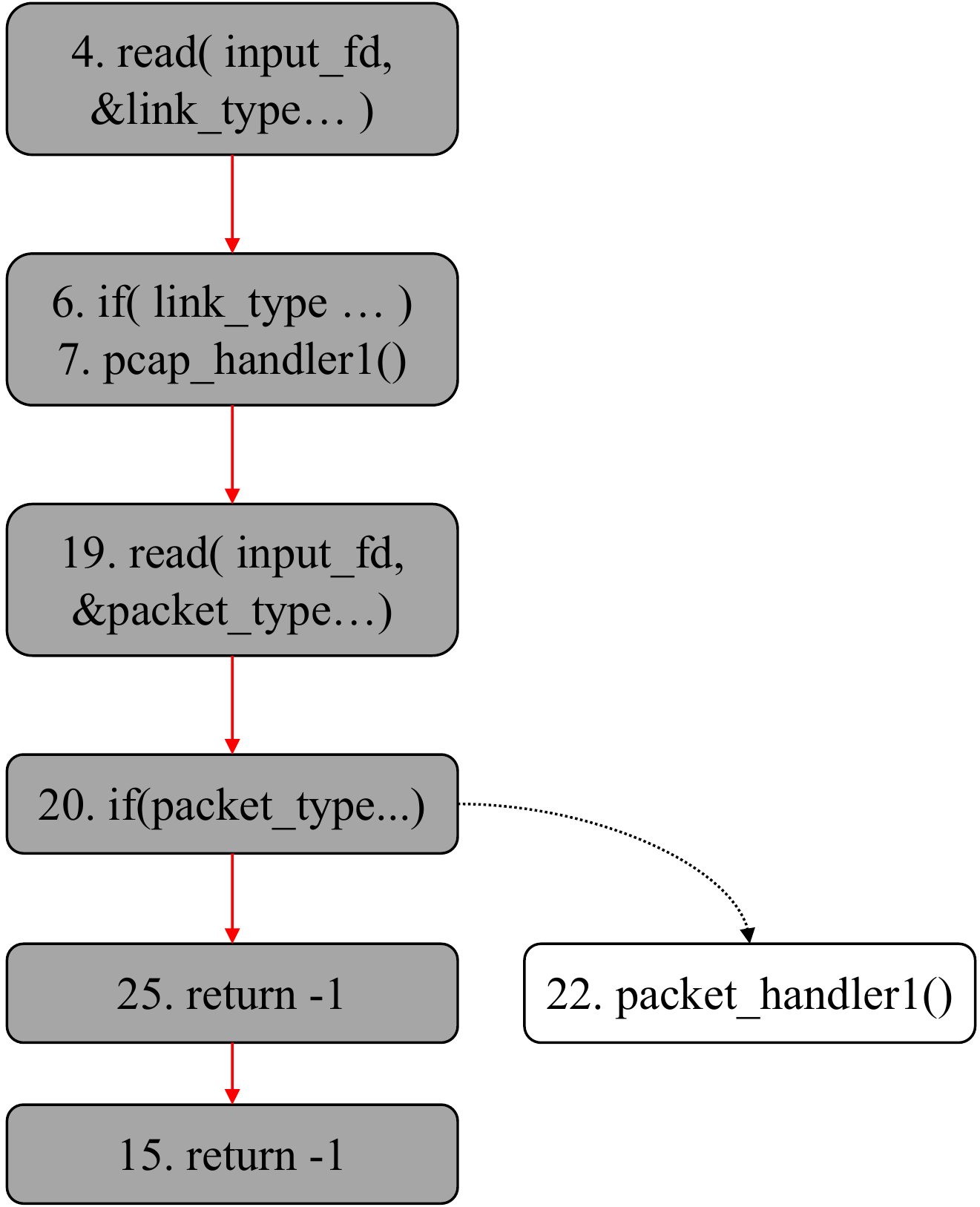}
        \caption{The path followed by a seed that matches {\tt LINKTYPE1} but mismatches {\tt PACKET1}.}
	\label{fuzz:seed1}
    \end{subfigure}~
    \begin{subfigure}[t]{0.28\textwidth}
        \centering
        \includegraphics[scale=0.3]{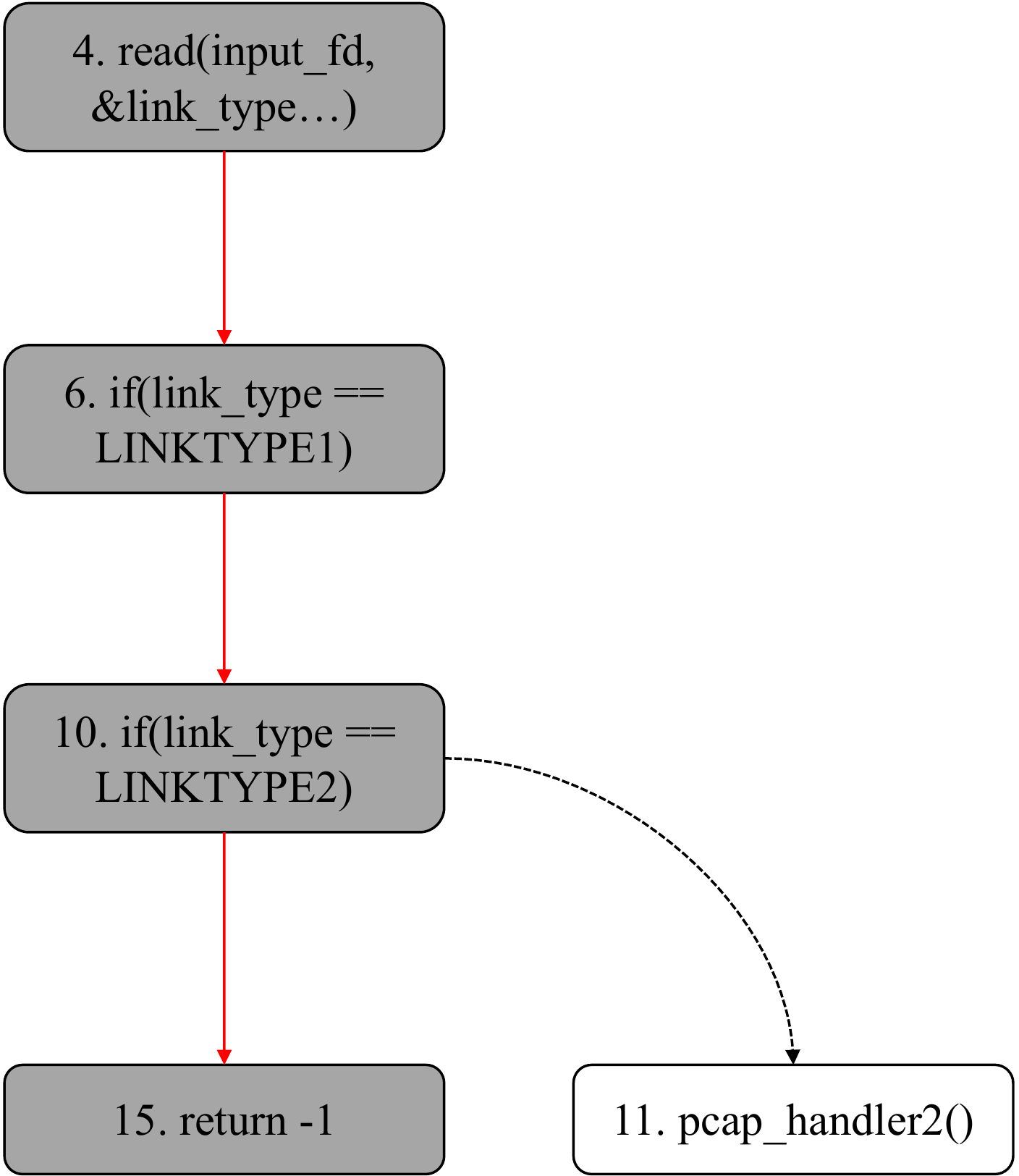}
        \caption{The path followed by a seed that matches neither {\tt LINKTYPE1} nor {\tt LINKTYPE2}.}
	\label{fuzz:seed2}
    \end{subfigure}
    \caption{A demonstrative example of hybrid testing. Figure~\ref{fuzz:code} presents the code under test. Figure~\ref{fuzz:seed1} and ~\ref{fuzz:seed2} are the paths followed by two seeds from the fuzzer. Their execution follows the red line and visits the grey boxes. Note that the white boxes connected by dotted lines are non-covered code.}
    \vspace{-2ex}
    \label{fig:hybrid}
\end{figure*}

In this work, we design and implement \savior (abbreviation for {\bf S}peedy-{\bf A}utomatic-{\bf V}ulnerability-{\bf I}ncentivized-{\bf OR}acle), a hybrid, {\em bug-driven}
testing method. To fulfill this goal, we use two novel techniques in 
\savior: 

\point{Bug-driven prioritization} Instead of running all seeds without distinction in concolic execution, \savior prioritizes those that have 
higher possibilities of leading to vulnerabilities. Specifically, 
before the testing, \savior analyzes the source code and statically labels the potentially vulnerable locations in 
the target program. Here \savior follows existing methods~\cite{ubsanlist, Dietz:2012} 
to conservatively label all suspicious locations. Moreover, \savior computes 
the set of basic blocks reachable from each branch. During dynamic 
testing, \savior prioritizes the concolic execution seeds that can visit more 
important branches (i.e., branches whose reachable code has 
more vulnerability labels). Intuitively, those branches may guard higher volumes 
of vulnerabilities and hence, prioritizing them could expedite 
the discovery of new vulnerabilities. As we will show in Section~\ref{sec:eval},
this prioritization enables \savior to outperform \driller~\cite{driller} 
and \qsym~\cite{qsyminsu} with a \MOREBUGDRILLER and \MOREBUGQSYM increase in 
bug discovering rate, respectively.

\point{\capbugguidedsearch} 
Aside from accelerating vulnerability detection, \savior also verifies the labeled 
vulnerabilities along the program path traversed by the concolic executor. 
Specifically, \savior synthesizes the faulty constraint of triggering each  vulnerability on the execution path. If such constraint under the current path condition is satisfiable, \savior solves the constraint to construct a test input 
as the proof. Otherwise, \savior 
proves that the vulnerability is infeasible on this path, regardless of the input. 
This SMT-solving based strategy, as demonstrated in Section~\ref{sec:eval}, 
enables \driller, \qsym, and \savior to disclose not only all the listed bugs but 
also an additional group of bugs in LAVA-M~\cite{lava}. Besides, 
it facilitates the three hybrid tools to find at least {\bf 22.2\%}, {\bf 25\%}, 
{\bf 4.5\%} more UBSan violations. 

This work is not the first one that applies hybrid testing to vulnerability detection. 
However, to the best of our knowledge, \savior is the first work that explores bug-driven hybrid 
testing. On one hand, \savior concentrates on software
code that contains more potential vulnerabilities. This design not only brings faster 
coverage of vulnerabilities but also decreases the testing cost of the code that is 
less likely vulnerable. On the other hand, \savior validates the vulnerabilities by 
the objective proofs of existence. In contrast, traditional hybrid testing methods can 
easily miss subtle cases. Moreover, the two proposed techniques are not limited 
to \savior itself since they are general enough for other systematic software analysis 
methods. We will discuss the details in Section~\ref{sec:design}. In summary, we make the 
following contributions.

\begin{itemize} 
\item We design \savior, a bug-driven hybrid testing technique. It substantially enhances 
hybrid testing with bug-driven prioritization and bug-guided verification. 

\item We build \savior and show that our implementation can scale to a diverse set of 
real-world software. 

\item We demonstrate the effectiveness of \savior by a comprehensive evaluation. In total, 
\savior discovers \TOTOALBUGNUM unique security violations in \BENCHNUM well-studied benchmarks. 
On average, \savior detects vulnerabilities \MOREBUGDRILLER
faster than \driller and \MOREBUGQSYM faster than \qsym, leading to 
the discovery of \BUGDIFFDRILLER and \BUGDIFFQSYM more security violations in 24 hours. 

\end{itemize}  

The rest of this paper is organized as follows. Section~\ref{sec:bcg} states the background 
of hybrid testing and motivates our research. Section~\ref{sec:design} and Section~\ref{sec:impl} 
present the design and implementation of \savior in detail. Section~\ref{sec:eval} evaluates the
core techniques of \savior. Section~\ref{sec:related} summarizes the related work. 
Finally, we conclude this work in Section~\ref{sec:conclude}.

%% file: background.tex
\section{Background and Motivation}
\label{sec:bcg}
This work is motivated by the limitations of hybrid testing in vulnerability detection. 
In this section, we first introduce the background of hybrid testing and then demonstrate 
the limitations by two examples. 

\subsection{Hybrid Testing}
\label{sec:hybrid}
Hybrid testing combines fuzz testing and concolic execution to achieve high code coverage. 
For the ease of understanding, we use the example in Figure~\ref{fig:hybrid} to explain 
how it works. The explanation is based on Driller~\cite{driller} since it has been the de 
facto implementation of hybrid testing.

The example in Figure~\ref{fig:hybrid} is taken from {\tt tcpdump-4.9.2}. Figure~\ref{fuzz:code} 
shows the code --- it first uses the \emph{link-layer} type from input to select a pcap 
handler and then uses the handler to dissect packets. Our objective is to test the entry 
function {\tt parse\_pcap} and reach the vulnerable function {\tt pcap\_handler2}. 

In the test, we assume hybrid testing starts with a seed that executes the path shown in 
Figure~\ref{fuzz:seed1}. After that, the fuzzer mutates the seed to run a second path shown in 
Figure~\ref{fuzz:seed2}. It then, however, fails to synthesize inputs that match the 
\emph{packet} type at line {\tt 20} and the \emph{link-layer} type at line {\tt 10}, due to 
the huge mutation space ({\tt $2^{32}$} possibilities). This situation prevents the fuzzer 
from testing the remaining code and makes  hybrid testing switch to concolic execution. 

After executing the seed that covers the path in Figure~\ref{fuzz:seed1}, 
the concolic executor backtracks to the branch statement at line {\tt 20}. 
Solving the input {\tt packet\_type} to 
{\tt PACKET1} by a SMT solver, the executor generates a new seed to cover that branch. Then, 
the hybrid controller suspends the concolic execution and resumes the fuzzer. Guided by the new 
seed, the fuzzer tests {\tt packet\_handler1} and switches back to concolic execution after that. 
This time, the concolic executor runs the seed, following the path in Figure~\ref{fuzz:seed2}.  
After solving the branch condition at line {\tt 10}, it generates a seed for the flow from line 
{\tt 10} to line {\tt 11}. Further fuzz testing can finally reach the vulnerable code in 
{\tt pcap\_handler2}.

Note that the testing processes by different hybrid tools may vary from the above 
description. For instance, QSYM~\cite{qsyminsu} keeps running concolic execution instead of 
invoking it in an interleaved manner. Despite those implementation differences, existing 
tools share a similar philosophy on scheduling the seeds to concolic execution. That is, they 
treat the seeds indiscriminately~\cite{qsyminsu,driller}, presumably assuming that these seeds 
have equal potentials in contributing to new coverage. 

\subsection{Motivation} 
\label{sec:motivation}
\point{Inefficiency in Covering Vulnerable Code} 
Although hybrid testing specializes in coverage-driven testing, it still needs substantial time 
to saturate hard-to-reach code compartments, which often overspends the time budget. To discover more 
vulnerabilities in a limited time frame, an intuitive way is to prioritize the testing of 
vulnerable code. However, the current hybrid testing method introduced in Section~\ref{sec:hybrid} 
does not meet this requirement.

Consider the example in Figure~\ref{fig:hybrid}, where concolic execution chronologically 
runs the seeds to explore the paths shown in Figure~\ref{fuzz:seed1} and Figure~\ref{fuzz:seed2}. 
This sequence indeed postpones the testing of the vulnerable function {\tt pcap\_handler2}. 
The delay can be significant, because concolic execution runs slowly and the fuzz testing on 
{\tt packet\_handler1} may last a long time. In our experiments\footnote{\savior is customized 
to do this test since \driller cannot run on {\tt tcpdump}. More details can be found in Section
~\ref{sec:eval}}, \driller spends minutes on reaching {\tt pcap\_handler2} with the aforementioned 
schedule. However, if it performs concolic execution first on the path in Figure~\ref{fuzz:seed2}, 
the time can reduce to seconds.

Not surprisingly, the delayed situations frequently happen in practice. As we will show in 
Section~\ref{sec:eval}, on average this defers \driller and \qsym to cover vulnerabilities
by \MOREBUGDRILLER and {\bf 44.3\%}, 
leading to reduced efficiency in vulnerability finding.

\begin{figure}[t!]
    \centering
    \includegraphics[scale=0.92]{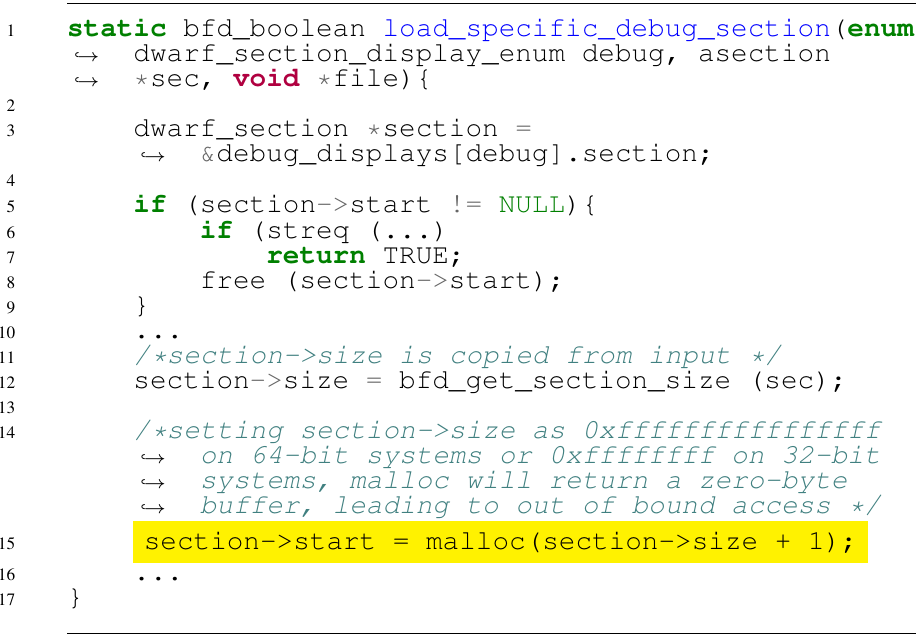}
    \caption{A demonstrative example of limitation in finding defects by existing hybrid testing. This defect comes from {\tt objdump-2.29}~\cite{sourcewa0:online}.}
    \vspace{-2ex}
    \label{fig:objbug}
\end{figure}

\point{ Deficiency in Vulnerability Detection} Hybrid testing often fails to identify 
a vulnerability even if it approaches the vulnerable location along the right path. Figure
~\ref{fig:objbug} demonstrates an integer overflow in {\tt objdump-2.29}. At line {\tt 12}, 
the program copies a value from {\tt sec} to {\tt section$\to$size}. Next, this value is used 
as the size of a memory allocation request at line {\tt 15}. By carefully handcrafting the 
input, an adversary can make {\tt section$\to$size} be the value {\tt $2^{32}$-1} on 32-bit 
systems or {\tt $2^{64}$-1} on 64-bit systems. This wraps {\tt section$\to$size+1} around to 
{\tt 0} and makes {\tt malloc} return a zero-byte buffer. When the buffer is further used, a segfault or a memory leak would occur. 

In this example, hybrid testing can quickly generate a seed to hit line {\tt 15}. However, it 
could barely trigger the integer overflow. As the program enforces no constraints on the input 
bytes that propagate to {\tt section$\to$size}, hybrid testing can only do random mutation to 
synthesize the extreme value(s). Taking into account the tremendous possibility space 
({\tt $2^{32}$} or {\tt $2^{64}$}), the mutation is unlikely to succeed.

%% file: design.tex
\section{Design}
\label{sec:design}

\subsection{Core Techniques}
\label{sec:overview}

The design of \savior is bug-driven, aiming to find bugs faster 
and more thoroughly. We propose two techniques to achieve the goal: \emph{bug-driven  
prioritization} and \emph{\bugguidedsearch}. Below we present an overview of 
our techniques. 
\point{Bug-driven prioritization} 
Recall that classic hybrid testing blindly schedules the seeds for concolic execution, 
without weighing their bug-detecting potentials. This can greatly defer the discovery 
of vulnerabilities. To remedy this limitation, \savior collects information from the target source code to prioritize seeds which have higher 
potentials to trigger vulnerabilities. This approach, however, needs to predict the amount 
of vulnerabilities that running concolic execution on a seed could expose. The prediction 
essentially depends on two prerequisites: \textbf{R1} -- \emph{A method to assess the 
reachable code regions after the concolic execution on a seed} and \textbf{R2} -- 
\emph{A metric to quantify the amount of vulnerabilities in a chunk of code}. \savior 
fulfills them as follows.  

To meet \textbf{R1}, \savior approximates the newly explorable code regions based on 
a combination of static and dynamic analysis. During compilation, \savior statically 
computes the set of reachable basic blocks from each branch. At run-time, 
\savior identifies the unexplored branches on the execution path of a seed and calculates 
the basic blocks that are reachable from those branches. We deem that these blocks become 
explorable code regions once the concolic executor runs that seed. 

To meet \textbf{R2}, \savior utilizes UBSan~\cite{ubsanlist} to annotate three types of 
potential bugs (as shown in Table~\ref{tab:ubsan-info}) in the program under testing. It 
then calculates the UBSan labels in each code region as the quantitative metric for 
\textbf{R2}. As UBSan's conservative instrumentation may generate dummy labels, \savior 
incorporates a static filter to safely remove useless labels. We discuss the details of 
this method in Section~\ref{subsec:compiling}.

\begin{figure}[t!]
    \centering
    \includegraphics[scale=0.42]{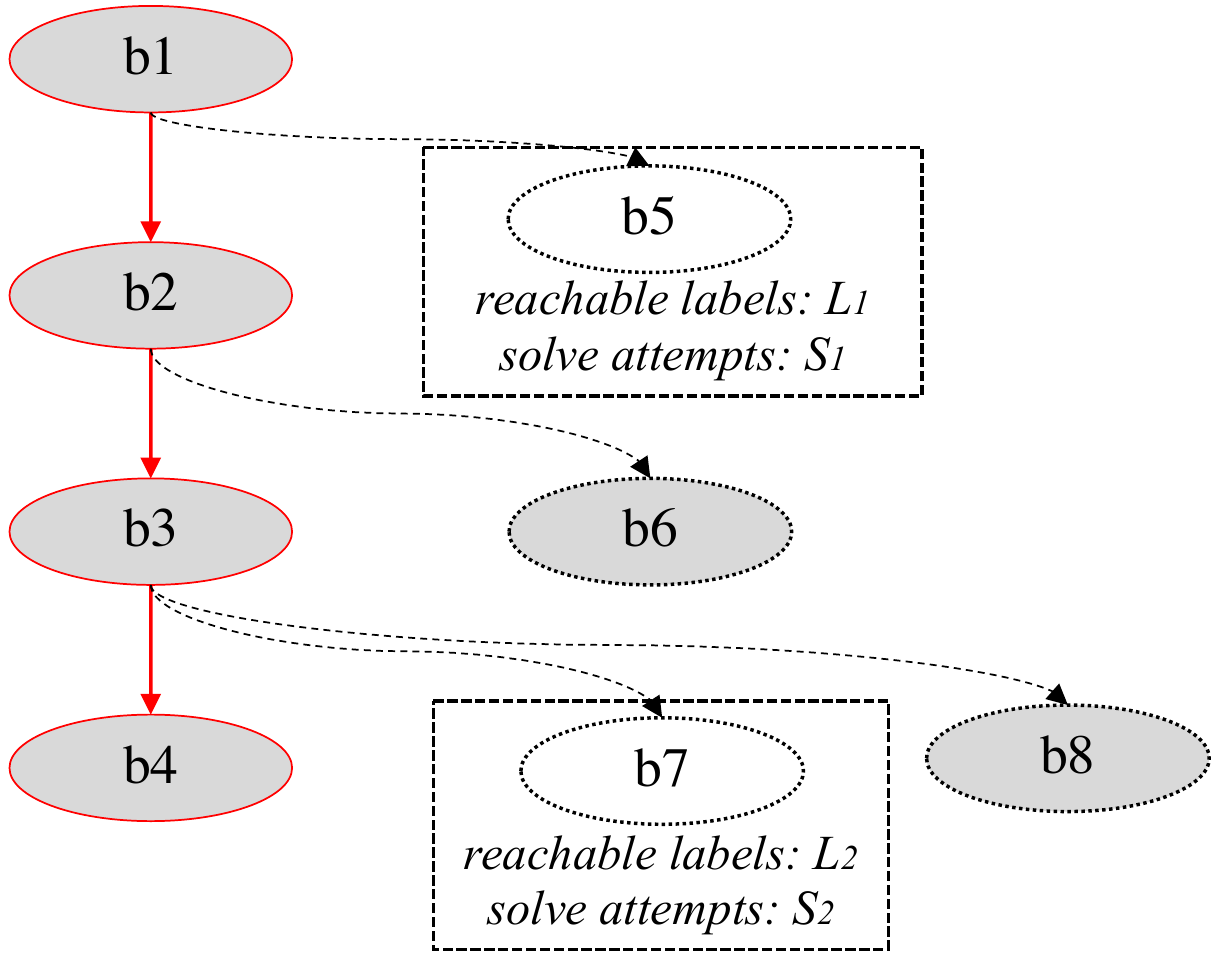}
    \caption{An example showing how to estimate the bug-detecting potential of a seed. 
		In this example, the seed follows the path {\tt b1->b2->b3->b4}. Basic block {\tt b5} and {\tt b7} are unexplored and they can reach $L_{1}$ and $L_{2}$ UBSan labels, respectively. They have been attempted by constraint solving for $S_{1}$ and $S_{2}$ times. The final score for this seed is $\frac{e^{-0.05S_{1}} \times L_{1}+e^{-0.05S_{2}} \times L_{2}}{2}$.}
    \vspace{-2ex}
    \label{fig:seedscore}
\end{figure}

The above two solutions together ensure a sound analysis for identifying potential bugs. First, 
our static reachability analysis, as described in Section~\ref{subsec:compiling}, is
built upon a sound algorithm. It over-approximates all the code regions that may be 
reached from a branch. Moreover, UBSan adopts a conservative design, which counts all 
the operations that may lead to the undefined behavior issues listed in Table~\ref{tab:ubsan-info}~\cite{Dietz:2012, ubsanlist}. 
Facilitated by the two aspects of soundness, we can avoid mistakenly underrating the 
bug-detecting potential of a seed. 

Following the two solutions, \savior computes the importance score for each seed as
follows. Given a seed with {\tt n} unexplored branches $\{e_{1}, e_{2}, \dots, e_{n}\}$, 
\savior calculates the UBSan labels in the code that are reachable from these branches, 
respectively denoted as $\{L_{1}, L_{2}, \dots, L_{n}\}$. Also note that, in the 
course of testing, \savior has made $\{S_{1}, S_{2}, \dots, S_{n}\}$ 
attempts to solve those branches. With these pieces of information, \savior evaluates the importance score of this seed with a weighted average 
$\frac{1}{n} \times \sum_{i=1}^n{e^{-0.05S_{i}}} \times L_{i}$. 
$L_{i}$ represents the potential of the $i_{th}$ unexplored branch. We 
penalize $L_{i}$ with $e^{-0.05S_{i}}$ to monotonically decrease its weight as the 
attempts to solve this branch grow. The rationale is that more failed attempts (usually 
from multiple paths) indicate a low success possibility on resolving the branch. Hence, 
we decrease its potential so that \savior can gradually de-prioritize hard-to-solve 
branches. Lastly, \savior takes the average score of each candidate branches in order 
to maximize the bug detection gain per unit of time. To better understand this scoring method, 
we show an example and explain the score calculation in Figure~\ref{fig:seedscore}. 

This scoring method is to ensure that \savior always prioritizes seeds leading to more
unverified bugs, while in the long run it would not trap into those with hard-to-solve 
branch conditions. First, it conservatively assesses a given seed by the results of  
sound reachability and bug labeling analysis. A seed which leads to more unexplored 
branches where more unverified bugs can be reached from will earn a higher score.
Second, it takes into account runtime information to continuously improve the precision 
of the assessment. This online refinement is important because statically \savior may hardly 
know whether a branch condition is satisfiable or not. Utilizing the history of constraint 
solving attempts, \savior can decide whether a seemingly high-score branch is worth more 
resources in the future. As shown by our evaluation in Section~\ref{sec:eval}, this 
scoring scheme significantly accelerates the detection of UBSan violations, which empirically 
supports the effectiveness of our design. 

Referring to our motivating example in Figure~\ref{fig:hybrid}, the function 
{\tt packet\_handler1} has few UBSan labels while {\tt pcap\_handler2} contains 
hundreds of labels. Hence, the seed following Figure~\ref{fuzz:seed1} has a 
lower score compared to the other seed which runs the path in 
Figure~\ref{fuzz:seed2}. This
guides \savior to prioritize the latter seed, which can significantly expedite the exploration of vulnerable code.


\begin{figure}[t!]
    \centering
    \includegraphics[scale=0.483]{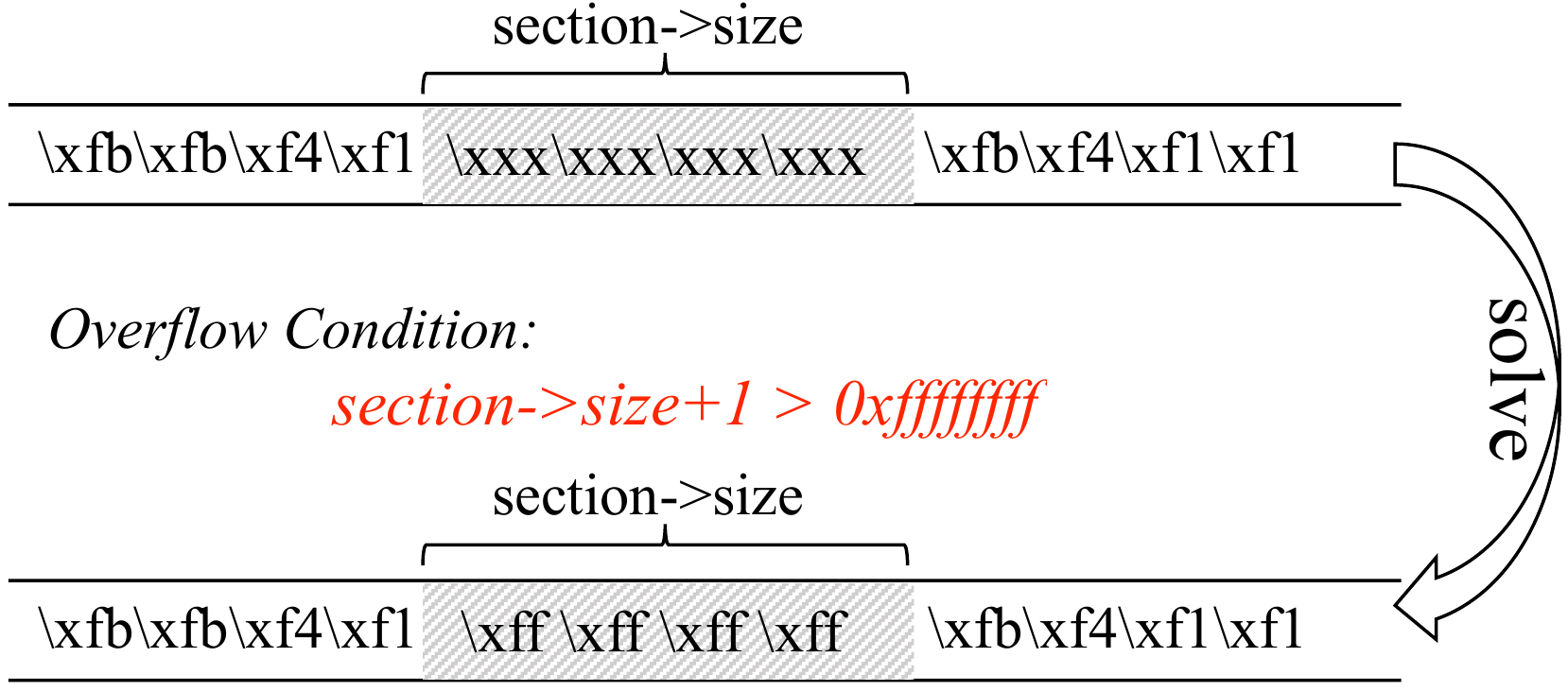}
    \caption{Solving the integer overflow in Figure~\ref{fig:objbug}. This shows the case 
		in a 32-bit system, but it applies to 64-bit as well.}
    \vspace{-2ex}
    \label{fig:solveobj}
\end{figure}

\point{\capbugguidedsearch} 
This technique also ensures a sound vulnerability detection on the explored paths 
that reach the vulnerable sites. Given a seed from fuzz testing, \savior executes
it and extracts the label of each vulnerability along the execution path. After 
that, \savior verifies the predicates implanted in each label by checking the 
satisfiability under the current path condition --- if the predicate is satisfiable 
then its corresponding vulnerability is valid.
This enables \savior to generate a proof of either vulnerability or non-existence
along a specific program path. Note that in concolic execution, many new states
with new branch constraints will be created. \savior will prioritize the constraint
solving for states who require \bugguidedsearch.

Going back to the example in Figure~\ref{fig:objbug}, classic hybrid testing misses 
the integer overflow at line {\tt 15}. In contrast, \savior is able to identify it 
with \bugguidedsearch. Aided by the Clang sanitizer~\cite{ubsanlist}, \savior instruments 
the potential overflows in a solver-friendly way (\ie the predicate of triggering 
this overflow is {\tt section->size + 1 > 0xffffffff}). Due to the limited space, 
we present the instrumented IR code in Figure~\ref{fig:bug:objllvm} at 
Appendix~\ref{sec:appendix1}.
As demonstrated in Figure~\ref{fig:solveobj}, following a seed to reach the integer 
overflow location, \savior tracks that the value of {\tt section->size} relies on a 
four-byte field in the input. By solving the vulnerability predicate, \savior generates 
a witness value {\tt 0xffffffff} and triggers the vulnerability.

%% file: design2.tex
\begin{figure}[t!]
    \centering
    \includegraphics[scale=0.36]{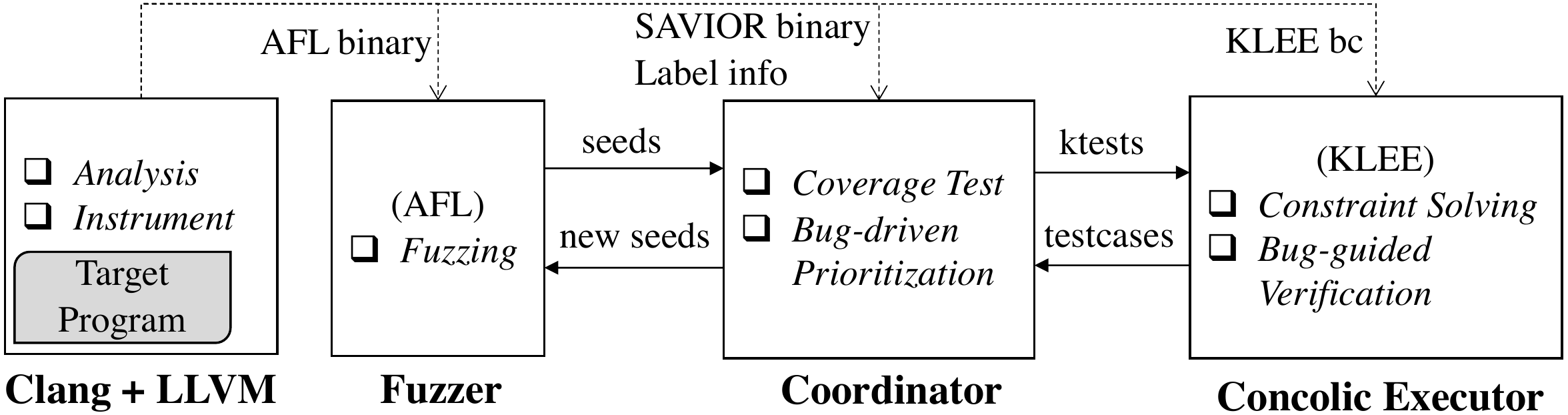}
    \caption{System architecture of \savior.}
    \vspace{-2ex}
    \label{fig:system}
\end{figure}

\subsection{System Design}
\label{subsec:system}
Figure~\ref{fig:system} depicts the overall architecture of \savior. It consists 
of a compiling tool-chain built upon Clang and LLVM, a fuzzer derived from \afl, 
a concolic executor ported from KLEE, and a hybrid coordinator responsible for 
the orchestration. We explain these components in details in the following 
sections.
\newline
\subsubsection{The Compilation Tool-chain} 
\label{subsec:compiling}
\savior's compilation tool-chain has multiple purposes including vulnerability 
labeling, control flow reachability analysis, and the targets building of 
different components. 

\point{Sound Vulnerability Labeling} In our design, we use Clang's Undefined 
Behavior Sanitizer (UBSan)~\cite{ubsanlist} to label different families of 
potential bugs\footnote{Clang supports enabling checks on each individual bug 
family.}. Table~\ref{tab:ubsan-info} summarizes those families used in \savior
and the operations pertaining to them. 

We ignore other bug types listed in UBSan (\eg misaligned reference) since they 
are less likely to cause security issues. For each inserted label, we patch the 
Clang front-end to attach a {\tt !saviorBugNum} metadata, aiding the reachability 
analysis that we will shortly discuss.

As explained in Section~\ref{sec:overview}, UBSan over-approximates the potential 
vulnerabilities. This approximation ensures soundness since it never misses true 
bugs. UBSan also models the conditional triggers of the labeled bugs as shown in 
Table~\ref{tab:ubsan-info}. E.g., \emph{out-of-bound} (OOB) array access happens when 
the index {\tt x} is not between \emph{zero} and \emph{array size minus 1}. At 
the time of \bugguidedsearch, \savior solves each triggering condition to produce 
a witness of the bug or, prove that the bug never happens on current path in terms
of the unsatisfiable condition. In Figure~\ref{fig:bug:objllvm} at Appendix
~\ref{sec:appendix1}, we present the IR with instrumented UBSan checks for the 
defect shown in Figure~\ref{fig:objbug}.

\savior uses UBSan by default, while other labeling methods may also apply if they 
meet the following two properties. First, they can comprehensively annotate the 
potential vulnerabilities. Second, they can synthesize the triggering condition 
of each labeled vulnerability. Note that such condition must have data dependency 
on the program input. Otherwise, our concolic execution cannot correlate the input 
with the vulnerable conditions and hence, has no guidance for \bugguidedsearch. 
For instance, the AddressSanitizer~\cite{serebryany2012addresssanitizer} builds 
checks upon the status of its own red-zone, which is not applicable to \savior 
at the moment.

UBSan's conservative approximation inevitably introduces false positives and 
might mislead \savior's prioritization. In practice, we incorporate a static 
counter-measure to reduce fake labels. Specifically, we trim a label when all 
the following requirements hold:
\emph{1)} The label's parent (basic block) is its immediate dominator~\cite{sreedhar1995incremental}; 
\emph{2)} The IR variables involved in the vulnerability conditions are not 
re-defined between the label and its parent;
\emph{3)} The parent basic block has constraints that conflict with the 
vulnerability conditions, and these constraints are enforced by constant values. 
The first two points ensure that the constraints added by the parent will 
persist upon reaching the label, and the third point indicates that the 
conflict always arises, regardless of the input and the execution path. 
Therefore, we can safely remove this label. 
\begin{lstlisting}
char array[MAX]; // 0 < MAX < INT_MAX
for(int i = 0; i < MAX;){
    array[i] = getchar();//LABEL: OOB access 
    i++;//LABEL: integer-overflow
}
\end{lstlisting}
For instance, the code above has two labels that meet the three requirements.
In this example, the variable {\tt i} ranges from {\tt 0} to {\tt MAX}, meaning 
that neither the array access at line {\tt 3} can be \emph{out-of-bound} nor the 
self increment at line {\tt 4} can cause an integer overflow. \savior hence 
removes the two labels. In Table~\ref{tab:labelremove} at Appendix~\ref{sec:appendix1}, 
we summarize the number of labels that are removed from each of our benchmark programs. 
On average, we can conservatively reduce 5.36\% of the labels. 

\definecolor{mygray}{gray}{0.9}

\begin{table}[t!]

\centering
\scriptsize
\begin{tabular}{l|cc}
\toprule[0.5pt]
\toprule[0.5pt]

\multirow{2}{*}{\bf{\emph{UB Families}}} & \multicolumn{2}{c}{\bf{\emph{UBSan Labeling Details}}}   
 \\ \cline{2-3}
 & {\tt Operation} & {\tt Condition} 
 \\ \hline

\rowcolor{mygray}
Out-of-bound array access  & $array[x]$  &  $x < 0 \lor x \ge size(array)$\\
Oversized shift & $x \ll y, x \gg y$ & $ y < 0 \lor y \ge n$  \\
\rowcolor{mygray}
Signed integer overflow  & $x$ $op_{s}$ $y$ & $x$ $op_{s}$ $y$ $\notin [-2^{n-1}, 2^{
n-1} - 1]$ \\
Unsigned integer overflow  & $x$ $op_{u}$ $y$ & $x$ $op_{u}$ $y$ $>$ $2^{n}-1$\\
\bottomrule[0.5pt]
\bottomrule[0.5pt]
\end{tabular}
\caption{Families of potential bugs that \savior enables UBSan to label. 
Here, $x$, $y$ are $n$-bit integers; $array$ is an array, the size of which is specified 
as $size(array)$; $op_{s}$ and $op_{u}$ refers to binary operators $+,-,\times, \div, \%$ 
over signed and unsigned integers, respectively.}
\label{tab:ubsan-info}
\vspace{-2ex}
\end{table}

\point{Reachability Analysis} This analysis counts the number of vulnerability labels that 
can be forwardly reached by each basic block in the program control flow graph (CFG). It 
proceeds with two phases. The first step constructs an inter-procedure CFG. The construction 
algorithm is close to the method implemented in SVF~\cite{sui2016svf}. It individually builds 
intra-procedure CFGs for each function and then bridges function-level CFGs by the caller-callee 
relation. To resolve indirect calls, our algorithm iteratively performs Andersen's point-to 
analysis and expands the targets of the calls. This prevents \savior from discarding aliasing 
information of indirect calls and therefore, our prioritization would not miscount the number 
of vulnerability labels. By examining the CFGs, we also extract the edge relations between a 
basic block and its children for further use in the hybrid coordinator.

\begin{wrapfigure}{r}{0.24\textwidth}
  \begin{center}
    \includegraphics[width=0.24\textwidth]{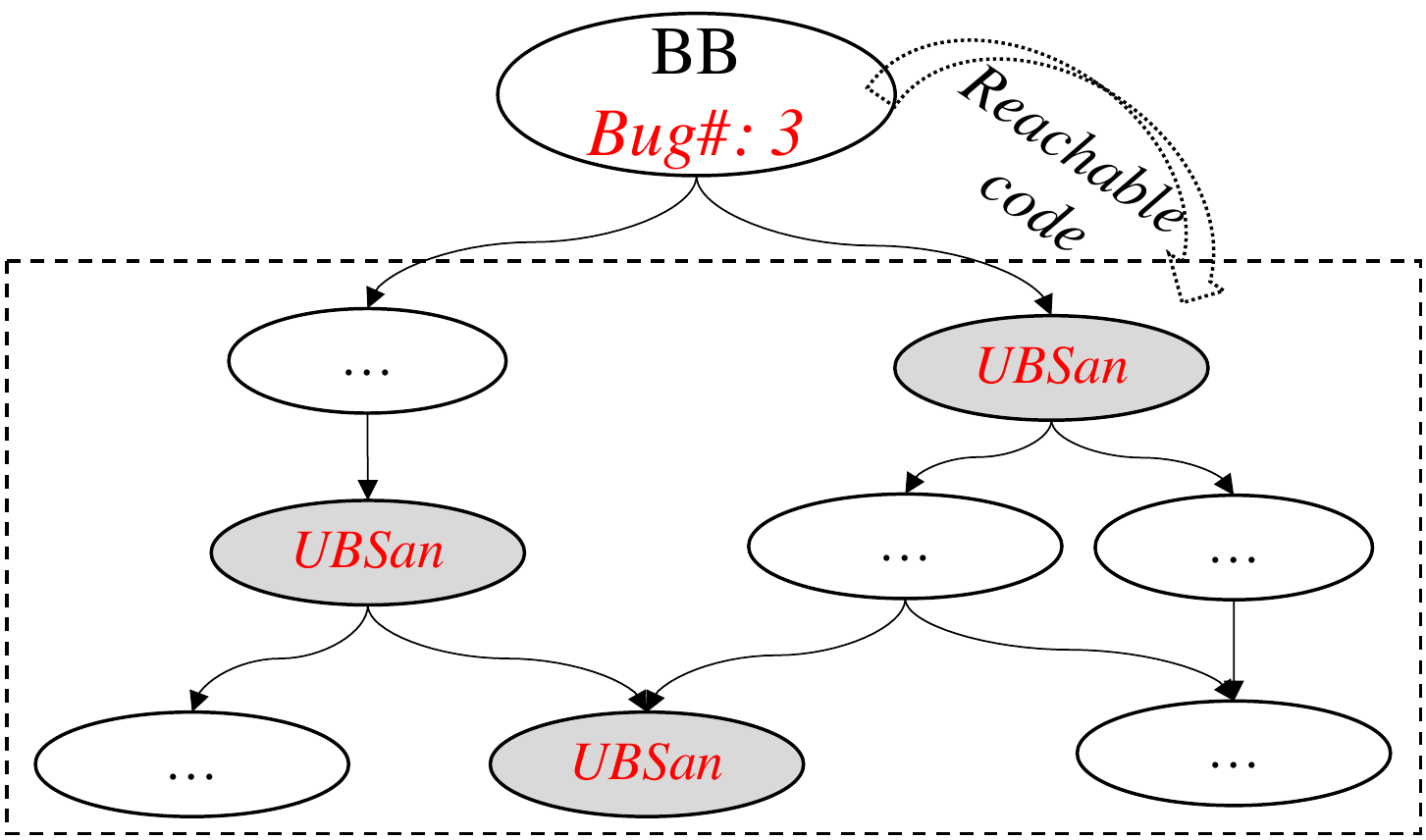}
  \end{center}
  \caption{A demonstrative example of reachability analysis. The target {\bf BB} can ``reach'' 3 UBSan labels.}
  \vspace{-2ex}
  \label{fig:reachability}
\end{wrapfigure}


The second step is to calculate the UBSan labels that are reachable from each basic block
in the constructed inter-procedure CFG. Specifically, we identify the regions of code that 
a basic block can reach and count the number of UBSan labels in those regions. In \savior, 
we deem this number as the importance metric of that basic block and use it for bug-driven 
prioritization. For example, in Figure~\ref{fig:reachability} the basic block {\bf BB} can
reach 8 other basic blocks while 3 of them have \emph{UBSan} labels. Thereby we output 3 as 
the number of reachable UBSan labels for {\bf BB}. Note that each basic block at most has one 
label after Clang's compilation.



\point{Target Building} After the labeling and the reachability analysis, \savior's compiling 
tool-chain begins its building process. It compiles three binaries from the source code --- 
a fuzzing-binary for the fuzzer, a \savior-binary for the coordinator, and a LLVM bitcode 
file for the concolic executor. In particular, the \savior-binary is instrumented to print 
the unique IDs of the executed basic blocks. With this design, \savior completely decouples 
the fuzzer, the concolic executor and the coordinator, thus it supports quick replacement 
of any components.
\newline
\subsubsection{The Coordinator} The coordinator bridges the fuzzer and the concolic executor. 
It keeps polling seeds from the fuzzer's queue and prioritizes those with higher importance 
for concolic execution. We explain the details as follows. 

\point{Bug-driven Prioritization} 
In a polling round, the coordinator operates the new seeds in the fuzzer's queue after last 
round. Each seed is fed to the \savior-binary and the coordinator updates two pieces of 
information based on the execution result. First, it updates the global coverage information. 
The coverage computation here follows \afl's original approach. That is,  we take the hit 
counts of an edge in the following ranges as different coverage: 
$[1], [2], [3], [4, 7], [8, 15], [16, 31], [32, 127], [128,\infty)$. 
Second, the coordinator records the sequence of basic blocks visited by each seed. 
Using the updated coverage information, the coordinator assigns a score to each seed following 
the scheme presented in Section~\ref{sec:overview}. Here, we re-score all the seeds except those 
already tested by our concolic executor, since the coverage information is dynamically adjusted. 

Finally, the coordinator selected the top-ranked seeds and feed them into the input queue of the 
concolic executor. If two seeds have the same score, the coordinator prefers the seed with the 
{\tt +cov} property. {\tt +cov} indicates that the seed brings new code coverage.

\point{Post-processing of Concolic Execution} Going beyond seed scheduling for concolic execution, 
the coordinator also need to triage the new seeds generated by the concolic executor for the fuzzer. 
First, it re-runs the new seeds and retains those who provide new coverage or can reach uncovered 
bug labels. As a result, \savior transfers the valuable test cases from the concolic executor to 
the fuzzer. 

Second, the coordinator updates the number of solving attempts upon uncovered branches. If a branch 
remains uncovered, its solving attempts would be increased by 1. As such, a branch having a much 
higher solving attempt value will be de-prioritized. 
\newline
\subsubsection{The Concolic Executor} The concolic executor replays the seeds scheduled by the 
coordinator and chooses to solve branch conditions based on coverage information. In addition,
it also performs \bugguidedsearch.

\point{Independent Coverage Scheme} When encountering a branch instruction the concolic executor 
needs to decide whether to solve this branch's condition. An intuitive design is to reuse the 
coverage information from the coordinator. However, since our coverage scheme is ID based, yet 
as KLEE invokes a group of transformations on the target bitcode, this leads to numerous mismatches 
between the edge IDs in the \savior-binary and the KLEE bitcode. To tackle this problem, we opt 
to use KLEE's internal coverage information to better decouple the concolic executor and other 
components. 


\begin{figure}[t!]
    \centering
    \includegraphics[scale=0.55]{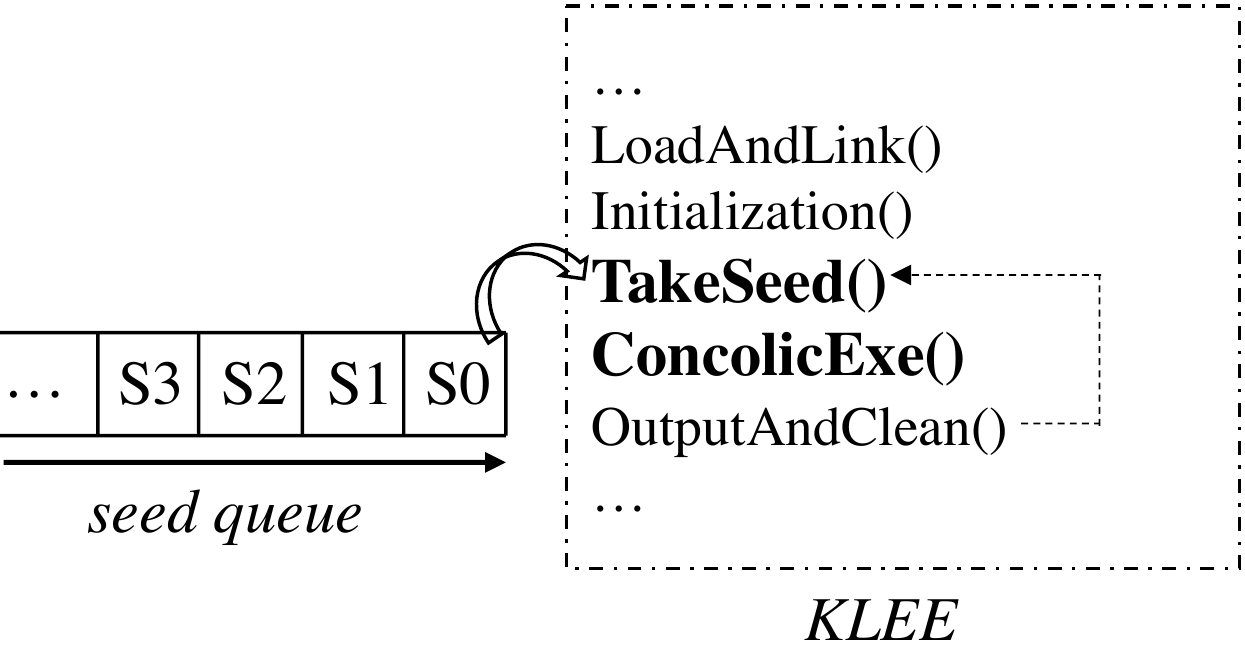}
    \caption{Fork server mode in KLEE. In this mode, KLEE only performs initialization once 
    and reuses the same executor for all the received seeds.}
    \vspace{-2ex}
    \label{fig:forkserver}
\end{figure}

\point{Fork Server Mode} Before running a seed, KLEE needs to perform a group of initialization, 
including bitcode loading, library bitcode linking, and global data preparation to place the program under
testing into the virtual machine. This initialization process, however, typically takes a long time 
on large bitcode files. For instance, the initialization time for {\tt tcpdump} is usually several 
times longer than the actual concolic execution time. To address this issue, we introduce an optimization named 
\emph{fork server} mode for the KLEE concolic executor (as shown in Figure~\ref{fig:forkserver}).
Technical details are explained in Section~\ref{sec:impl}.


\point{Bug-guided Verification} Our concolic executor also performs \bugguidedsearch. Once an 
non-covered vulnerability label is reached, we endeavor to solve the triggering constraint following the current path. If the solving succeeds, KLEE generates a seed as the proof 
of the vulnerability. 

In certain cases, the path constraints may conflict with the vulnerability triggering conditions, 
while that vulnerability can indeed happen following the same path (with fewer constraints). \qsym
~\cite{qsyminsu} summarizes this issue as the over-constraint problem. We adopt \qsym's optimistic 
solving strategy only on solving the vulnerability conditions. However, the relaxed-constraint may 
also produce a false positive, and we do not count a vulnerable label as being covered through 
relaxed-constraint solving.

\point{Timeout on Concolic Execution} To prevent the concolic execution from hanging on localized 
code regions (\eg, deep loops and blocking IO), the concolic executor usually needs a time threshold while running a seed. \qsym adjusts this timing budget  by watching \afl's 
status. If the number of hanging seeds increases, \qsym increases the timeout 
(up to {\tt 10} minutes). 
We set the timeout to be proportional to the number of uncovered branches that a 
seed can reach. The rationale is that those seeds need more time for constraint solving and such 
setting benefits higher bug coverage.

%% file: impl.tex
\section{Implementation}
\label{sec:impl}

\definecolor{mygray}{gray}{0.9}

\begin{table*}[t!]

\centering
\begin{tabular}{l|c|c|c}
\toprule[0.5pt]
\toprule[0.5pt]

\multirow{2}{*}{\bf{\emph{Fuzzers}}} & \multicolumn{3}{c}{\bf{\emph{Setup}}}   
 \\ \cline{2-4}
 & {\tt Source} & {\tt Instances} & {\tt Note} 
 \\ \hline

\rowcolor{mygray}
\afl  & ~\cite{tool-afl} & 1 AFL master; 2 AFL slaves & N/A \\
\aflgo  &  ~\cite{tool-aflgo} & 1 AFLGo master; 2 AFLGo slaves & Use in-lined {\tt lava\_get} as target locations of guided fuzzing \\
\rowcolor{mygray}
\tfuzz  & ~\cite{tool-tfuzz} & 3 AFL jobs (adjust default argument to Fuzzer) & Use the docker environment prepared at ~\cite{tool-tfuzz} for evaluation \\
\angora  &  ~\cite{tool-angora} & 3 Angora threads (with option {\tt "-j 3"}) & Patch Lava to support Angora, as suggested by the developers~\cite{tool-angora1} \\
\rowcolor{mygray}
\driller  & Self-developed &  1 concolic executor; 1 AFL master; 1 AFL slave & Follow the original {\tt Driller} in scheduling concolic execution~\cite{tool-driller1}  \\
\qsym  &  ~\cite{tool-qsym} &  1 concolic executor; 1 AFL master; 1 AFL slave & N/A \\
\rowcolor{mygray}
\savior  & Self-developed &  1 concolic executor; 1 AFL master; 1 AFL slave & Use in-lined {\tt lava\_get} as labels of vulnerabilities \\

\bottomrule[0.5pt]
\bottomrule[0.5pt]
\end{tabular}
\caption{Fuzzer specific settings in evaluation with Lava-M.}
\label{tab:lava-setup}
\vspace{-2ex}
\end{table*}


We have implemented \savior, which can be applied to software as sophisticated as 
Baidu's Apollo Autonomous Driving System~\cite{fan2018baidu, apollo}. 
\savior consists of four major components: a compiling tool-chain built on top of Clang and LLVM-4.0, a fuzzing component based on AFL-2.5b~\cite{tool-afl}, 
a concolic executor built atop KLEE~\cite{klee} (with LLVM-3.6), and a python middle-ware which coordinates the fuzzing component and the concolic executor.
In total, our implementation has about 3.3K lines of python code and 4K lines of C/C++ code. 
\savior can run on both 32-bit and 64-bit systems, and it can support 
both 32-bit and 64-bit targets. In the following, we discuss the important implementation 
details.


\point{Concolic Executor} 
We develop our concolic executor based on KLEE-3.6. The original KLEE aims at full symbolic 
execution, and it does not support concolic execution. We port a concolic 
executor from KLEE's symbolic executor. Specifically, the concolic executor attaches the concrete input 
as the {\tt assignment} property in the initial state. It then symbolically 
interprets each instruction as KLEE originally does. On 
reaching a conditional statement, it always follows the branch 
that matches the concrete input. For the other branch, if not covered, 
the concolic executor solves the conditions and generate
a corresponding testcase. The state following that branch is 
then immediately terminated. When generating the seed, our concolic 
executor copies the un-constrained bytes from the input, instead of padding 
with random values. 

Another limitation of KLEE is that the initialization phase is notoriously time-consuming. To overcome this, we introduce a fork server mode.  
In a run, KLEE first sets up the environments with bitcode loading, library linking, 
and preparing for globals and constants. These are then followed by the initialization of an {\tt Executor}. By default, the {\tt Executor} executes one seed and 
then destructs itself. In our implementation, after the execution of one seed, 
we clean up any stateful changes introduced in last execution 
(including destructing the memory manager, clearing the global data objects, 
and erasing all the remaining states). Then we reuse the {\tt Executor} 
to run a new seed from the input queue. In this mode, we 
avoid repeating the lengthy environments setup.

Recall that we invoke UBSan to label potentially vulnerable operations. 
At the IR level, UBSan replaces those operations with LLVM intrinsic functions, 
which are incomprehensible by KLEE. We replace those intrinsic functions 
with general LLVM IR so that KLEE can execute without exceptions. 
The replacements follow those that KLEE already enforced~\cite{kleeIntr6:online}.

By default, KLEE redirects un-modeled external functions 
(\eg system calls) to the native code.
This causes two issues. 
First, KLEE is unaware of their effects 
on the symbolic address space, which can interrupt memory operations. 
For instance, the function {\tt strdup} 
allocates a new buffer and copies data from the source to this buffer.
However, KLEE cannot capture this allocation due to the lack of modeling. 
On future accesses to this buffer, KLEE will throw 
an out-of-bound access error. There are many similar cases, such as {\tt  getenv}. 
We extend KLEE's environment model to include the symbolic versions of those functions. 
Second, KLEE concretizes the data passed to the external functions and 
adds constant constraints on such data for future execution. However,
this may over-constraint the concretized variables. 
For instance, KLEE concretizes the data written to standard output or files. 
This leads to over-constraints -- When the concretized data is later used in constraint solving,
KLEE will not be able to find a satisfying solution. To address this issue, 
we prevent KLEE from adding constraints on concretization. This scheme, 
following the design of S2E~\cite{s2e} and \qsym~\cite{qsyminsu}, 
ensures that we never miss solutions for non-covered branches.


Last but not least, stock KLEE provides limited support for software written in {\tt C++}. 
Since a lot of the C++ programs rely on the standard {\tt C++} library 
(\eg libstdc++ on Linux) but KLEE neither models this library nor supports the 
semantics of calls to this library. Therefore, KLEE frequently aborts the 
execution in the early stage of running a {\tt C++} program. 
We customize the GNU libstdc++ library to make it compilable 
and linkable to KLEE. Considering that many libstdc++ functions also access 
in-existent devices (\eg Random), we also build models of those devices. 

%% file: eval.tex
\section{Evaluation}
\label{sec:eval}

\definecolor{mygray}{gray}{0.9}

\savior approaches bug-driven hybrid testing with the key techniques of bug-driven prioritization and \bugguidedsearch. In this section, we evaluate these techniques and our evaluation centers around two questions:
\begin{itemize}
\item \emph{With bug-driven prioritization, can hybrid testing find vulnerabilities quicker?}
\item \emph{With \bugguidedsearch, can hybrid testing find vulnerabilities more thoroughly?}
\end{itemize}

To support our evaluation goals, we prepare two groups of widely-used benchmarks. 
The first group is the LAVA-M data-set~\cite{lava}.
This data-set comes with artificial vulnerabilities, and the ground truth is provided. 
The second group includes a set of \PROGNUM real-world programs. 
Details about these programs are summarized in Table~\ref{tab:eval-setup}. 
All these programs have been extensively tested in both industry~\cite{ossfuzz} and academia~\cite{driller,qsyminsu,vuzzer}. In addition, they represent a higher level of diversity in functionality and complexity.

Using the two benchmarks, we compare \savior with the most effective tools from 
related families. To be specific, we take \afl~\cite{tool-afl} as 
the baseline of coverage-based testing. As \savior performs testing in a directed manner, we also include the state-of-the-art directed fuzzer, \aflgo~\cite{aflgo}. 
To handle complex conditions, recent fuzzing research introduces a group of 
new techniques to improve code coverage. From this category, 
we cover \tfuzz~\cite{tfuzz} and \angora~\cite{angora}, because they are open-sourced
and representatives of the state-of-the-art. Finally, we also consider the existing 
implementations of hybrid testing, \driller~\cite{driller} and 
\qsym~\cite{qsyminsu}. 

Note that the original \driller has problems of running 
many of our benchmarks, due to lack of system-call modeling or failure to 
generate test cases (even with the patch~\cite{driller-patch} 
to support input from files). This aligns with the observations in~\cite{qsyminsu}. 
In the evaluation, we re-implement \driller on the top of \savior. 
More specifically, it runs \afl as the fuzzing component and it invokes the concolic executor 
once the {\tt pending\_favs} attribute in \afl drops to 0. These implementations 
strictly follow the original \driller~\cite{tool-driller1}. 
Similar to the Angr-based concolic executor in \driller, 
our KLEE-based concolic executor focuses on generating new seeds to cover 
untouched branches. In addition, we keep the relaxed constraint solving and 
the fork-server mode. These two features increase the effectiveness and efficiency of \driller without introducing algorithmic changes. 

In the following, we will explain the experimental setups and evaluation results 
for the two groups of benchmarks. 

\setlength{\belowcaptionskip}{0pt}
\begin{figure}[t]
    \centering   
    \begin{subfigure}[b]{0.24\textwidth}
        \centering
        \includegraphics[width=1\textwidth]{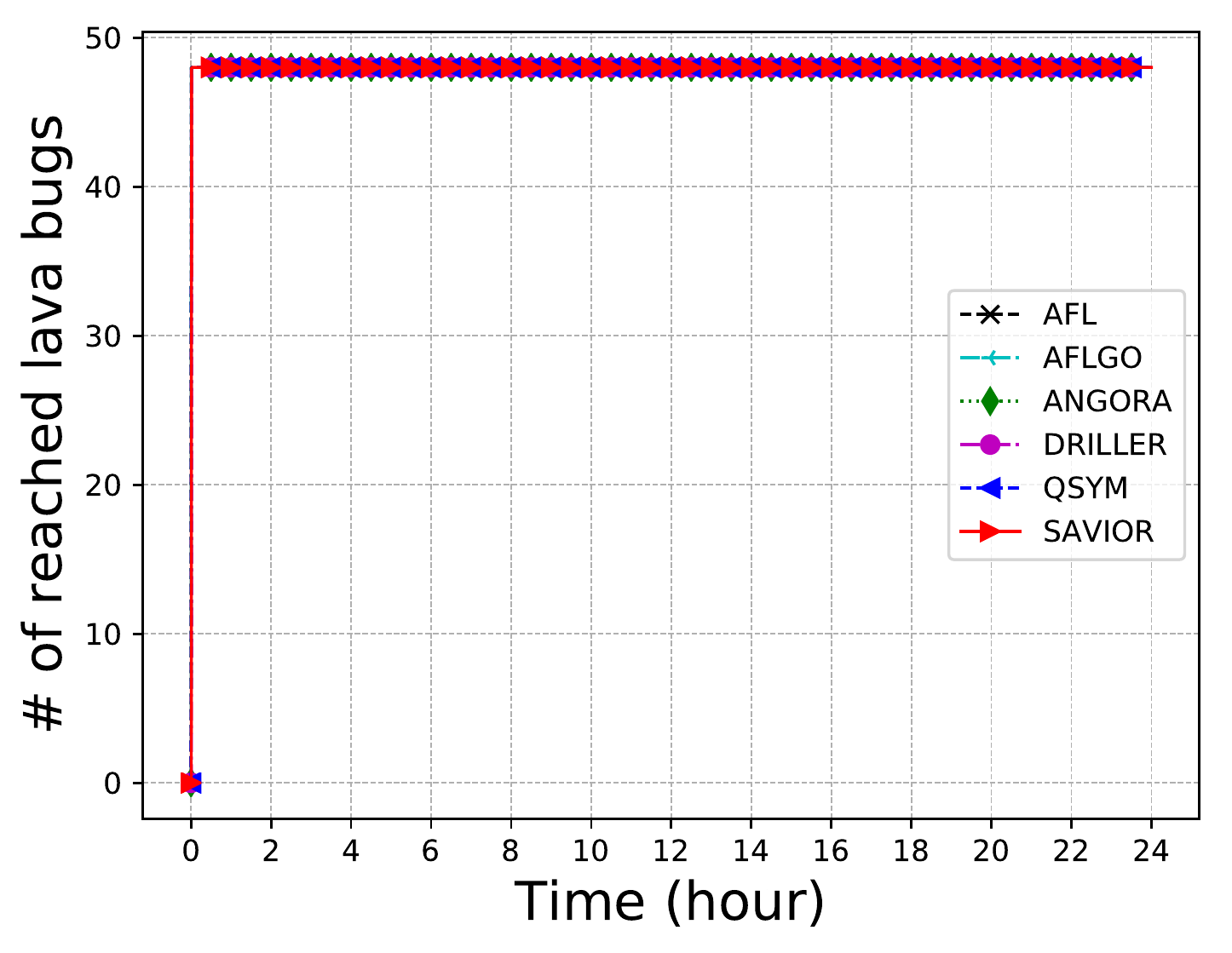}
        \vspace*{-15pt}
        \caption{\scriptsize{Number of bugs reached in base64}}
        \label{fig:eval:lava:base641}
    \end{subfigure}
        \begin{subfigure}[b]{0.24\textwidth}
        \centering
        \includegraphics[width=1\textwidth]{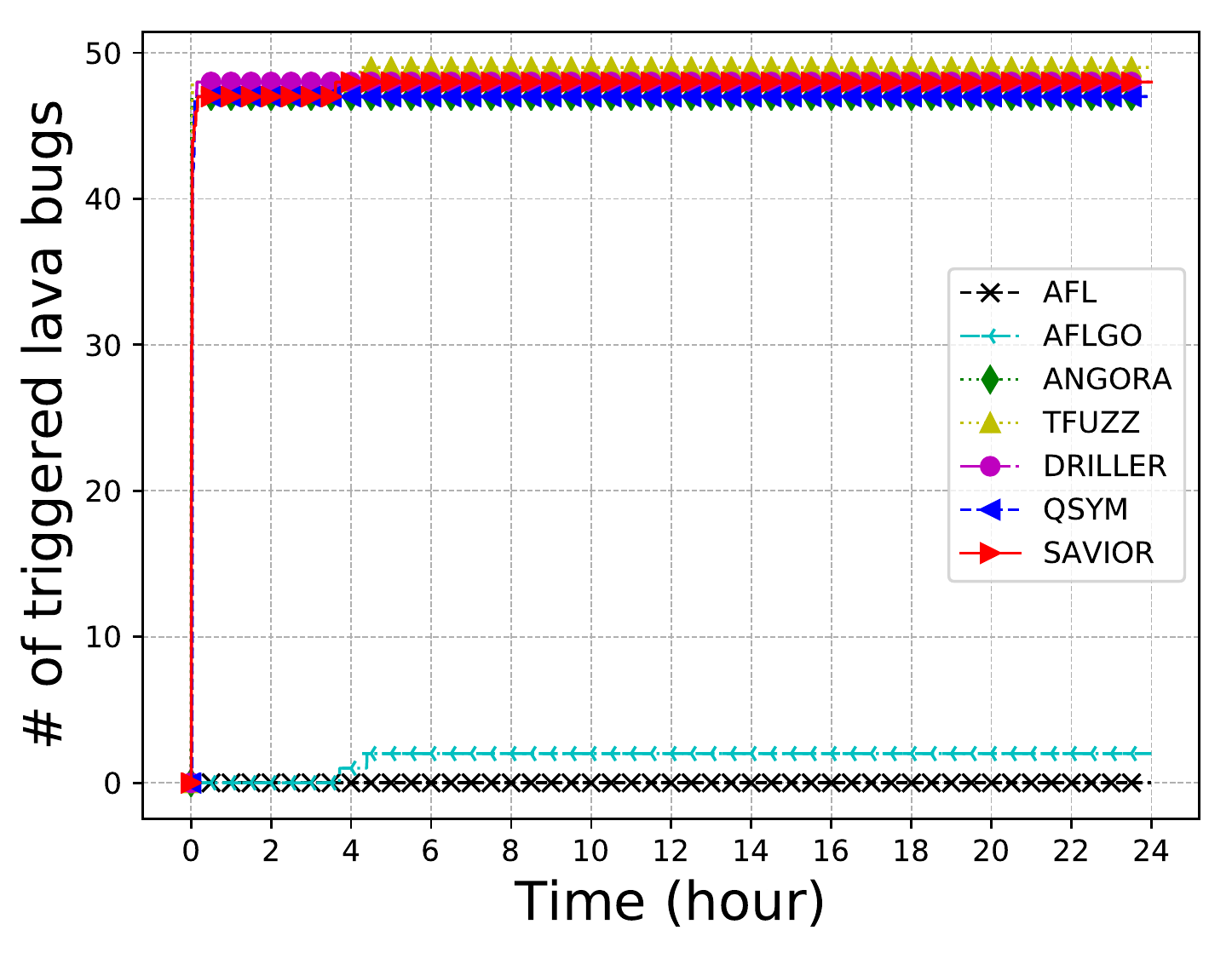}
         \vspace*{-15pt}
        \caption{\scriptsize{Number of bugs triggered in base64}}
        \label{fig:eval:lava:base642}
    \end{subfigure}\\
    \begin{subfigure}[b]{0.24\textwidth}
        \centering
        \includegraphics[width=1\textwidth]{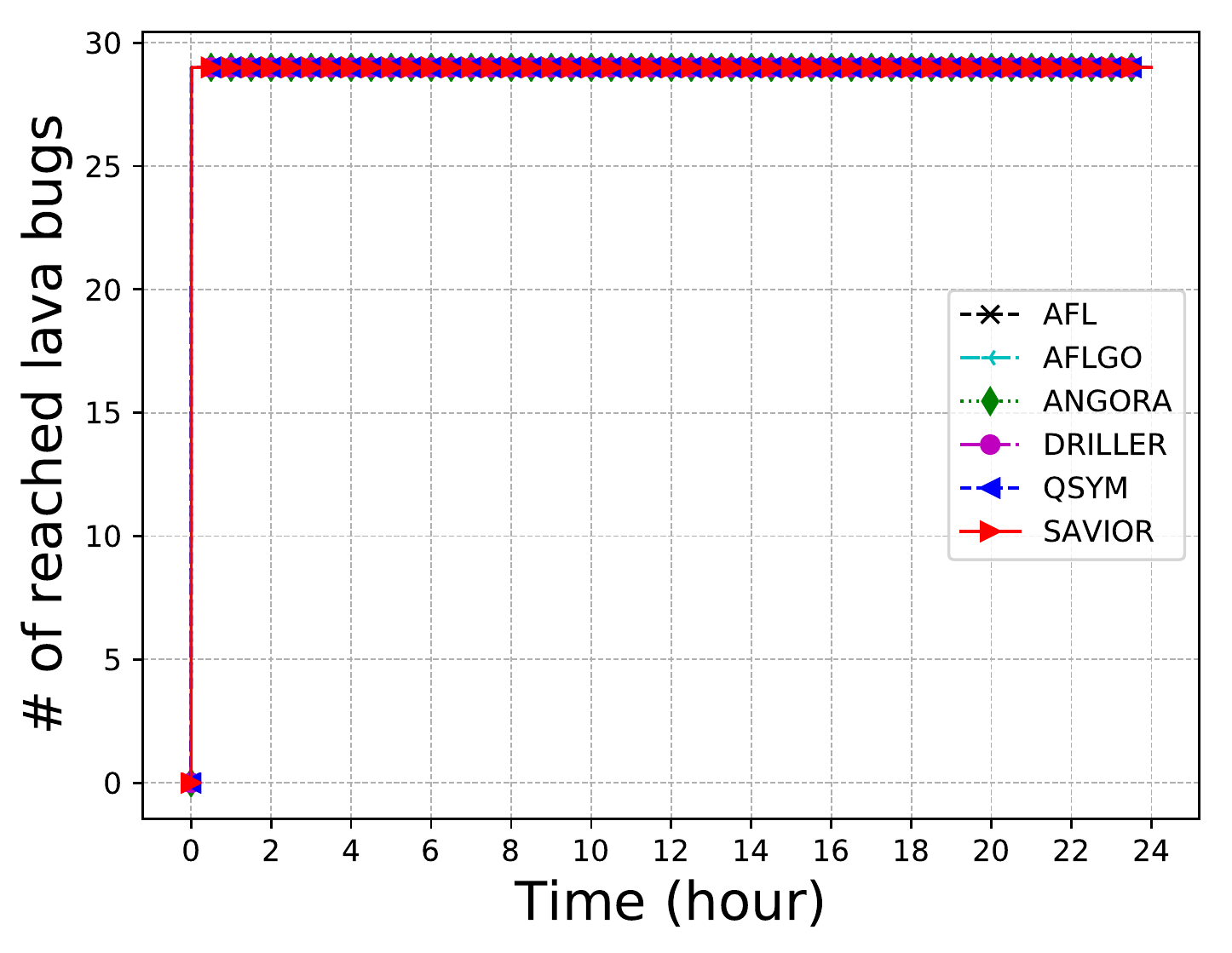}
         \vspace*{-15pt}
        \caption{\scriptsize{Number of bugs reached in uniq}}
        \label{fig:eval:lava:uniq1}
    \end{subfigure}
    \begin{subfigure}[b]{0.24\textwidth}
        \centering
        \includegraphics[width=1\textwidth]{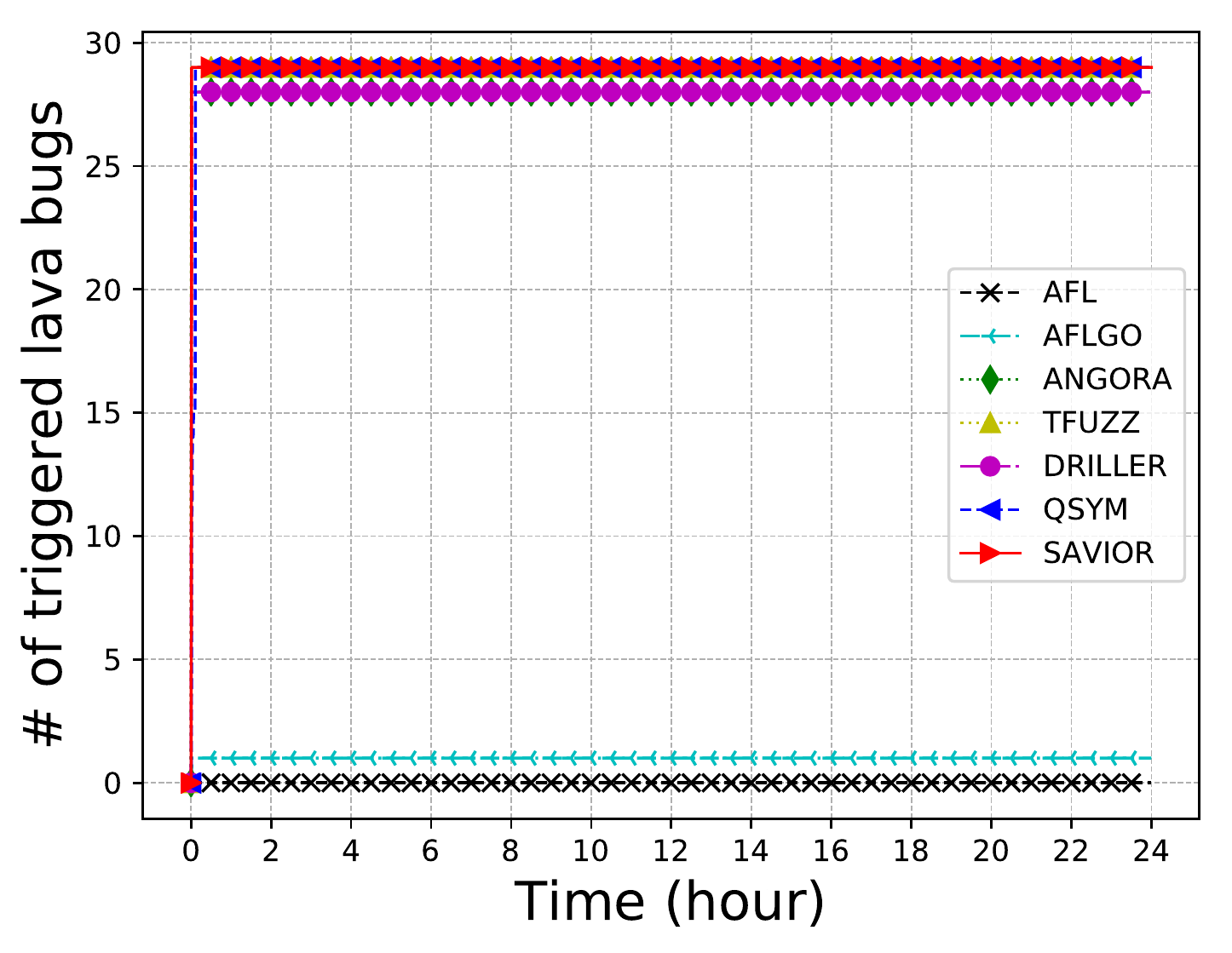}
       \vspace*{-15pt}
        \caption{\scriptsize{Number of bugs triggered in uniq}}
        \label{fig:eval:lava:uniq2}
    \end{subfigure}\\
    \begin{subfigure}[b]{0.24\textwidth}
        \centering
        \includegraphics[width=1\textwidth]{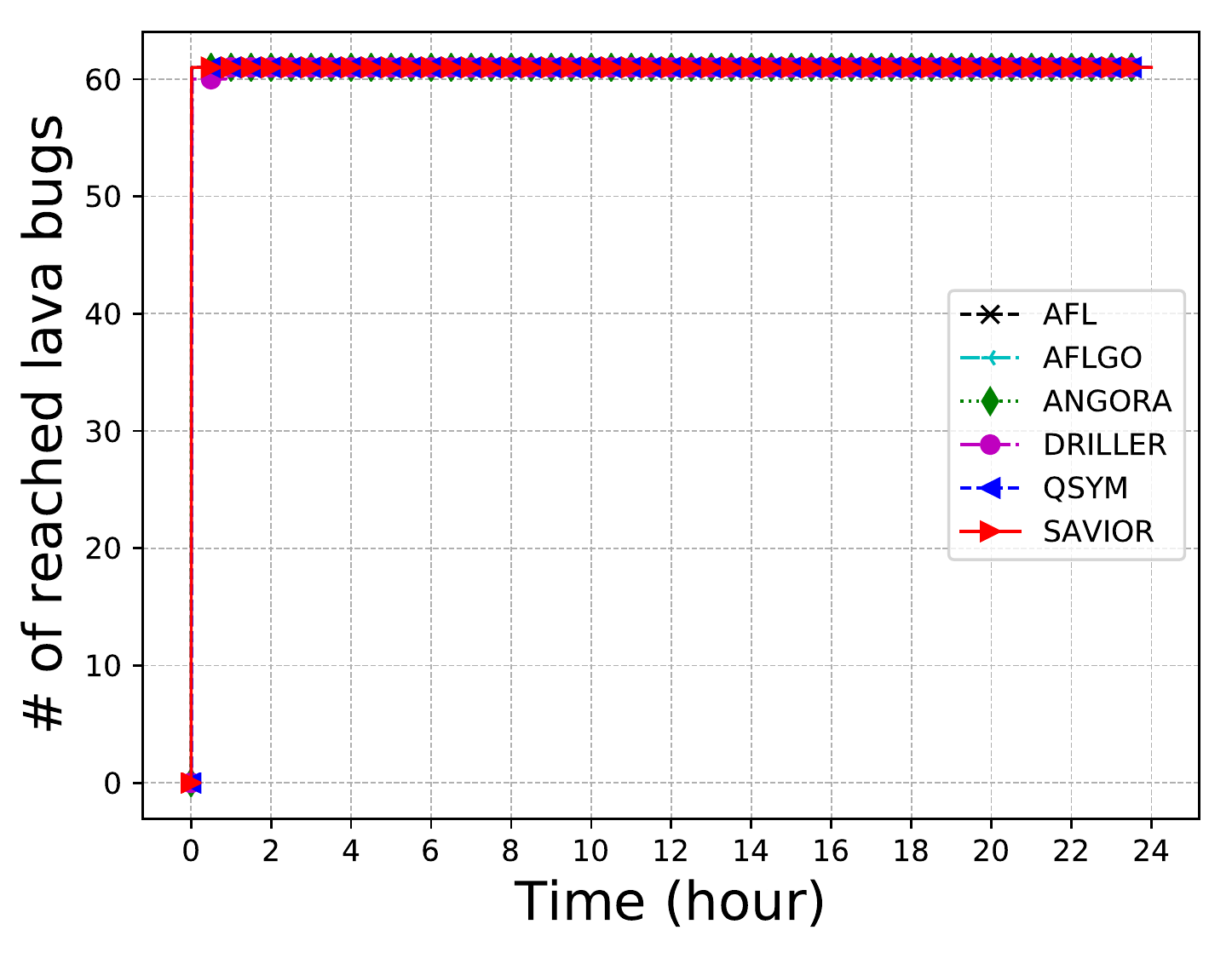}
        \vspace*{-15pt}
        \caption{\scriptsize{Number of bugs reached in md5sum}}
        \label{fig:eval:lava:md51}
    \end{subfigure}  
    \begin{subfigure}[b]{0.24\textwidth}
        \centering
        \includegraphics[width=1\textwidth]{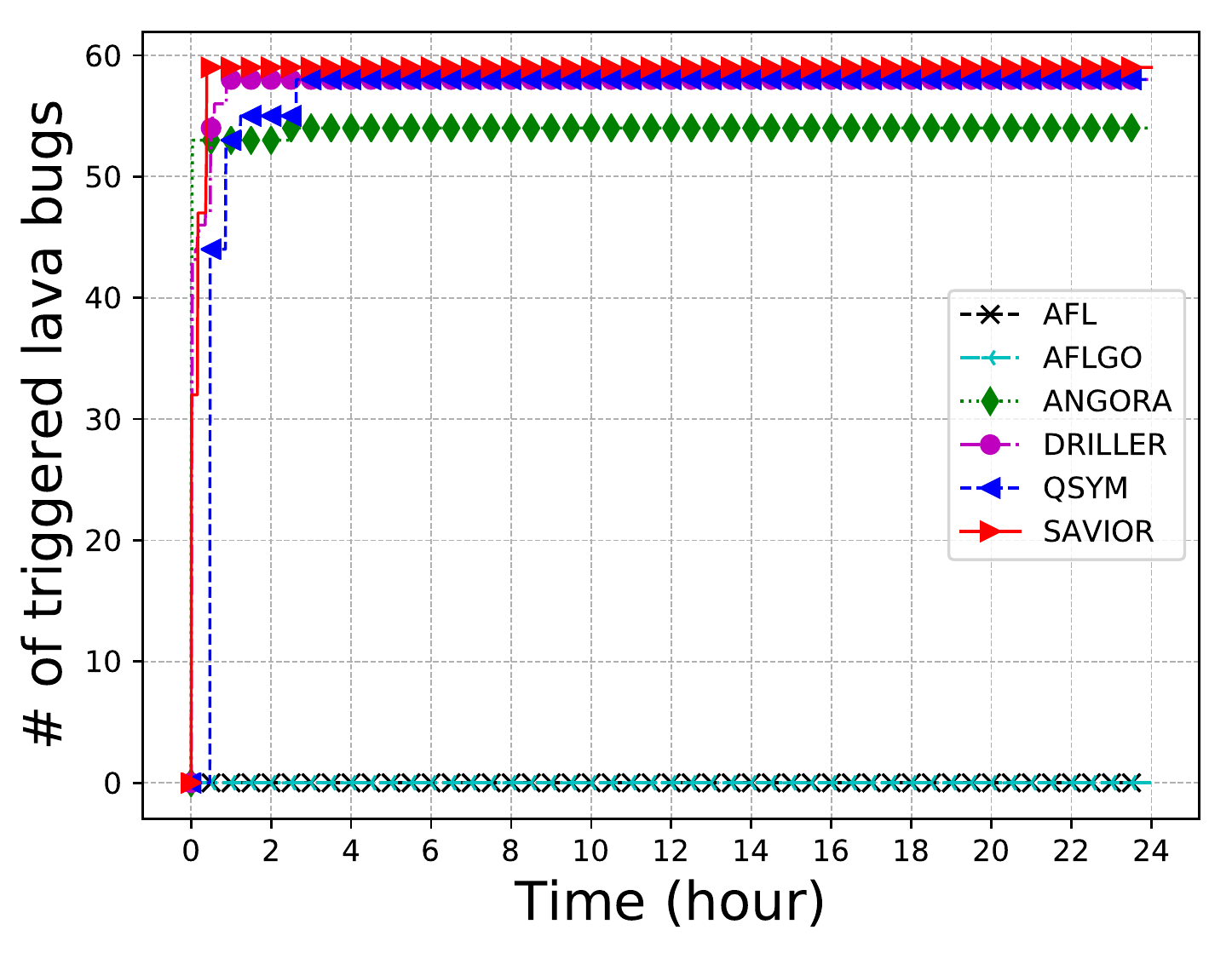}
       \vspace*{-15pt}
        \caption{\scriptsize{Number of bugs triggered in md5sum}}
        \label{fig:eval:lava:md5sum2}
    \end{subfigure}\\
    \begin{subfigure}[b]{0.24\textwidth}
        \centering
        \includegraphics[width=1\textwidth]{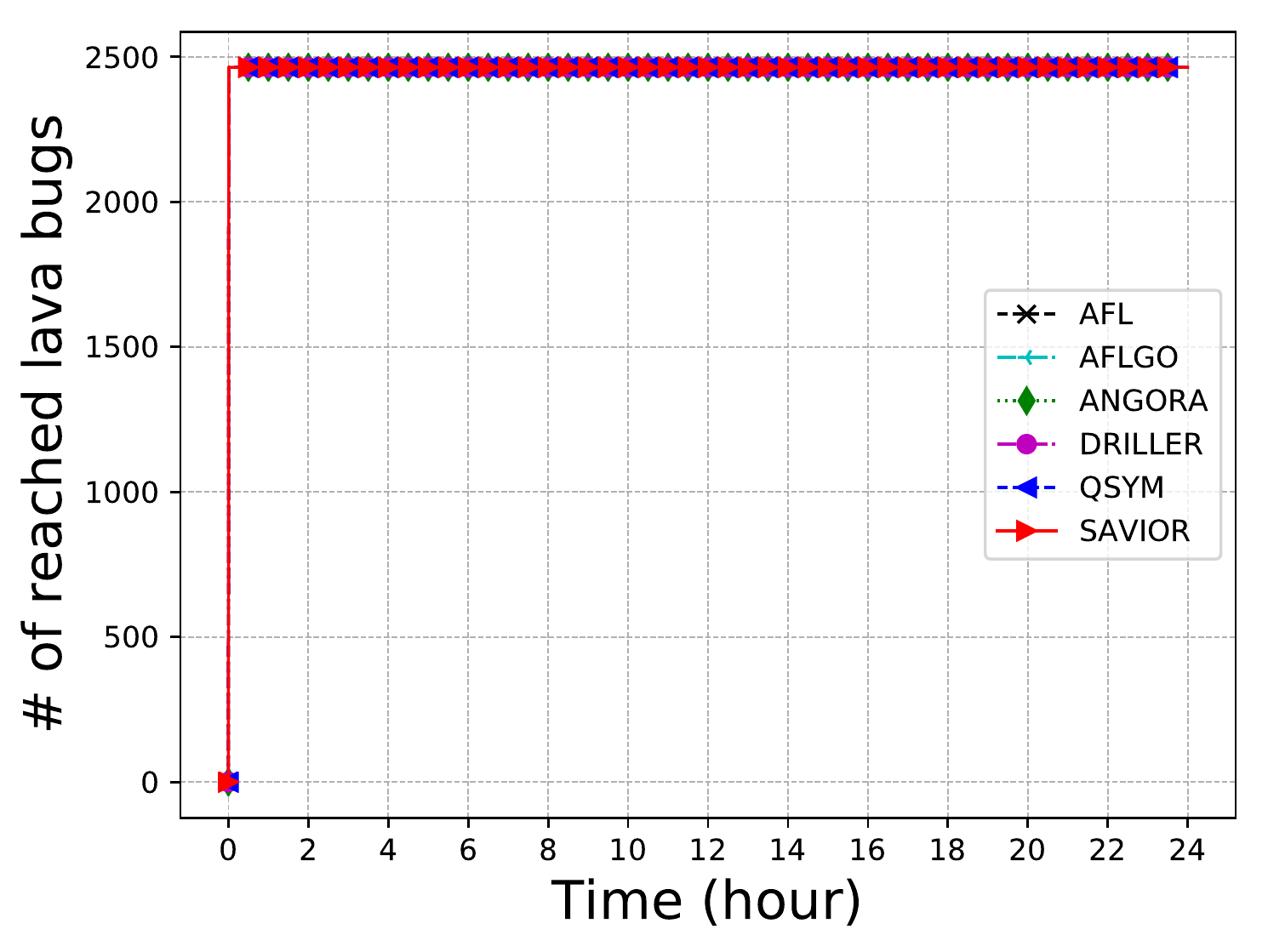}
       \vspace*{-15pt}
        \caption{\scriptsize{Number of bugs reached in who}}
        \label{fig:eval:lava:who1}
    \end{subfigure}
    \begin{subfigure}[b]{0.24\textwidth}
        \centering
        \includegraphics[width=1\textwidth]{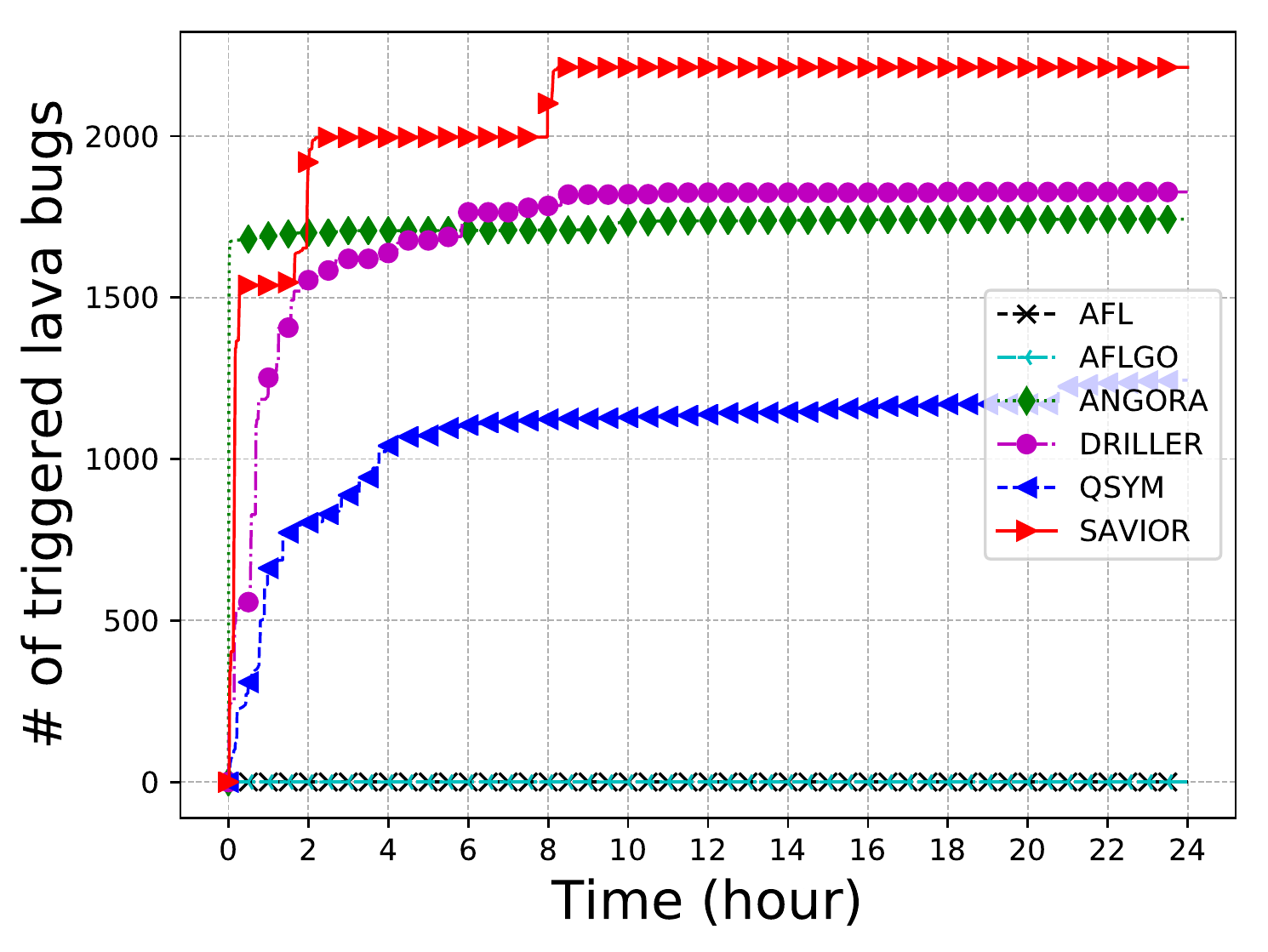}
       \vspace*{-15pt}
        \caption{\scriptsize{Number of bugs triggered in who}}
        \label{fig:eval:lava:who2}
    \end{subfigure}
    \caption{Evaluation results with LAVA-M. The left column shows the number of Lava bugs reached by different fuzzers and the right column shows the number of LAVA bugs triggered by the fuzzers. For \tfuzz, we only present the number of triggered bugs in {\tt base64} and {\tt uniq}, as the other results are not reliable due to a broken third-party dependency.}
    \vspace{-2ex}
    \label{fig:lava_cfa}
\end{figure}
\setlength{\belowcaptionskip}{-4pt}

\subsection{Evaluation with LAVA-M}
\label{sec:lava-eval}
\subsubsection{Experimental Setup}
In this evaluation, we run each of the fuzzers in Table~\ref{tab:lava-setup} with the four 
LAVA-M programs and we use the seeds shipped with the benchmark.  
For consistency, we conduct all the experiments on Amazon EC2 instances (Intel Xeon E5 Broadwell 64 cores, 256GB RAM, and running Ubuntu 16.04 LTS), and we sequentially run all the experiments to avoid interference. 
In addition, we assign each fuzzer 3 free CPU cores to ensure fairness in terms of 
computation resources. Each test is run for 24 hours.
To minimize the effect of randomness in fuzzing, we repeat each test 5 times and 
report the average results.

In Table~\ref{tab:lava-setup}, we also summarize the settings specific to each fuzzer, 
including how we distribute the 3 CPU cores and the actions we take to accommodate those fuzzers. 
In LAVA-M, each artificial vulnerability is enclosed and checked in a call to {\tt lava\_get} (in-lined in our evaluation).
We use these calls as the targets to guide \aflgo and we mark them as vulnerability labels 
to enable bug-driven prioritization in \savior. In addition, as the vulnerability condition is 
hard-coded in the {\tt lava\_get} function, we naturally have support for \bugguidedsearch.
Finally, for \angora, we adopt the patches as suggested by the developers~\cite{tool-angora1}.

\subsubsection{Evaluation Results} In the left column of Figure~\ref{fig:lava_cfa}, 
we show how many vulnerabilities are reached over time by different fuzzers. The results demonstrate that all 
the fuzzers can instantly cover the code with LAVA vulnerabilities. However, as presented 
in the right column of Figure~\ref{fig:lava_cfa}, \tfuzz, \angora, \driller, \qsym, and \savior are able to 
trigger most (or all) of the vulnerabilities while \afl and \aflgo can trigger few. 
The reason behind is that the triggering conditions of LAVA vulnerabilities are all in the form of 
32-bit magic number matching. Mutation-based fuzzers, including AFL
and AFLGo, can hardly satisfy those conditions while the other fuzzers are all featured 
with techniques to solve them. 

\point{Vulnerability Finding Efficiency} Despite \tfuzz, \angora, \driller, \qsym, and \savior all trigger large numbers of LAVA vulnerabilities, 
they differ in terms of efficiency. \tfuzz quickly covers 
the listed vulnerabilities in {\tt base64} and {\tt uniq}. This is attributable to 
that (1) \tfuzz can reach all the vulnerabilities with several initial seeds and (2) \tfuzz 
can transform the program to immediately trigger the encountered vulnerabilities.
Note that we do not show the results of \tfuzz on {\tt md5sum} and {\tt who},
because \tfuzz gets interrupted because of a broken dependency~\footnote{The broken component is the QEMU based tracer in Angr~\cite{angrtrac7:online}. 
This has been confirmed with the 
developers.}. For all the cases, \angora triggers the vulnerabilities
immediately after its start. The main reason is that the ``black-box function'' 
pertaining to all LAVA vulnerabilities 
is {\tt f(x) = x} and the triggering conditions are like {\tt f(x) == CONSTANT}. 
\angora always starts evaluating such functions with {\tt x = CONSTANT} and hence, 
it can instantly generate seeds that satisfy the vulnerability conditions. 
In the case of {\tt who}, \angora does not find all the vulnerabilities 
because of its incomplete dynamic taint analysis. 

Regarding the three hybrid tools, they trigger every vulnerability 
that their concolic executors encounter. In the cases of {\tt base64}, {\tt uniq}, and {\tt md5sum}, 
their concolic executors can reach all the vulnerabilities with initial seeds. 
This explains why they all quickly trigger the listed vulnerabilities, 
regardless of their seed scheduling. 

In the case of {\tt who}, even though the fuzzing component 
quickly generates seeds to cover the vulnerable code, 
the concolic executor takes much longer to run those seeds.
For instance, while executing the inputs from \afl, \qsym needs 
over 72 hours of continuous concolic execution to reach 
all the LAVA bugs in {\tt who}. Differing from \driller and \qsym,
\savior prioritizes seeds that have a higher potential 
of leading to Lava bugs. As demonstrated by the results of {\tt who} 
in Table~\ref{tab:lava-bug-num}, our technique of 
\emph{bug-driven prioritization indeed advances the 
exploration of code with more vulnerabilities}.  
Note that \driller (with a random seed scheduling) 
moves faster than \qsym. This is because \qsym prioritizes concolic execution 
on small seeds, while reaching the vulnerabilities in {\tt who} needs 
seeds with a larger size.

\begin{table}[t!]

\centering
\scriptsize
\begin{tabular}{l|cccc}
\toprule[0.5pt]
\toprule[0.5pt]

\multirow{2}{*}{\bf{\emph{Fuzzers}}} & \multicolumn{4}{c}{\bf{\emph{Fuzzing results}}}   
 \\ \cline{2-5}
 & {\tt base64} & {\tt uniq} & {\tt md5sum} & {\tt who}
 \\ \hline

\rowcolor{mygray}
\afl  & 0 (0\%) &  0 (0\%) & 0 (0\%) & 0 (0\%) \\
\aflgo  & 2 (5\%) & 1 (4\%) & 0 (0\%) & 0 (0\%) \\
\rowcolor{mygray}
\tfuzz  & 47 (100\%) &  29 (100\%) & N/A & N/A\\
\angora  & 47 (100\%) &  28 (100\%) & 54 (95\%) & 1743 (79\%)\\
\rowcolor{mygray}
\driller & 48 (100\%) &  28 (100\%) & 58 (100\%) & 1827 (78\%)\\
\qsym  & 47 (100\%) &  29 (100\%) & 58 (100\%) & 1244 (53\%)\\
\rowcolor{mygray}
\savior  & 48 (100\%) &  29 (100\%) & 59 (100\%) & 2213 (92\%)\\ \hline
{\bf Listed}  & {\bf 44} & {\bf 28} & {\bf 57} & {\bf 2136}\\

\bottomrule[0.5pt]
\bottomrule[0.5pt]
\end{tabular}
\caption{LAVA-M Bugs triggered by different fuzzers (before \bugguidedsearch). ``X\%'' indicates that X\% of the listed LAVA bugs are triggered.}
\label{tab:lava-bug-num}
\vspace{-2ex}
\end{table}

\begin{table}[t!]

\centering
\scriptsize
\begin{tabular}{l|cccc}
\toprule[0.5pt]
\toprule[0.5pt]

\multirow{2}{*}{\bf{\emph{Fuzzers}}} & \multicolumn{4}{c}{\bf{\emph{Fuzzing results}}}  
 \\ \cline{2-5}
 & {\tt base64} & {\tt uniq} & {\tt md5sum} & {\tt who}
 \\ \hline

\rowcolor{mygray}
\afl  & 48 (100\%) &  29 (100\%) & 59 (100\%) & 2357 (96.3\%) \\
\aflgo  & 48 (100\%) & 29 (100\%) & 59 (100\%) & 2357 (96.3\%) \\
\rowcolor{mygray}
\tfuzz  & 47 (100\%) &  29 (100\%) & N/A & N/A\\
\angora  & 48 (100\%) &  29 (100\%) & 59 (100\%) & 2357 (96.3\%)\\
\rowcolor{mygray}
\driller & 48 (100\%) &  29 (100\%) & 59 (100\%) & 2357 (96.3\%)\\
\qsym  & 48 (100\%) &  29 (100\%) & 59 (100\%) & 2357 (96.3\%)\\
\rowcolor{mygray}
\savior  & 48 (100\%) &  29 (100\%) & 59 (100\%) & 2357 (96.3\%)\\ \hline
{\bf Listed}  & {\bf 44} & {\bf 28} & {\bf 57} & {\bf 2136}\\

\bottomrule[0.5pt]
\bottomrule[0.5pt]
\end{tabular}
\caption{LAVA-M Bugs triggered by different fuzzers (after \bugguidedsearch). ``X\%'' indicates that X\% of the listed LAVA bugs are triggered.}
\label{tab:lava-search-bug}
\vspace{-2ex}
\end{table}

\point{Vulnerability Finding Thoroughness} 
We further evaluate our \bugguidedsearch design. Specifically, we run 
the seeds generated by all the fuzzers with our concolic executor. 
In this experiment, we only perform constraint solving when 
a vulnerability condition is encountered. 
As shown in Table~\ref{tab:lava-search-bug}, \bugguidedsearch 
facilitates all the fuzzers to not only cover the listed LAVA bugs but 
also disclose an extra group of Lava bugs. Due to limited space, 
those additionally identified bugs are summarized in Table~\ref{tab:new-laval-bug-search}
at Appendix. Such results \emph{strongly demonstrate the promising potential 
of \bugguidedsearch to benefit fuzzing tools in vulnerability findings}.  
 
\begin{table}[t!]
\centering
\scriptsize
\begin{tabular}{lccc|cc}
\toprule[0.5pt]
\toprule[0.5pt]

\multicolumn{4}{c|}{\bf{\emph{Programs}}} & \multicolumn{2}{c}{\bf{\emph{Settings}}}   
 \\ \hline
 {\tt Name} & {\tt Version} & {\tt Driver} & {\tt Source} & {\tt Seeds} & {\tt  Options}
 \\ \hline

\rowcolor{mygray}
libpcap & 4.9.2/1.9.0 & tcpdump & ~\cite{Indexofr65:online} & build-in & -r @@ \\
libtiff & 4.0.10 & tiff2ps & ~\cite{Indexofl62:online} & \afl &  @@ \\
\rowcolor{mygray}
libtiff & 4.0.10 & tiff2pdf & ~\cite{Indexofl62:online} & \afl &  @@ \\
binutils & 2.31 & objdump & ~\cite{Indexofg63:online} & \afl &  -D @@ \\
\rowcolor{mygray}
binutils & 2.31 & readelf & ~\cite{Indexofg63:online} & \afl &  -A @@ \\
libxml2 & 2.9.7 & xmllint & ~\cite{libxml2src} & \afl &  @@ \\
\rowcolor{mygray}
libjpeg & 9c & djpeg & ~\cite{libjpegsrc} & \afl &   \\
jasper & master & jasper & ~\cite{jaspersrc} & \afl &  -f @@ -T pnm \\
\bottomrule[0.5pt]
\bottomrule[0.5pt]
\end{tabular}
\caption{Real-world benchmark programs and evaluation settings. 
In the column for {\tt Seeds}, 
{\tt AFL} indicates we reuse the testcases provided in \afl and {\tt build-in} indicates that we reuse the test cases shipped with the program.}
\label{tab:eval-setup}
\vspace{-2ex}
\end{table}

\subsection{Evaluation with Real-world Programs}

\subsubsection{Experimental Setup} In this evaluation, we prepare \PROGNUM programs. 
Details about these programs and the test settings are summarized in Table~\ref{tab:eval-setup}. 
All the programs have been extensively tested by both industry~\cite{ossfuzz} and academic researches~\cite{driller,qsyminsu,vuzzer}. 
Since different seed inputs and execution
options could lead to varying fuzzing results~\cite{seedoptimize, evaluatefuzz},
we follow existing works to use the seeds shipping with \afl or the vendors, as well as to configure the fuzzing options. 
Similar to our evaluation with LAVA-M, we conduct all the experiments 
on Amazon EC2 instances. To reduce randomness during testing, we run each
test 5 times and report the average results. 
In addition, we leverage Mann Whitney U-test~\cite{mcknight2010mann} 
to measure the significance of our improvements, 
following the suggestion by George etc~\cite{evaluatefuzz}. 

In this evaluation, we also prepare the setups that are specific to each fuzzing tool. 
These setups mostly follow Table~\ref{tab:lava-setup} except the following. 
First, we use UBSan labels as the target locations for \aflgo and as the guidance of bug-driven prioritization in \savior. Second, to prevent \angora from terminating the fuzzing process once it encounters un-instrumented library functions, we follow suggestions from the developers and add the list of un-instrumented functions 
into \angora's {\tt dfsan\_abilist.txt} configuration file. 
Third, we do not include \tfuzz, 
because it does not function correctly on our benchmark programs due to issues in the aforementioned 
 third-party component. 
Furthermore, we prepare these benchmark programs such that they are instrumented with UBSan for all fuzzers to ensure a fair comparison. This also means that \bugguidedsearch is enabled 
by default in \driller, \qsym, and \savior.

\subsubsection{Evaluation Results} 
\begin{figure*}[t!]
    \centering
    \begin{subfigure}[b]{0.24\textwidth}
        \centering
        \includegraphics[width=1\textwidth]{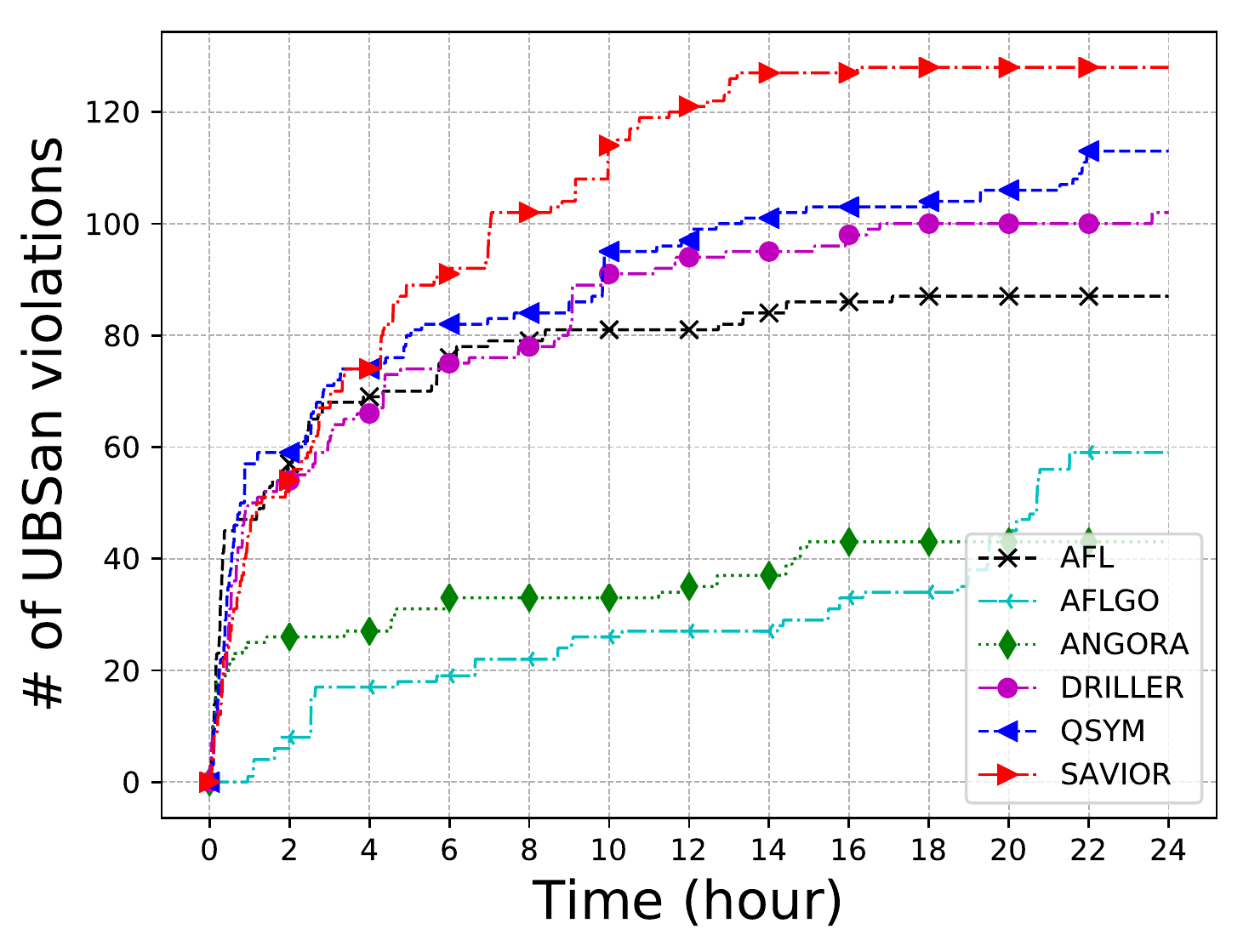}
        \captionsetup{margin=0.5cm}
         \vspace*{-15pt}
        \caption{\scriptsize{Number of UBSan violations triggered in \emph{\bf tcpdump} ($p_{1}$=$0.029$, $p_{2}$=$0.047$).}}
        \label{fig:eval:fast:tcpdump1}
    \end{subfigure}
     \begin{subfigure}[b]{0.24\textwidth}
        \centering
        \includegraphics[width=1\textwidth]{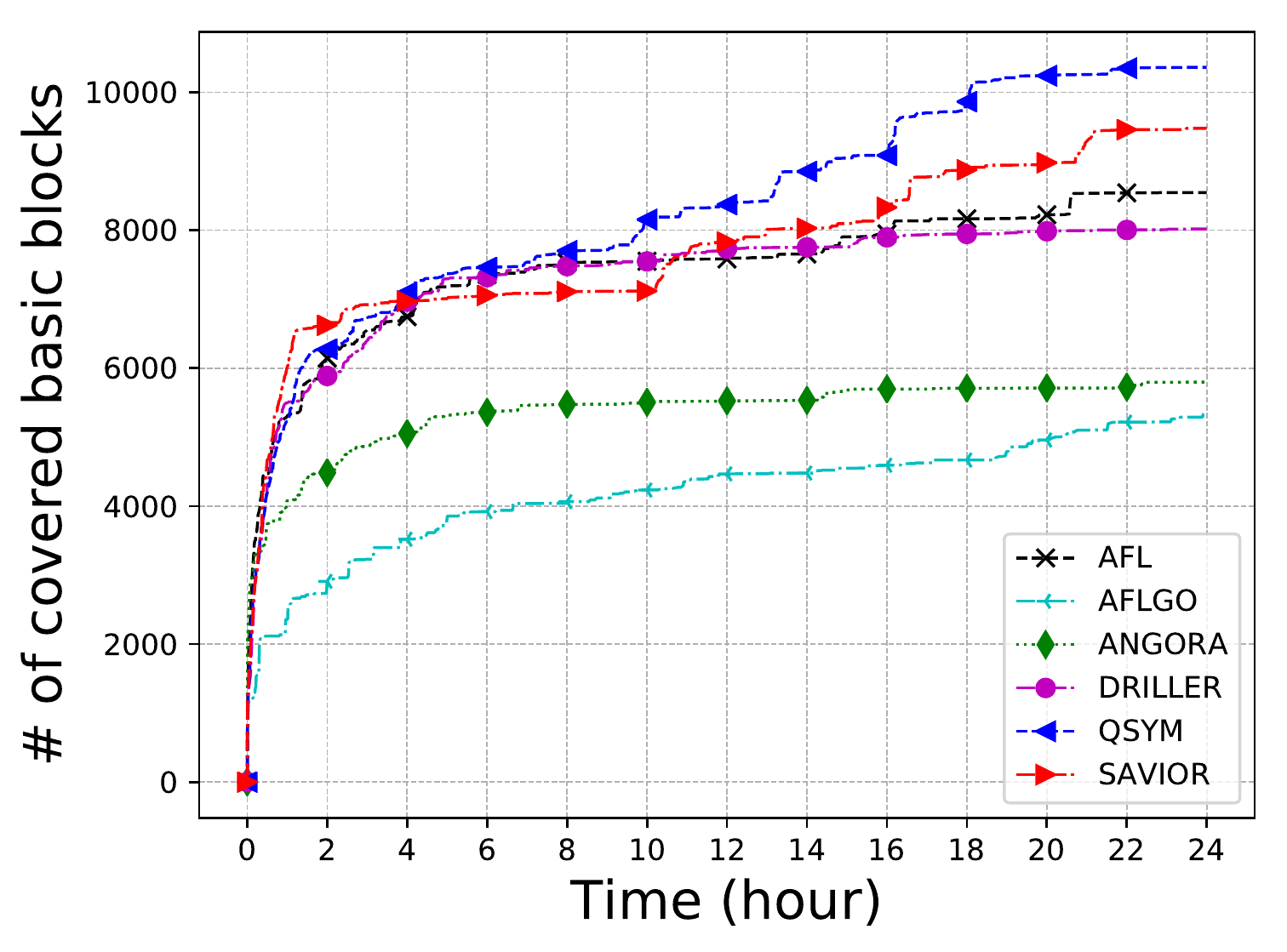}
        \captionsetup{margin=0.5cm}
        \vspace*{-15pt}
        \caption{\scriptsize{Number of basic blocks reached in \emph{\bf tcpdump} ($p_{1}$=$0.106$,$p_{2}$=$0.999$).}}
        \label{fig:eval:fast:tcpdump2}
    \end{subfigure}\rulesep
     \begin{subfigure}[b]{0.24\textwidth}
        \centering
        \includegraphics[width=1\textwidth]{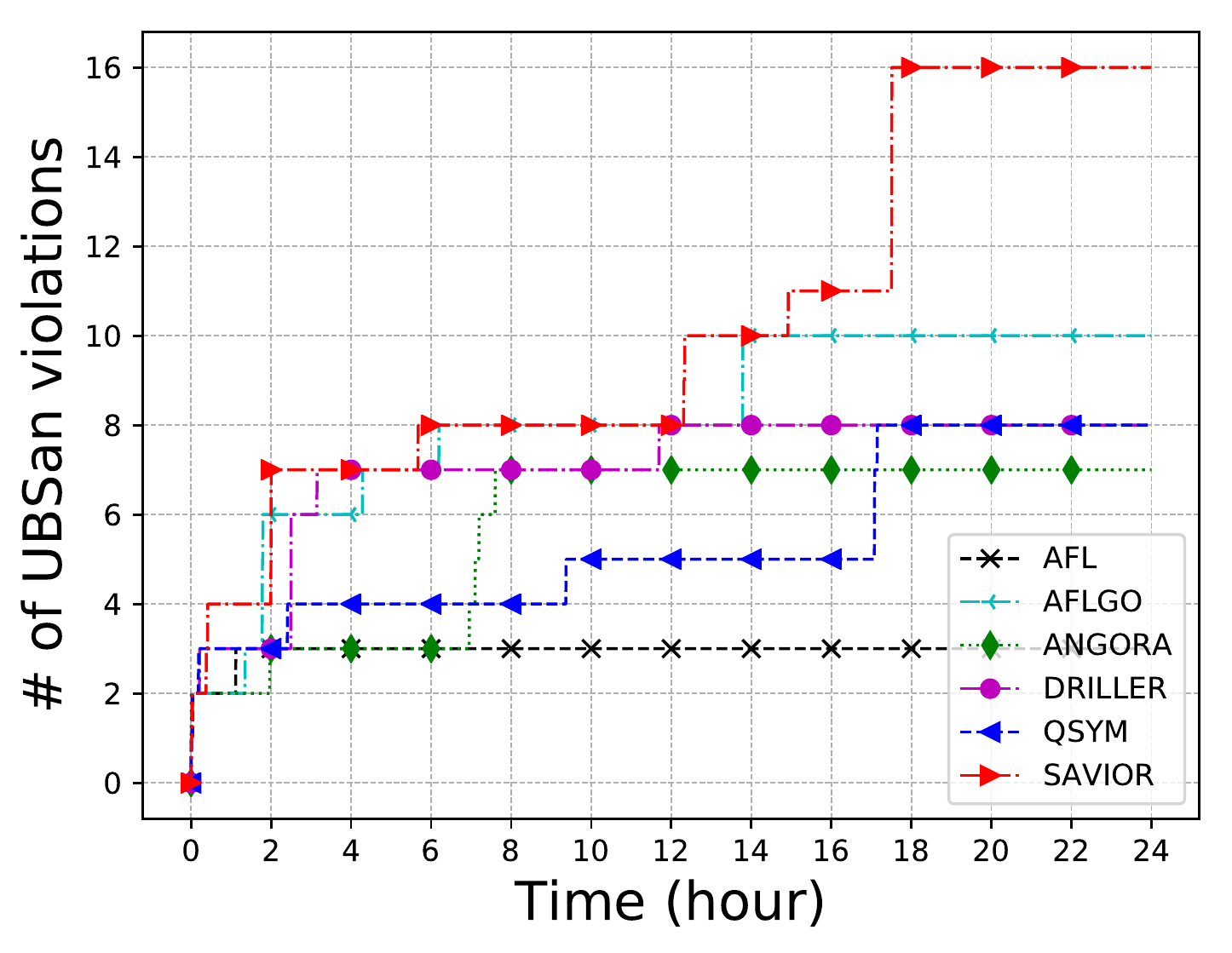}
        \captionsetup{margin=0.5cm}
         \vspace*{-15pt}
        \caption{\scriptsize{Number of UBSan violations triggered in \emph{\bf tiff2ps} ($p_{1}$=$0.005$, $p_{2}$=$0.046$).}}
        \label{fig:eval:fast:tiff2ps1}
    \end{subfigure}
     \begin{subfigure}[b]{0.24\textwidth}
        \centering
        \includegraphics[width=1\textwidth]{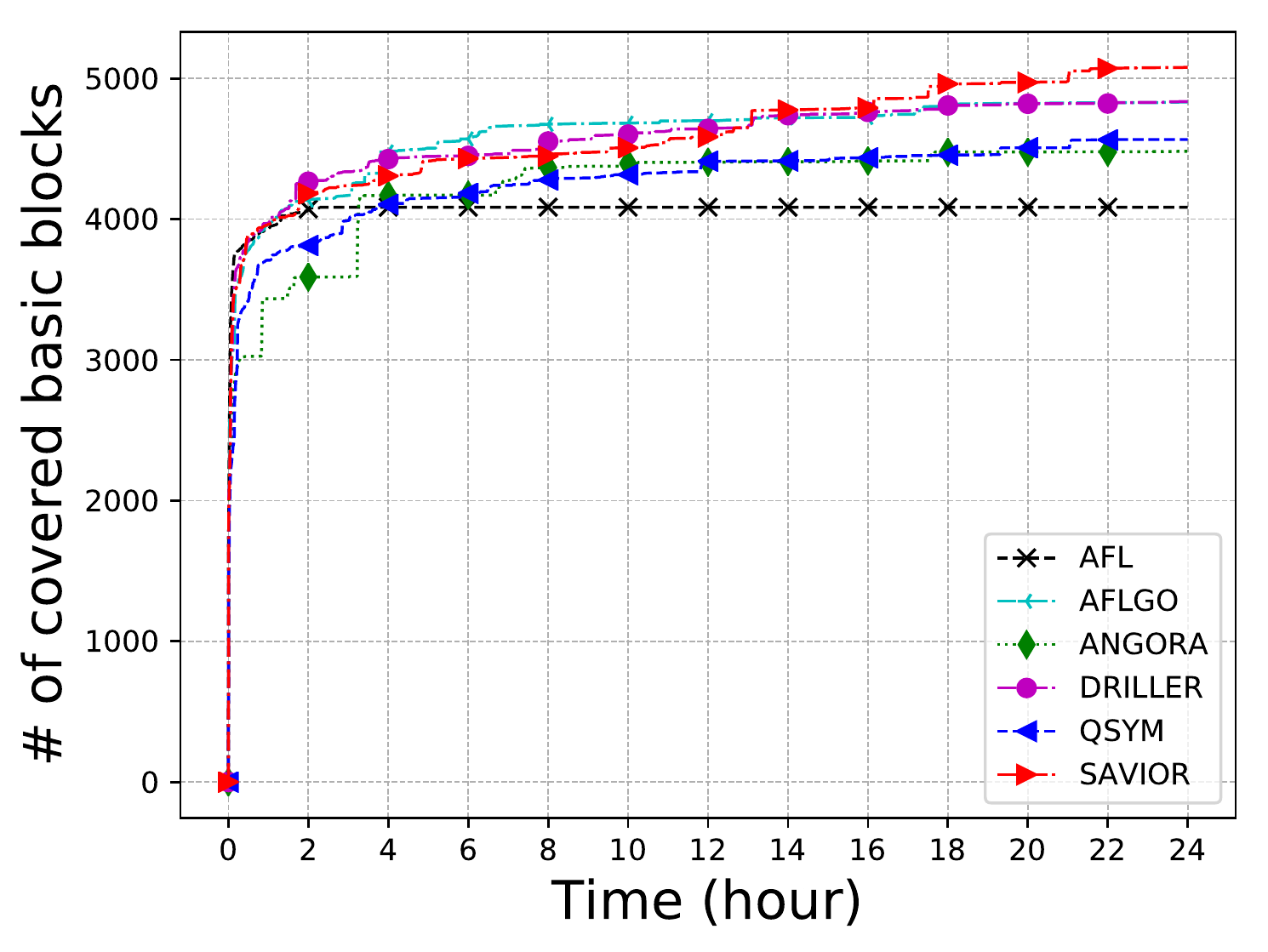}
        \captionsetup{margin=0.5cm}
          \vspace*{-15pt}
        \caption{\scriptsize{Number of basic blocks reached in \emph{\bf tiff2ps} ($p_{1}$=$0.049$, $p_{2}$=$0.073$).}}
        \label{fig:eval:fast:tiff2ps2}
    \end{subfigure}\\
     \begin{subfigure}[b]{0.24\textwidth}
        \centering
        \includegraphics[width=1\textwidth]{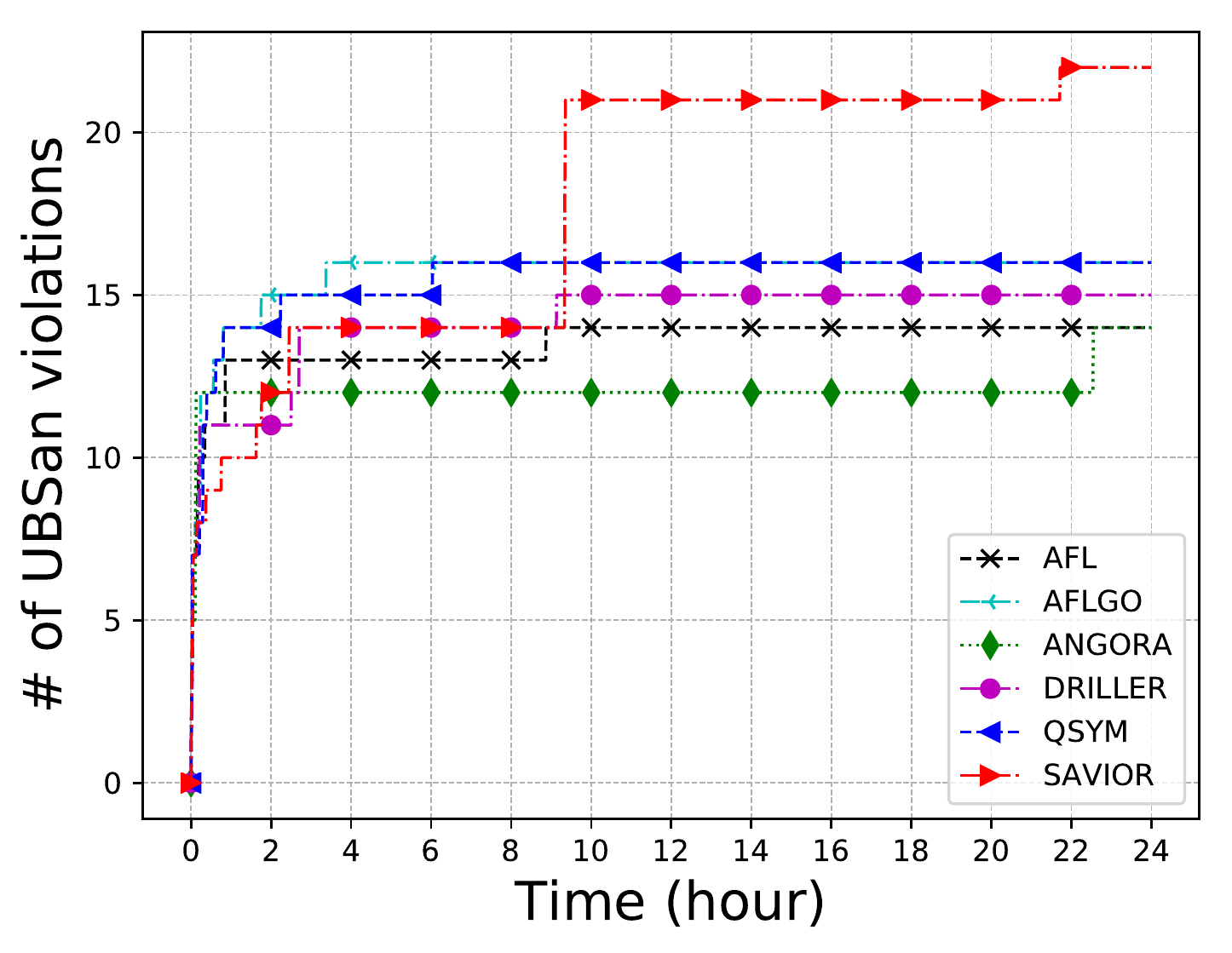}
        \captionsetup{margin=0.5cm}
         \vspace*{-15pt}
        \caption{\scriptsize{Number of UBSan violations triggered in \emph{\bf readelf} ($p_{1}$=$0.098$, $p_{2}$=$5.63*e^{-5}$).}}
        \label{fig:eval:fast:readelf1}
    \end{subfigure}
     \begin{subfigure}[b]{0.24\textwidth}
        \centering
        \includegraphics[width=1\textwidth]{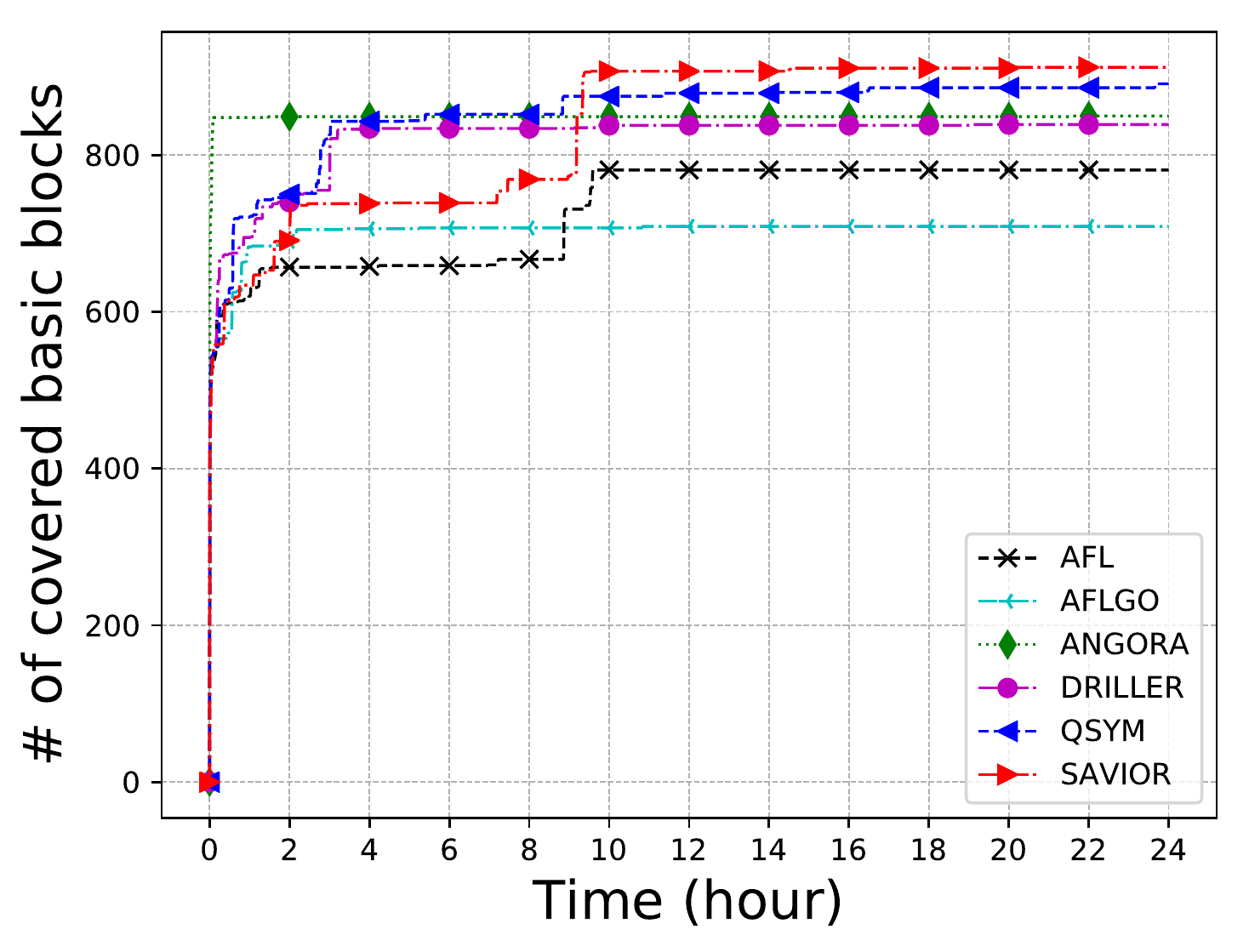}
        \captionsetup{margin=0.5cm}
          \vspace*{-15pt}
        \caption{\scriptsize{Number of basic blocks reached in \emph{\bf readelf} ($p_{1}$=$0.042$, $p_{2}$=$0.726$).}}
        \label{fig:eval:fast:readelf2}
    \end{subfigure} \rulesep
     \begin{subfigure}[b]{0.24\textwidth}
        \centering
        \includegraphics[width=1\textwidth]{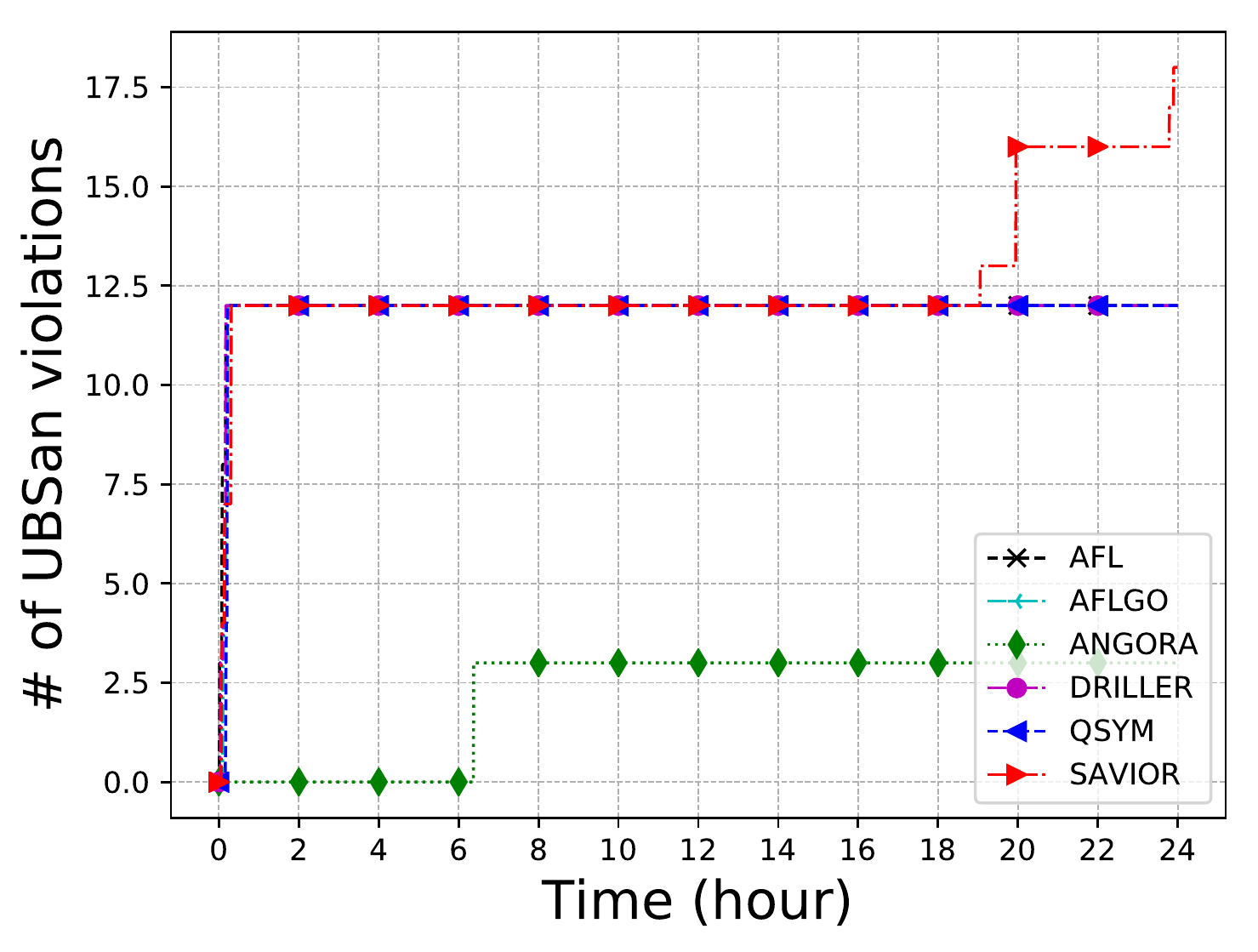}
        \captionsetup{margin=0.5cm}
         \vspace*{-15pt}
        \caption{\scriptsize{Number of UBSan violations triggered in \emph{\bf libxml} ($p_{1}$=$7.04*e^{-5}$, $p_{2}$=$2.15*e^{-7}$).}}
        \label{fig:eval:fast:libxml1}
    \end{subfigure} 
     \begin{subfigure}[b]{0.24\textwidth}
        \centering
        \includegraphics[width=1\textwidth]{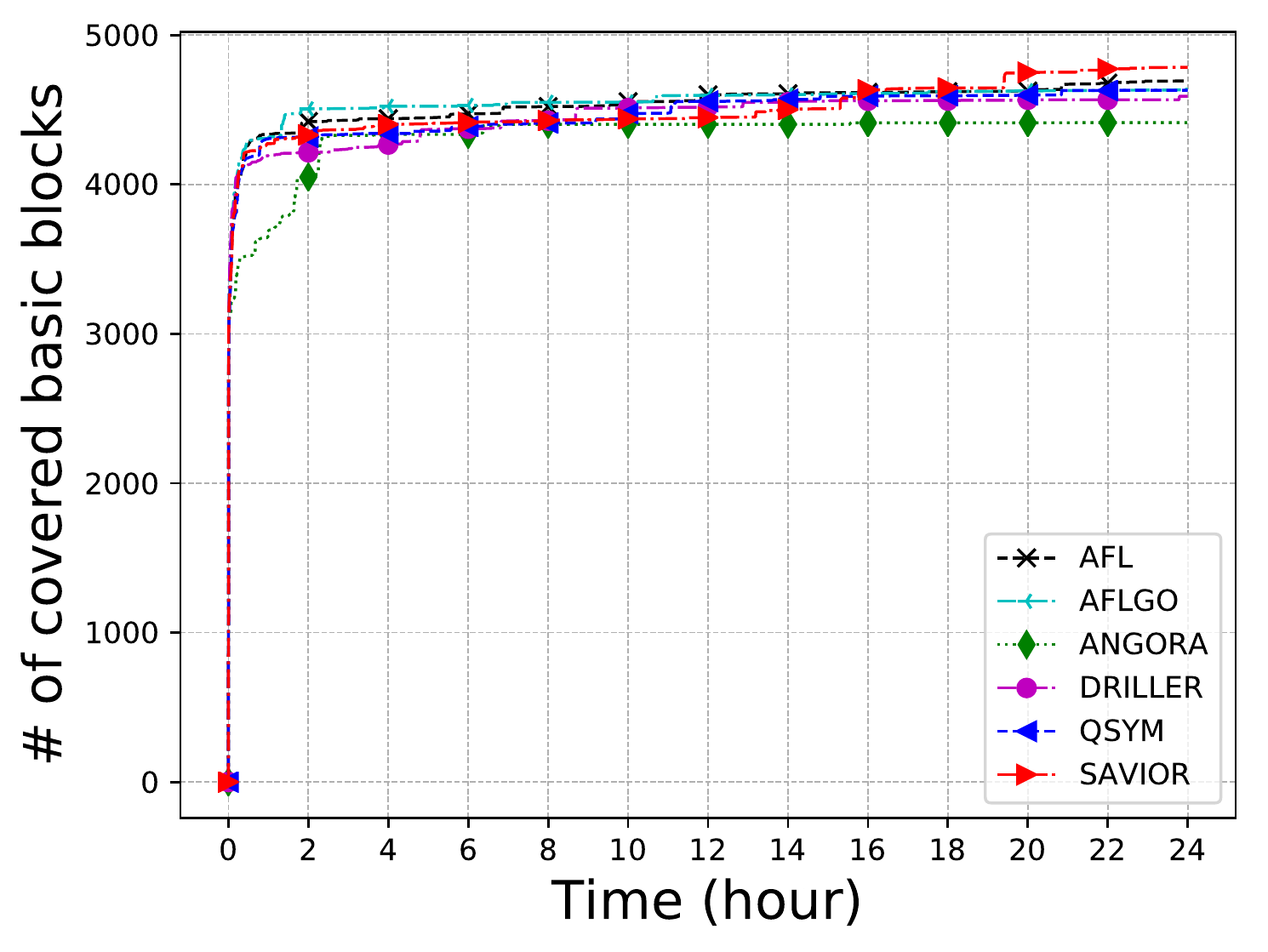}
        \captionsetup{margin=0.5cm}
        \vspace*{-15pt}
        \caption{\scriptsize{Number of basic blocks reached in \emph{\bf libxml} ($p_{1}$=$0.042$,$p_{2}$=$0.094$).}}
        \label{fig:eval:fast:libxml2}
    \end{subfigure}\\
    \begin{subfigure}[b]{0.24\textwidth}
        \centering
        \includegraphics[width=1\textwidth]{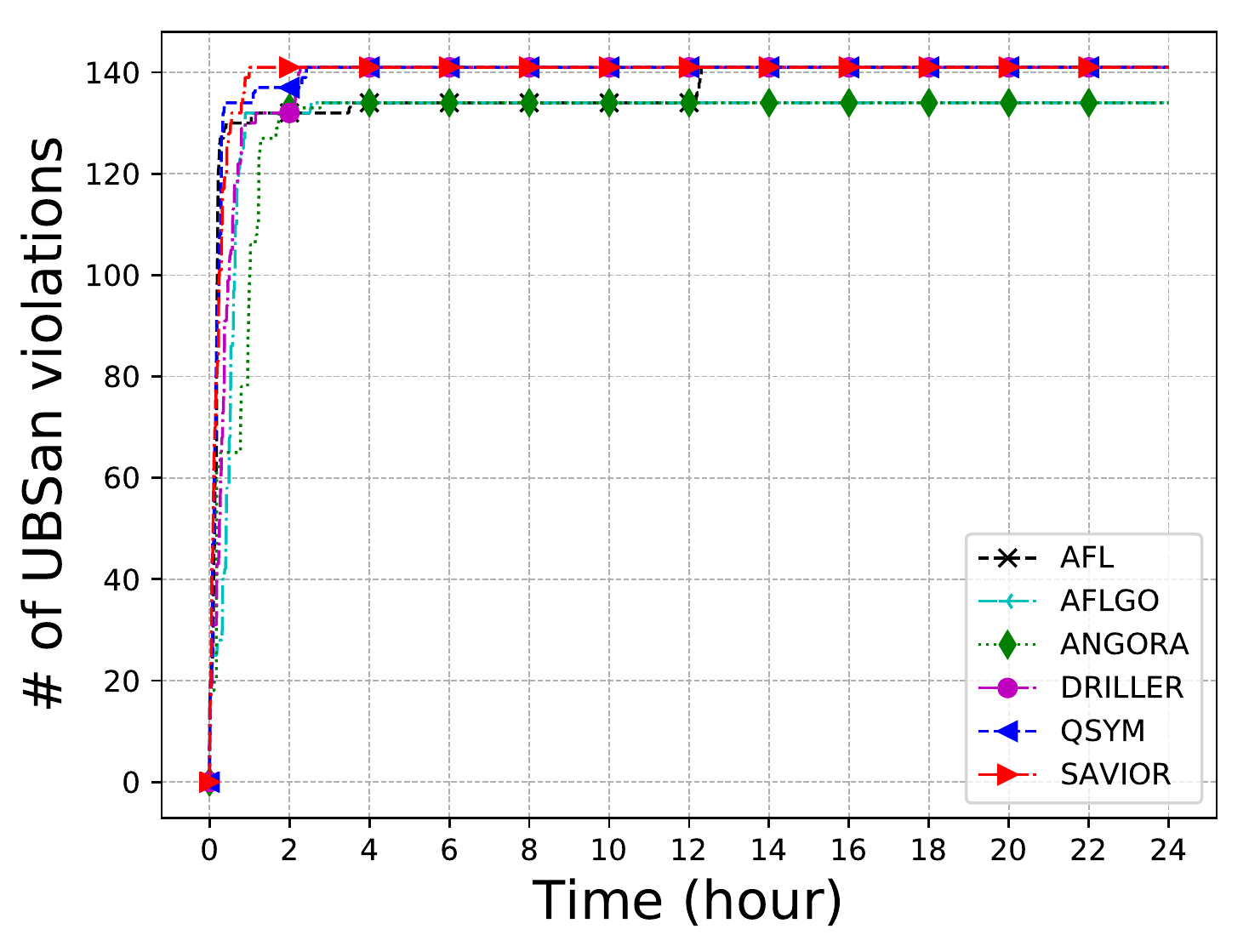}
        \captionsetup{margin=0.5cm}
       \vspace*{-15pt}
        \caption{\scriptsize{Number of UBSan violations triggered in \emph{\bf djpeg} ($p_{1}$=$0.777$,$p_{2}$=$0.203$).}}
        \label{fig:eval:fast:jpeg1}
    \end{subfigure}
     \begin{subfigure}[b]{0.24\textwidth}
        \centering
        \includegraphics[width=1\textwidth]{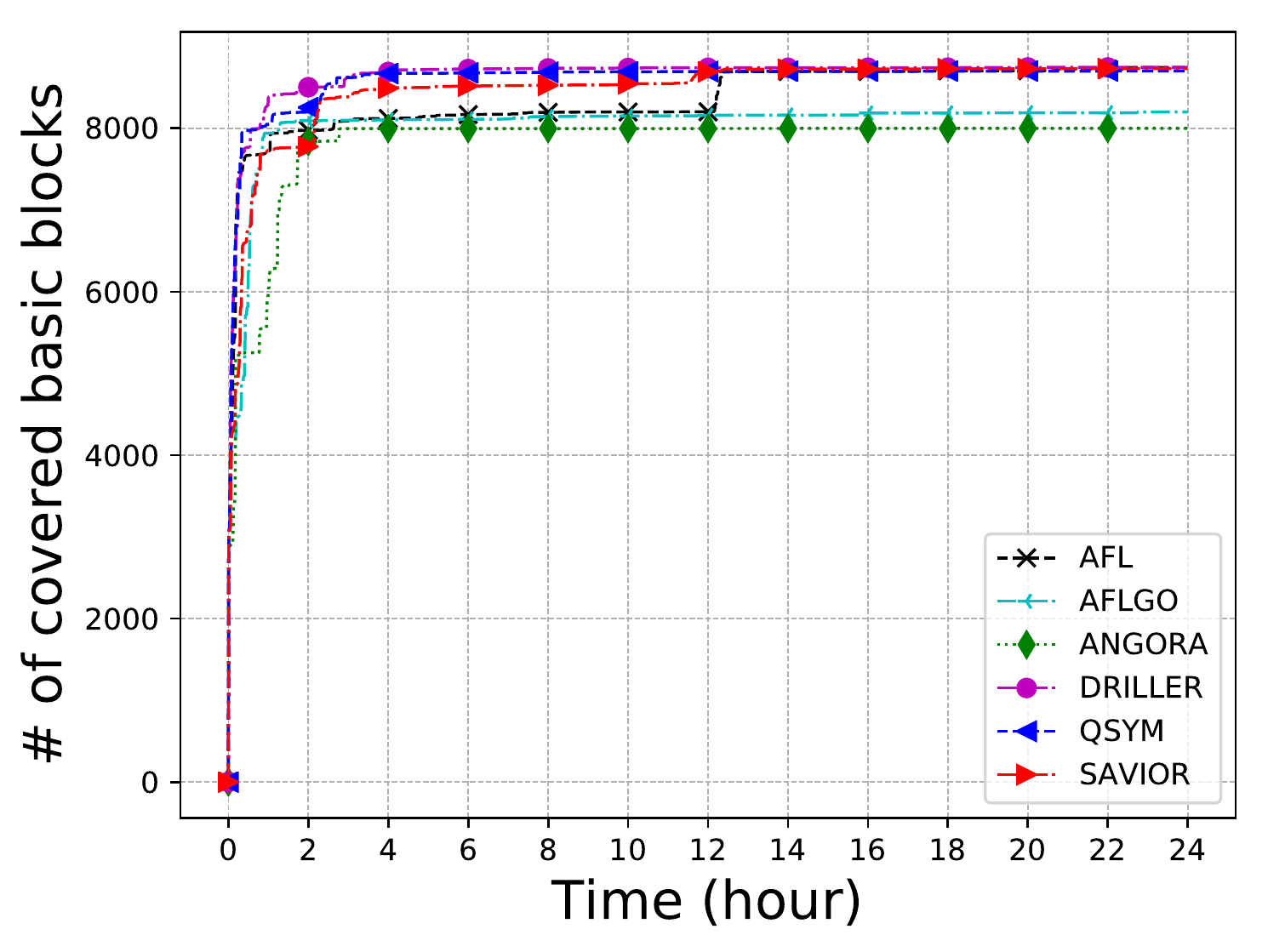}
        \captionsetup{margin=0.5cm}
        \vspace*{-15pt}
        \caption{\scriptsize{Number of basic blocks reached in \emph{\bf djpeg} ($p_{1}$=$3.28*e^{-7}$,$p_{2}$=$3.79*e^{-6}$).}}
        \label{fig:eval:fast:jpeg2}
    \end{subfigure}\rulesep
     \begin{subfigure}[b]{0.24\textwidth}
        \centering
        \includegraphics[width=1\textwidth]{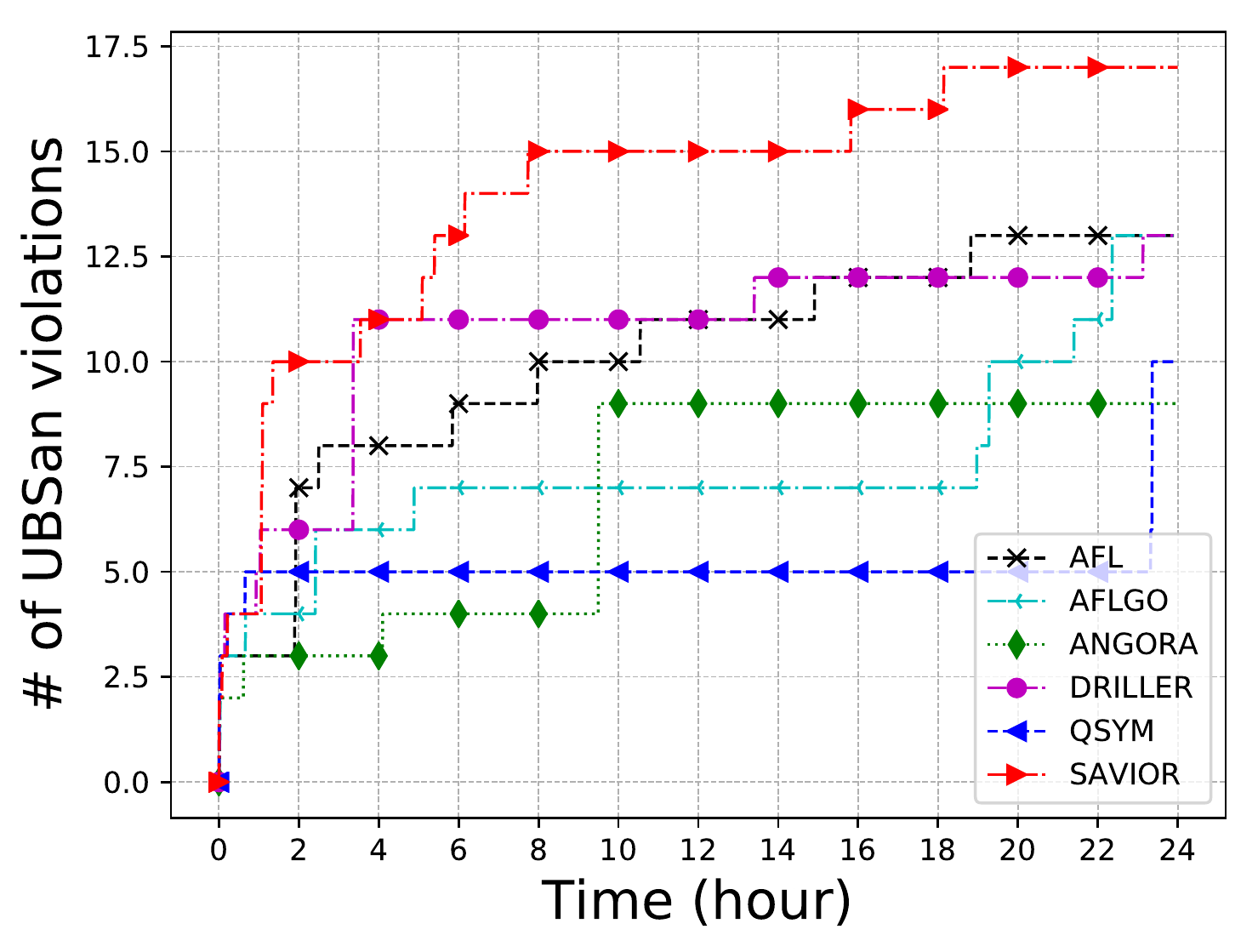}
        \captionsetup{margin=0.5cm}
         \vspace*{-15pt}
        \caption{\scriptsize{Number of UBSan violations triggered in \emph{\bf tiff2pdf} ($p_{1}$=$0.002$,$p_{2}$=$3.95*e^{-6}$).}}
        \label{fig:eval:fast:tiff2pdf1}
    \end{subfigure}
     \begin{subfigure}[b]{0.24\textwidth}
        \centering
        \includegraphics[width=1\textwidth]{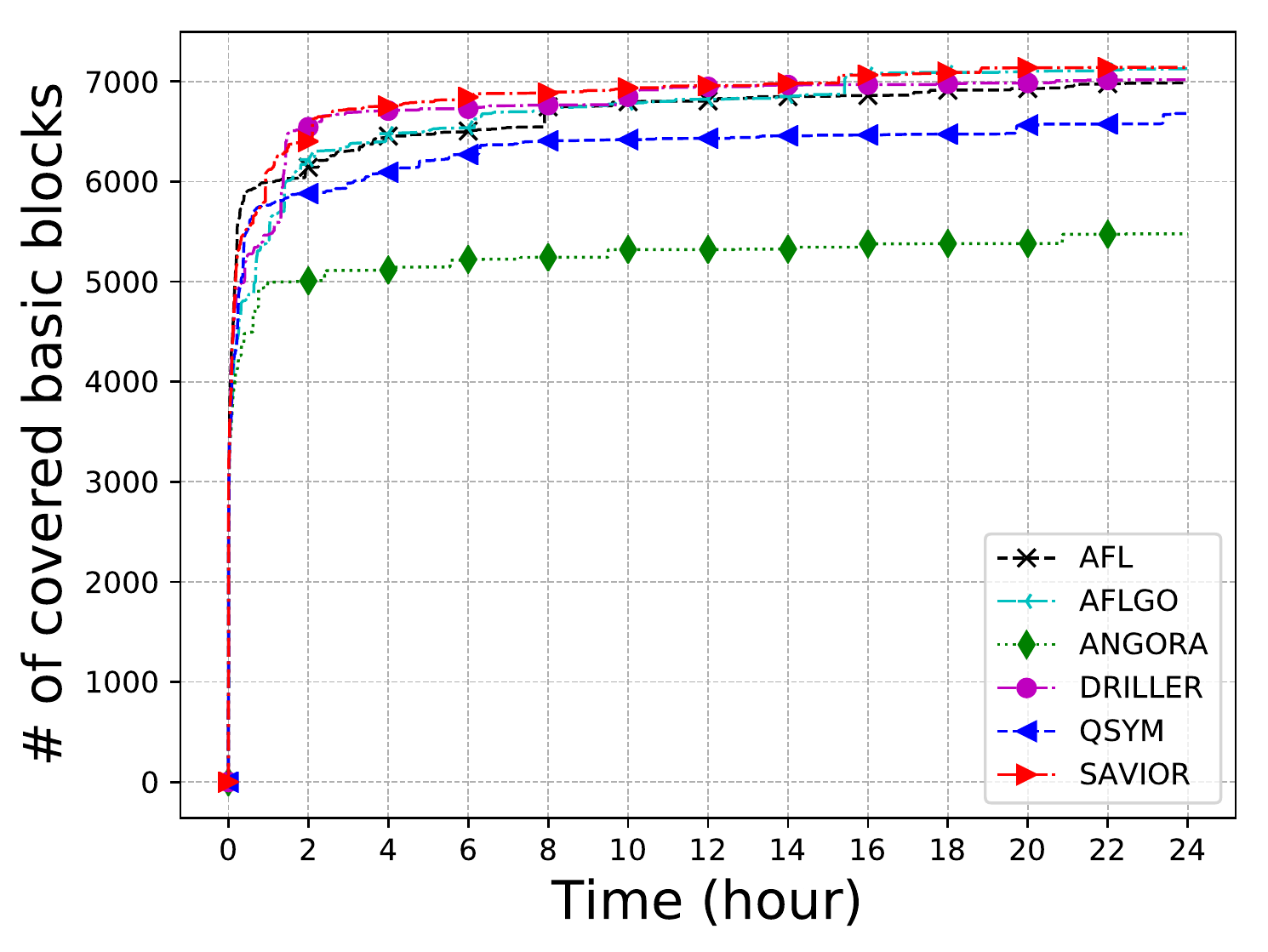}
        \captionsetup{margin=0.5cm}
       \vspace*{-15pt}
        \caption{\scriptsize{Number of basic blocks reached in \emph{\bf tiff2pdf} ($p_{1}$=$0.009$,$p_{2}$=$0.807$).}}
        \label{fig:eval:fast:tiff2pdf2}
    \end{subfigure} \\
     \begin{subfigure}[b]{0.24\textwidth}
        \centering
        \includegraphics[width=1\textwidth]{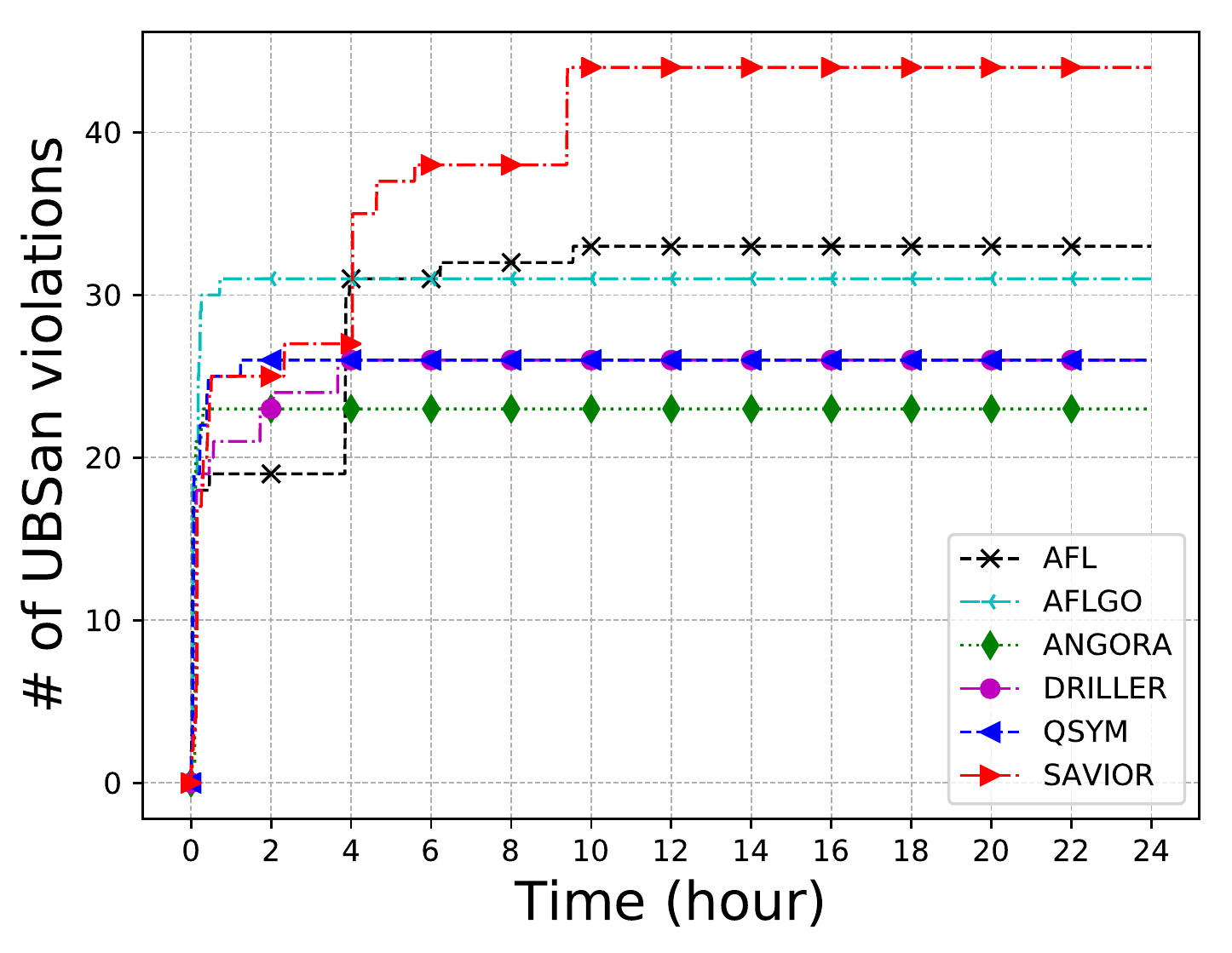}
        \captionsetup{margin=0.5cm}
         \vspace*{-15pt}
        \caption{\scriptsize{Number of UBSan violations triggered in \emph{\bf jasper} ($p_{1}$=$0.010$,$p_{2}$=$0.002$).}}
        \label{fig:eval:fast:jasper1}
    \end{subfigure}
     \begin{subfigure}[b]{0.24\textwidth}
        \centering
        \includegraphics[width=1\textwidth]{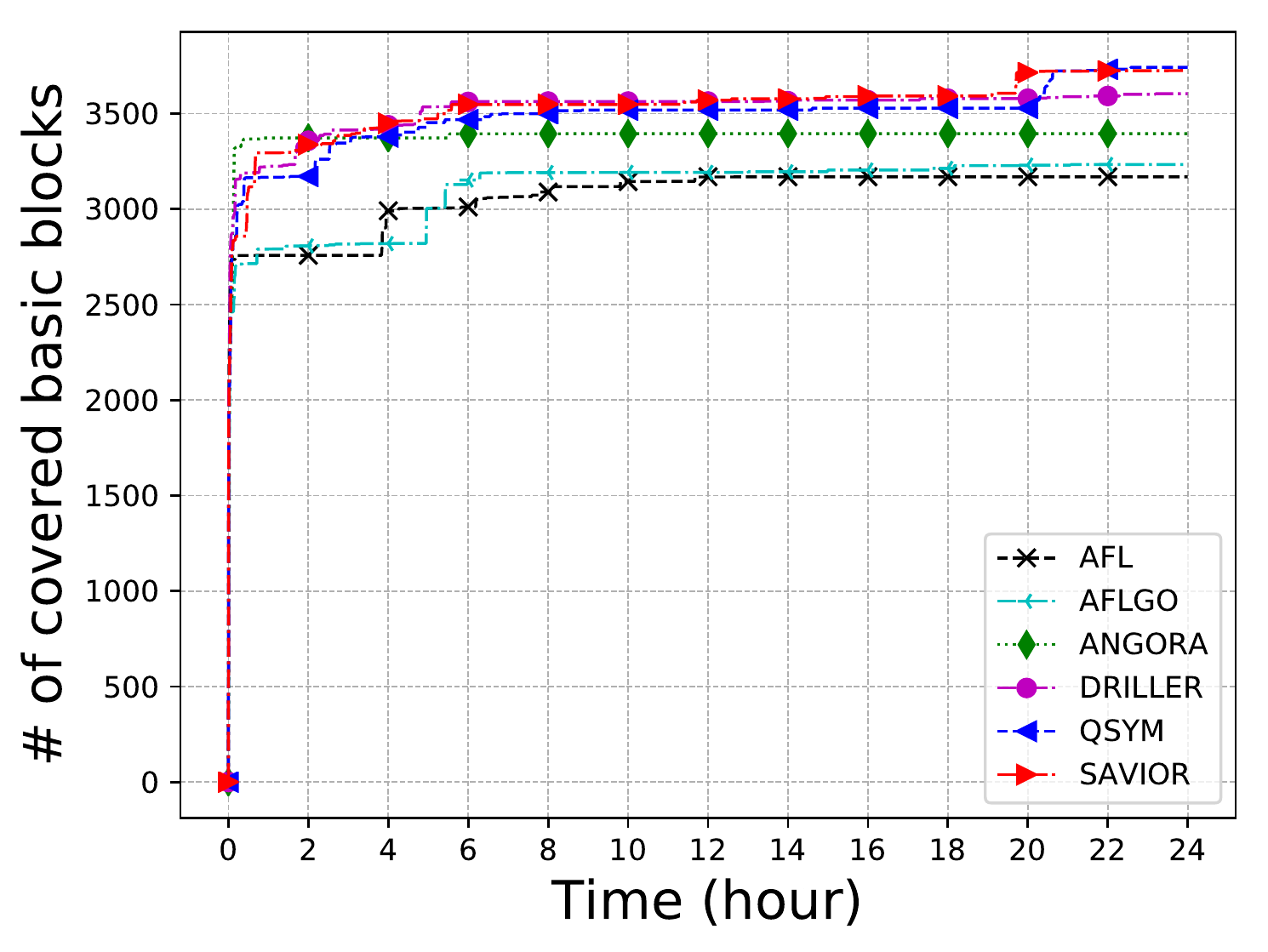}
        \captionsetup{margin=0.5cm}
         \vspace*{-15pt}
        \caption{\scriptsize{Number of basic blocks reached in \emph{\bf jasper} ($p_{1}$=$0.287$,$p_{2}$=$0.653$).}}
        \label{fig:eval:fast:jasper2}
    \end{subfigure}\rulesep
    \begin{subfigure}[b]{0.24\textwidth}
        \centering
        \includegraphics[width=1\textwidth]{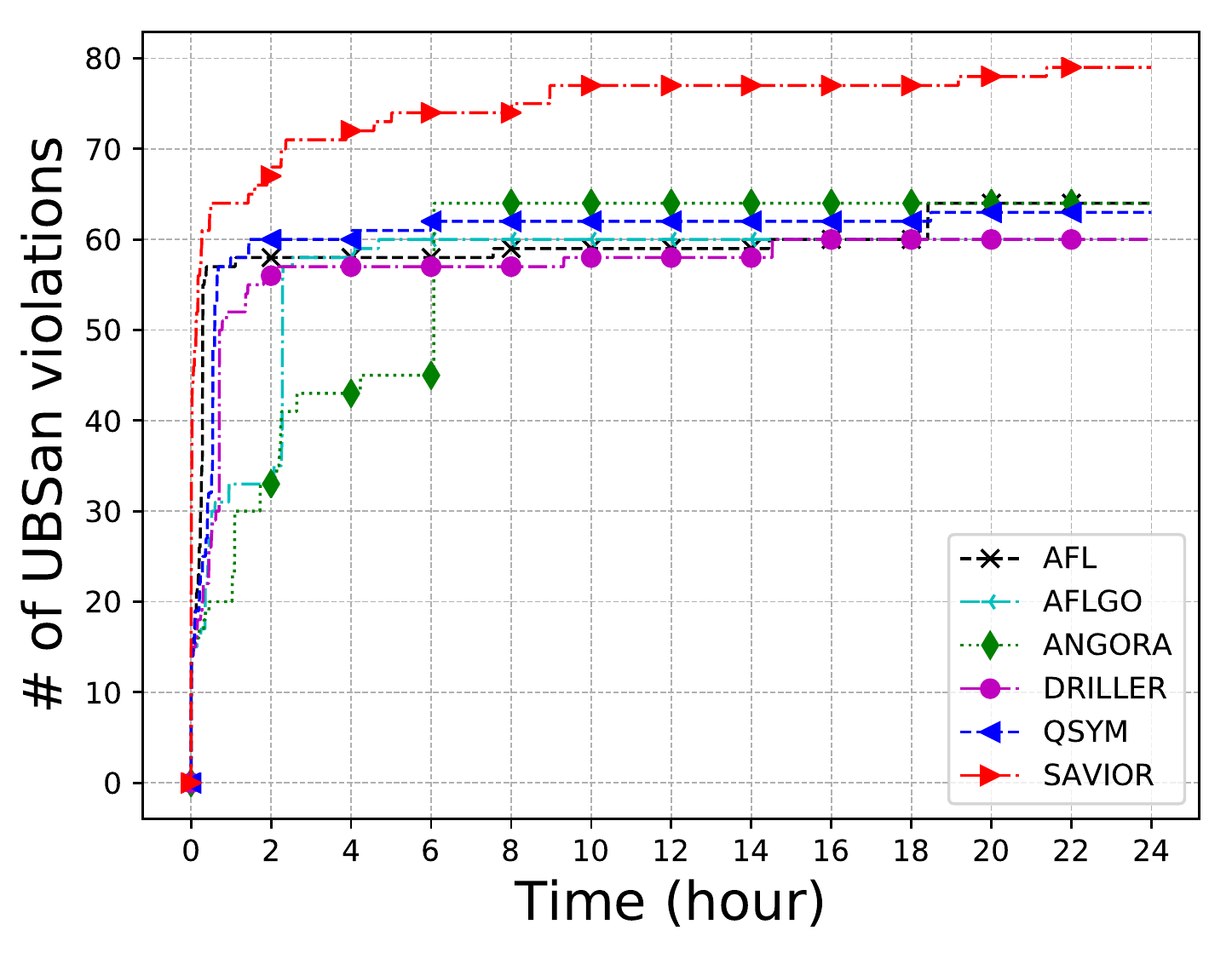}
        \captionsetup{margin=0.5cm}
         \vspace*{-15pt}
        \caption{\scriptsize{Number of UBSan violations triggered in \emph{\bf objdump} ($p_{1}$=$0.096$, $p_{2}$=$0.003$).}}
        \label{fig:eval:fast:objdump1}
    \end{subfigure}
     \begin{subfigure}[b]{0.24\textwidth}
        \centering
        \includegraphics[width=1\textwidth]{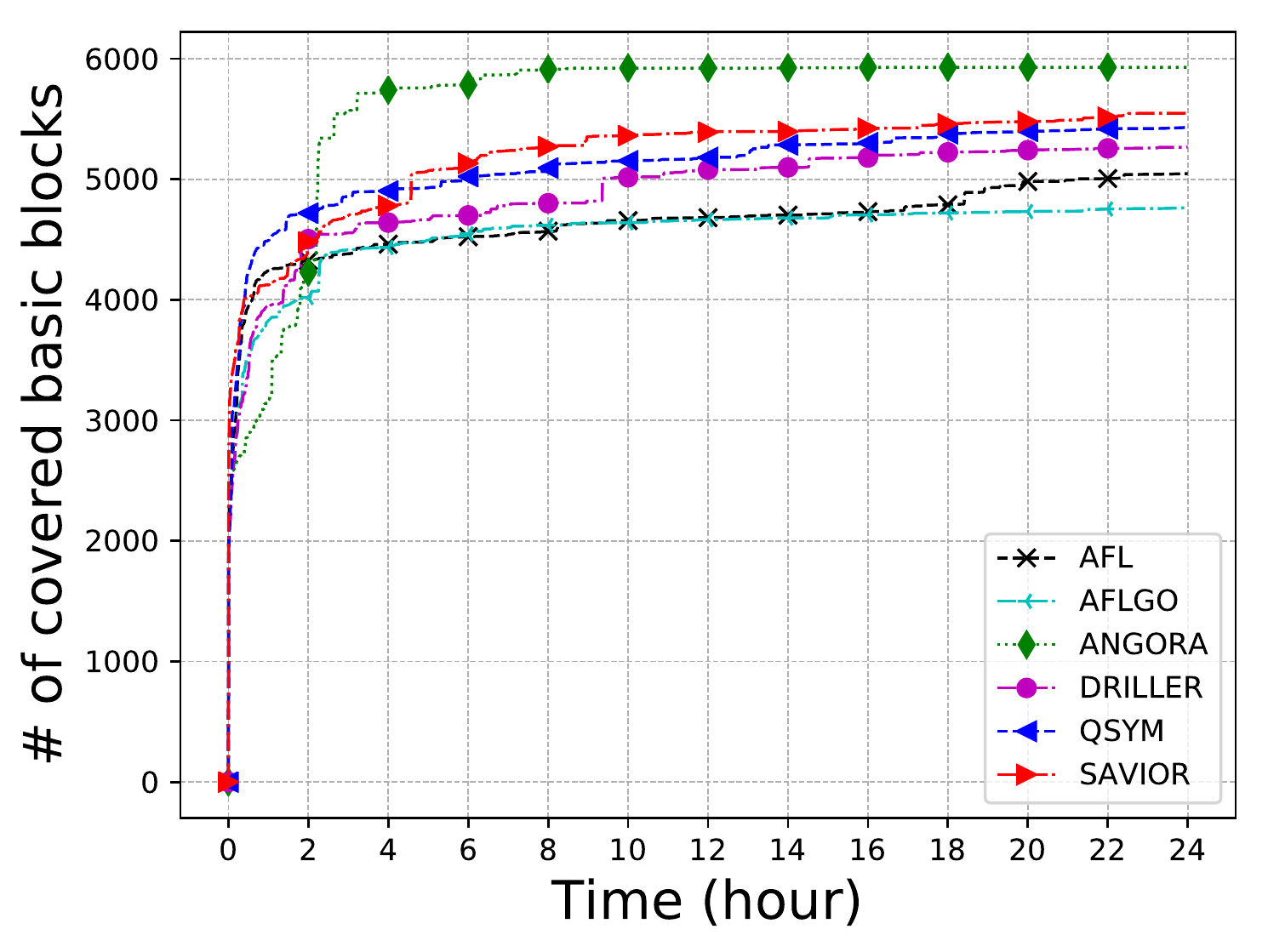}
        \captionsetup{margin=0.5cm}
         \vspace*{-15pt}
        \caption{\scriptsize{Number of basic blocks reached in \emph{\bf objdump} ($p_{1}$=$0.0001$, $p_{2}$=$0.125$).}}
        \label{fig:eval:fast:objdump2}
    \end{subfigure}
    \caption{Evaluation results with real-world programs. Each program takes two columns, respectively 
    showing the number of triggered UBSan violations and the amount of covered basic blocks by the fuzzers over 24 hours. $p_{1}$ and $p_{2}$ are the p-values for the Mann Whitney U-test of \savior vs. \driller and \savior vs. \qsym, respectively.}
    \vspace{-2ex}
    \label{fig:eval-res-real}
\end{figure*}

In Figure~\ref{fig:eval-res-real}, we summarize the results
of our second experiment. It shows the outputs over time from two metrics, 
including the number of triggered UBSan bugs and basic block coverage.
In addition, we calculate the p-values for Mann Whitney U-test of  \savior vs. \driller and 
\savior vs. \qsym.  Note that we use the IDs of UBSan labels for de-duplication 
while counting the UBSan bugs, as each UBSan label is associated with
a unique potential defect. In the following, we delve into the details and explain how 
these results testify our design hypotheses. 
 \setlength{\belowcaptionskip}{-0.5pt}
\setlength{\belowcaptionskip}{-4pt}

\point{Vulnerability Finding Efficiency} 
As shown in Figure~\ref{fig:eval-res-real} (the left column of each program), 
\savior triggers UBSan violations with a pace generally faster than all the other 
fuzzers. In particular, it outperforms \driller and \qsym in 
all the cases except {\tt djpeg}. 
On average, \savior discovers vulnerabilities \MOREBUGDRILLER
faster than \driller and \MOREBUGQSYM faster than \qsym. 
The low p-values ($< 0.05$)\footnote{The p-values of {\tt readelf} and {\tt objdump} 
are larger than $0.05$ but they are at the level of quasi-significance. In the two programs, the variances are mainly due to randomness.} 
of Mann Whitney U-test well support that 
these improvements are statistically significant. 
Since the three hybrid tools only differ in the way of seed scheduling,
these results strongly demonstrate that \emph{the scheduling scheme in \savior ---
bug-driven prioritization --- accelerates vulnerability finding}. 
In the case of {\tt djpeg}, all six fuzzers trigger the same group of UBSan violations. 
This is because {\tt djpeg} has a tiny code base, with which these fuzzers 
quickly saturate on code exploration. In addition, 
the conditions of those UBSan violations are 
simple that even mutation-based approaches can solve. For a better reference, 
we also summarize the number of triggered violations at the end of 24 hours 
in Table~\ref{tab:ubsanviolations} at Appendix~\ref{subsec:appendix:bugdiff}. 

Going beyond, we examine the number of labels that are reached 
by different fuzzers.
In Table~\ref{tab:findbugspeed}, we list the average results 
from our 24-hour tests. Not surprisingly, the hybrid tools cover 
higher volumes of UBSan labels than the ordinary fuzzers. 
This is likely because a hybrid tool can solve complex conditions, 
enabling the coverage on the code and labels behind. 
Among the hybrid tools, \savior reaches \MORECOVDRILLER and \MORECOVQSYM  
more labels than \driller and \qsym, respectively.
Such results are consistent with the number of triggered UBSan violations. 
This also signifies that our bug-driven prioritization guides \savior to spend more resources on code with 
richer UBSan labels. In the case of {\tt djpeg}, \savior nearly ties with the other tools. This is due to a similar reason as explained above. 

We further find that the efficiency boost of \savior in vulnerability finding is 
not due to high code coverage. 
As shown in Figure~\ref{fig:eval-res-real} (the right column for each program), 
we compare the code coverage of the six fuzzers. 
As demonstrated by the results, the efficiency of code coverage and 
UBSan violation discovery are not positively correlated. 
Particularly, in the case of {\tt tcpdump}, {\tt libxml}, {\tt tiff2pdf}, 
{\tt objdump} and {\tt jasper}, 
\savior covers code in a similar or even slower pace than \driller 
and \qsym (the high p-values also support that \savior is not quicker). 
However, \savior 
triggers UBSan violations significantly quicker in these cases. 
Such results validate the above hypothesis with high confidence. 

\begin{table}[t!]
\scriptsize
\centering
\begin{tabular}{l|p{8mm}p{8mm}p{8mm}p{8mm}p{8mm}p{8mm}}
\toprule[0.5pt]
\toprule[0.5pt]

\multirow{2}{*}{\bf{\emph{Prog.}}} & \multicolumn{6}{c}{\bf{\emph{Number of reached UBSan labels}}}

 \\ \cline{2-7}
 & \afl  & \aflgo & \angora & \driller  & \qsym & \savior \\\hline
\rowcolor{mygray}
tcpdump & 2029  &1235 & 1333 & 1906  & 2509 & 2582 \\
tiff2ps & 748  & 927 & 770 & 931  & 852 & 970 \\
\rowcolor{mygray}
readelf & 91  & 79 & 102 & 104  & 106 & 183 \\
xmllint & 588  & 580 & 456 & 567  & 568 & 597\\
\rowcolor{mygray}
djpeg & 2746  & 2588 & 2546 & 2713  & 2707 & 2746\\
tiff2pdf & 1488  & 1467 & 919 & 1448  & 1369 & 1478\\
\rowcolor{mygray}
jasper & 649  & 660 & 679 & 691  & 731 & 752\\
objdump & 780  & 715 & 844 & 835  & 906 & 1039\\
\hline {\bf Avg.} & {\bf 1139} & {\bf 1031} & {\bf 956} & {\bf 1149} & {\bf 1218} & {\bf 1289}\\

\bottomrule[0.5pt]
\bottomrule[0.5pt]
\end{tabular}
\caption{Number of unique UBSan labels reached by different fuzzers in 24 hours. On average \savior reaches \MORECOVDRILLER and \MORECOVQSYM more labels than \driller and \qsym.}
\label{tab:findbugspeed}
\vspace{-2ex}
\end{table}

\point{Vulnerability Finding Thoroughness} 
In this experiment, we also measure the performance of \bugguidedsearch in enhancing the thoroughness of vulnerability finding.
Specifically, we re-run the seeds from all the fuzzers with our 
concolic executor. In this test, we enable \savior to do constraint solving 
only when encountering un-solved UBSan labels. 

In Table~\ref{tab:savior-search-bug}, we summarize the comparison results. 
For all the \PROGNUM programs, \bugguidedsearch facilitates different fuzzers 
to trigger new violations. The average increase ranges from {\bf 4.5\%} (\savior) 
to {\bf 61.2\%} (\angora). In particular, it aids \angora to trigger 82 new 
UBSan bugs in total. 
In the case of {\tt djpeg} \bugguidedsearch does not help 
much. This is because {\tt djpeg} has a
relatively smaller code base and contains fewer vulnerability labels,
making \bugguidedsearch less utilized. These results are further evidence
that \emph{\bugguidedsearch can truly benefit fuzzing in 
terms of vulnerability finding thoroughness}. 

\begin{table}[h!]
\centering
\scriptsize
\begin{tabular}{p{10mm}|p{8mm}p{8mm}p{8.5mm}p{8.5mm}p{8mm}p{8.5mm}}
\toprule[0.5pt]
\toprule[0.5pt]

\multirow{2}{*}{\bf{\emph{Prog.}}} & \multicolumn{6}{c}{\bf{\emph{Improvements by \bugguidedsearch}}}   
 \\ \cline{2-7}
 & \afl & \aflgo & \angora & \driller & \qsym & \savior
 \\ \hline

\rowcolor{mygray}
tcpdump  & +10/11\% & +22/41.5\% & +29/76.3\% & +9/9.9\% & +4/4\% & +8/7\%\\
tiff2ps  & +4/133\% & +0/0\% & +3/42.9\% & +0/0\% & +0/0\% & +0/0\%\\
\rowcolor{mygray}
readelf  & +10/82\% & +9/72.2\% & +16/107\% & +9/68.4\% & +8/63.2\%& +7/29.2\%\\
libxml  & +4/33.3\% & +4/33.3\% & +5/166.7\% & +4/33.3\% & +4/33.3\% & +0/0\%\\
\rowcolor{mygray}
tiff2pdf  & +5/50\% & +1/7.7\% & +4/44.4\% & +3/27.2\% & +5/62.5\% & +0/0\%\\
djpeg  & +0/0\% & +7/5.2\% & +7/5.2\% & +0/0\% & +0/0\% & +0/0\%\\
\rowcolor{mygray}
objdump  & +7/10.9\% & +7/11.7\% & +11/17.2\% & +7/11.7\% & +6/9.5\% & +0/0\%\\
jasper  & +0/0\% & +0/0\% & +7/30.4\% & +7/26.9\% & +7/26.9\% & +0/0\%\\
\hline
\rowcolor{mygray}
{\bf Ave.}  & {\bf +5/40.1\%} &  {\bf +6/21.5\%}&  {\bf +10/61.2\%} &  {\bf +5/22.2\%} &   {\bf +4.3/25\%} &  {\bf +1.8/4.5\%}\\

\bottomrule[0.5pt]
\bottomrule[0.5pt]
\end{tabular}
\caption{New UBSan violations triggered with \bugguidedsearch in the evaluation with real-world programs. \emph{``+X/Y\%''} means \emph{``X''} new violations are triggered, increasing the total number by \emph{``Y\%''}. 
}
\label{tab:savior-search-bug}
\vspace{-2ex}
\end{table}

\subsection{Vulnerability Triage} The UBSan violations triggered by \savior could lead to various consequences and some of them might be harmless. Therefore, we manually examine all the UBSan violations triggered by \savior. 
These violations include those triggered in the 8 programs in Table~\ref{tab:eval-setup} and also those from {\tt mjs}, {\tt catdoc}, 
and {\tt c++filt}. We do not include the results of  {\tt mjs}, {\tt catdoc}, 
and {\tt c++filt} in the evaluation above, as all fuzzers trigger fewer than 10 UBSan violations. A small difference would result in 
a big variance in comparison. 

\point{Triage Result} In total, we collect \ALLNUM UBSan violations and 
we manually classify them based on their consequences and present the 
results in Table~\ref{tab:ubsan-bug}. 
Specifically, \OOBNUM of 
them lead to OOB reads/writes and \LOGICNUM of them result in logic errors. 
Those logic errors consist of different categories, such as 
incorrect computation,
wrong outputs, and polluted conditional variables. 
Among the {\bf \REALBUGNUM} OOB and logic errors, 
\CONFIRMNUM of them have been confirmed by the developers. 
Our further analysis so far reveals at least \EXPNUM of them are 
exploitable for goals such as information leak and control flow 
manipulation. 

The remaining {\bf 238} cases 
are likely harmless according to our triage result. They 
mainly consist of the following categories: (1) the variables 
triggering UBSan violations are used as storage (\eg {\tt int} as 
{\tt char[4]}) instead of computation-related objects; 
(2) the affected variables expire immediately after the violations; 
(3) the program already considers the case of UBSan violations 
and has handlers. 

\begin{table}[t!]

\centering
\scriptsize
\begin{tabular}{l|cc|cc}
\toprule[0.5pt]
\toprule[0.5pt]

\multirow{2}{*}{{\bf{\emph{Program}}}} & 
\multicolumn{2}{c|}{\bf{\emph{Defect categories}}}   &
\multicolumn{2}{c}{\bf{\emph{Note}}}   
 \\ \cline{2-5}
 & {\tt OOB}  & {\tt Logic Error} & {\tt Exploitable*} & {\tt Confirmed} 
 \\ \hline 
 \rowcolor{mygray}
tcpdump &  6 & 102 & 6+   &7 \\
libjpeg &  8 & 23 & 0+   & N/A \\
\rowcolor{mygray}
objdump &  41 & 4 & 4+  & N/A \\
readelf &  1  & 9& 10+  & 3 \\
\rowcolor{mygray}
libtiff &  20 & 0 & 0+  & N/A \\
jasper & 21 &  2 & 2+  & 2\\
\rowcolor{mygray}
mjs & 1 & 0 & 0+ &  1\\
catdoc & 3 & 0 & 3+  & 1\\
\rowcolor{mygray}
c++filt & 1 & 1 & 0  & 2 \\\hline
{\bf Total} & {\bf \OOBNUM} & {\bf \LOGICNUM} & {\bf 25+}   & {\bf 16} \\

\bottomrule[0.5pt]
\bottomrule[0.5pt]
\end{tabular}
\caption{Triage of UBsan violations triggered by \savior in 24 hours.}
\label{tab:ubsan-bug}
\vspace{-2ex}
\end{table}

\point{Case Studies} From each of the three categories (OOB, logic errors, 
and those without harm), we pick a case and explain the details here. 
All the cases have been fixed. 

The first case is an OOB in {\tt readelf}. The code is shown below. 
The variable {\tt inote.namesz} is copied from input. By making it 
equal to 0, ({\tt inote.namesz} $-$ 1) under-flows to the maximal 
unsigned value. It causes an OOB access to {\tt inote.namedata}.
\begin{lstlisting}
static bool process_notes_at(...){
 //readelf.c:18303
 if(inote.namedata[inote.namesz-1] != '\0') 
   ...
}
\end{lstlisting}

The second case is a logic error in {\tt libtiff}. Variable {\tt twobitdeltas[delta]} 
is controlled by user. With a specially 
crafted input, one can cause an overflow in the result of  {\tt lastpixel + twobitdeltas[delta]},
making {\tt SETPIXEL} set the wrong pixel value to the decoded image. 

\begin{lstlisting}
static int ThunderDecode(...){
 //tif_thunder.c:125
 if((delta = ((n >> 4) & 3)) != DELTA2_SKIP)
    SETPIXEL(op, lastpixel + twobitdeltas[delta]);
  ...
}
\end{lstlisting}

The last case is harmless, as the program already considers overflow. 
This case locates in {\tt libxml}. As shown below, with a special input, 
the variable {\tt okey} can be overflowed. However, the program modulo 
{\tt okey} with {\tt dict->size} before using it, making the overflow harmless. 

\begin{lstlisting}
static int xmlDictGrow(...) { 
  // dict.c:417
  okey = xmlDictComputeQKey(...);
  key = okey % dict->size;
  ...
}
\end{lstlisting}


%
%


%% file: related.tex
\section{Related Works}
\label{sec:related}
The lines of works mostly related to 
our work include advanced fuzzing, concolic execution, the state-of-the-art hybrid testing techniques,
and those that facilitate guided testing. 

\subsection{Advanced Fuzzing}
Many recent works focus on improving the capability of code exploration in fuzzing. 
CollAFL~\cite{collafl} aims to reduce hash collision in coverage feedback 
to decrease false negatives. PTrix~\cite{ptrix} enables path-sensitive fuzzing based on efficient hardware tracing.
\tfuzz~\cite{tfuzz} transforms tested programs to bypass complex conditions and 
improve code coverage, and later uses a validator to reproduce the inputs that work for 
the original program. To generate high-quality seeds, ProFuzzer~\cite{profuzzer} infers 
the structural information of the inputs. 
Along the line of seed generation, Angora~\cite{angora} assumes a black-box function 
at each conditional statement and applies gradient descent to find satisfying input bytes. 
This method is later improved by NEUZZ~\cite{neuzz} with a smooth surrogate function to approximate the behavior 
of the tested program. Compared with these approaches, \savior takes the bug-driven guidance
to maximize bug coverage and verifies the (non-)existence of these bugs in the explored paths. 

\subsection{Concolic Execution}
Symbolic execution, a systematic approach introduced in the
1970s~\cite{King76, Howden77} for program testing, has attracted new attention
due to the advances in satisfiability modulo theory~\cite{GaneshD07, MouraDS07,MouraB11}.  
However, classic symbolic execution has the problems of high computation cost and 
path explosion. To tackle these issues, Sen proposes concolic execution~\cite{Sen07a},
which combines the constraint solving from symbolic execution and the fast execution 
of concrete testing. 
Concolic execution increases the coverage of random testing~\cite{sage,GodefroidKS05} 
while also scales to large software. Hence, it has been adopted in various 
frameworks~\cite{SenA06,SenMA05,BurnimS08,s2e}.
Recently, concolic execution is also widely applied in automated vulnerability detection and 
exploitation, in which the concolic component provides critical inputs 
by incorporating security-related predicates ~\cite{ChaARB12,AEG}.

However, concolic execution operates based on emulation or heavy instrumentation, incurring tremendous execution overhead. Purely relying on concolic execution for code exploration is less practical for large software that involves large amounts of operations. 
In contrast, hybrid testing runs fuzzing for code exploration and invokes concolic execution only on hard-to-solve branches. This takes advantage 
of both fuzzer's efficiency and concolic executor's constraint solving. 


\subsection{Hybrid Testing}
Majundar et al.~\cite{majumdar2007hybrid} introduce the idea of hybrid concolic
testing a decade ago. This idea offsets the deficiency of both random testing 
and concolic execution. Specifically, their approach interleaves random testing and 
concolic execution to deeply explore a wide program state space. 
Subsequent development reinforces hybrid testing by replacing random testing with guided 
fuzzing~\cite{pak2012hybrid}. This approach could rapidly contributing more high-quality seeds to concolic execution.

Recently, \driller~\cite{driller} engineers the state-of-the-art hybrid testing system. It more 
coherently combines fuzzing and concolic execution and can seamlessly test various
software systems. Despite the advancement, 
\driller still achieves unsound vulnerability detection. DigFuzz~\cite{DigFuzz} is a
more recent work that tries to better coordinate the fuzzing and concolic execution components.
Using a Monte Carlo algorithm, DigFuzz predicts the difficulty for a fuzzer 
to explore a path and prioritizes to explore seeds with a higher difficulty score.  

Moreover, motivated by the growing demands in software testing, researchers 
have been reasoning the performance of hybrid testing. As commonly understood, 
hybrid testing is largely restricted by the slow concolic execution. 
To this end, QSYM~\cite{qsyminsu} implements a concolic executor that tailors 
the heavy but unnecessary computations in symbolic interpretation and constraint solving.
It leads to times of acceleration. 

Differing from the above works that bring code-coverage improvement,
\savior changes the philosophy of hybrid testing. It drives the concolic 
executor on seeds with higher potential and
guides the verification of the encountered vulnerabilities. 
This leads to quicker and better bug coverage. 



\subsection{Guided Software Testing} This line of research~\cite{GuoKWYG15,MarinescuC13,aflgo,christakis} aims to guide the testing towards
exploring specific code locations. Katch~\cite{MarinescuC13} prioritizes the seeds that approach patches to guide the symbolic executor. Together with three other guiding schemes, Katch can efficiently 
cover the target code.
With a similar goal, \aflgo~\cite{aflgo} calculates the distance from 
each code region to the targets 
(\eg vulnerable code regions or patches). In fuzz testing, \aflgo favors 
seeds that exercise code regions with smaller distances.
Christakis et al.~\cite{christakis} proposes to prune paths in dynamic 
symbolic execution. It discards paths that carry properties that have been verified. 
However, the existing works generally prefer seeds that approach the targets quicker, 
which oftentimes carry shallow contexts. Instead, \savior values all the seeds with 
high potential, creating various contexts to exercise the target code.
This enables \savior to outperforms these existing guided testing techniques in bug finding. 
Some other works use static 
analysis to label potential vulnerabilities, such as using data flow analysis to 
pinpoint data leaks~\cite{ArztRHB15}, using slicing to mark use-after-free paths~\cite{FeistMBDP16}, and using taint analysis to mark possible races~\cite{LiLG14}. they then rely on subsequent symbolic execution to confirm detection. These analyses are complementary to \savior. In addition, \savior relies on fuzz testing to stably approach the to-be-verified paths, while others use heuristic based approaches to guide symbolic execution towards the marked label.

%% file: conclude.tex
\section{Conclusion}
\label{sec:conclude}
We introduce \savior, a new hybrid testing approach in this work. Unlike the mainstream hybrid 
testing tools which follow the coverage-driven design, \savior moves towards being a bug-driven. 
We accordingly propose in \savior two novel techniques, named \emph{bug-driven prioritization} 
and \emph{\bugguidedsearch}, respectively. On one hand, \savior prioritizes the concolic
execution to run seeds with more potentials of leading to vulnerabilities. On the other hand, 
\savior examines all vulnerable candidates along the running program path in concolic execution. 
By modeling the unsafe conditions in SMT constraints, it solves for proofs of valid vulnerabilities 
or proves that the corresponding vulnerabilities do not exist. 
\savior significantly outperforms the existing coverage-driven tools. On average, it detects 
vulnerabilities \MOREBUGDRILLER faster than \driller and \MOREBUGQSYM faster than \qsym, resulting 
in the discovery of \BUGDIFFDRILLER and \BUGDIFFQSYM more security violations in 24 hours.

%% file: ack.tex
\section*{Acknowledgments}
\label{sec:acks}

We would like to thank our shepherd Mathias Payer and the anonymous reviewers 
for their feedback. This project was supported by the Office 
of Naval Research (Grant\#: N00014-17-1-2891, N00014-18-1-2043, 
and N00014-17-1-2787). Any opinions, findings, and conclusions or
recommendations expressed in this paper are those of the authors and do not
necessarily reflect the views of the funding agency.

%% file: appendix.tex
\begin{appendices}
\section{Supplementary Figures and Evaluation Data}
\label{sec:appendix1}

\begin{figure}[h]
    \centering
    \includegraphics[width=0.45\textwidth]{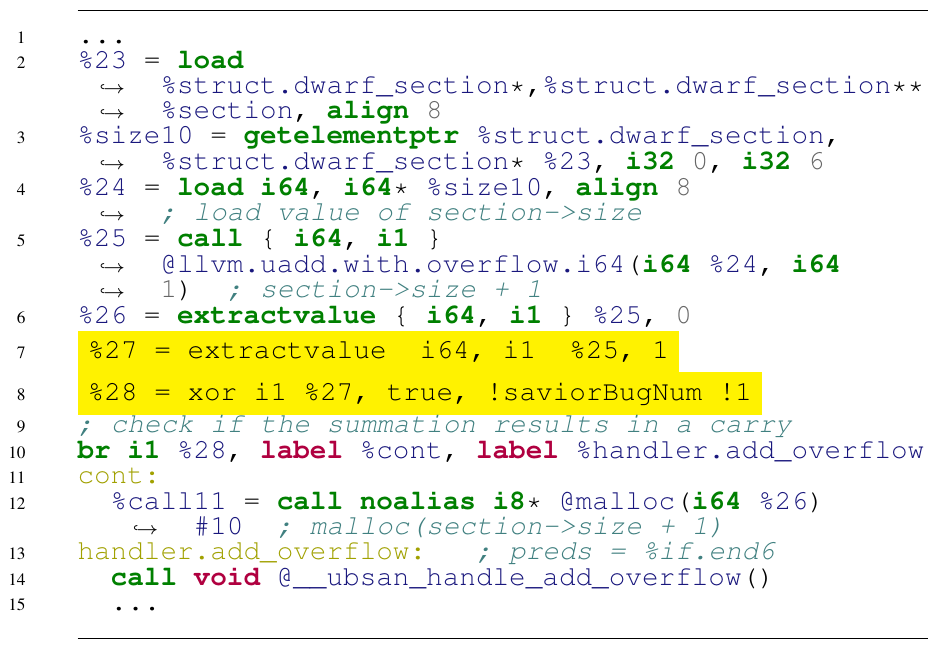}
    \caption{\savior instrumentation of UBSan label.}
    \label{fig:bug:objllvm}
\end{figure}
\subsection{Program Instrumentation}

Figure~\ref{fig:bug:objllvm} shows the UBSan-instrumented 
LLVM IR for the {\tt objdump} defect in our motivating example, 
of which source code is presented in Figure~\ref{fig:objbug}. 
In Figure~\ref{fig:bug:objllvm}, we highlight the instrumentation with 
{\tt !saviorBugNum} metadata for bug-driven prioritization.

\subsection{UBSan Label Reduction}
In the process of vulnerability labelling, \savior also reduces labels that can be 
confirmed as false positives. Table ~\ref{tab:labelremove} shows
the results of label reduction on our benchmark programs. 

\begin{table}[t!]
\centering
\begin{tabular}{l|ccc}
\toprule[0.5pt]
\toprule[0.5pt]

\multirow{2}{*}{\bf{\emph{Prog.}}} & \multicolumn{3}{c}{\bf{\emph{Label reduction results}}}

 \\ \cline{2-4}
 & Total UBSan Labels  & Removed UBSan Labels & Percentage  \\\hline
\rowcolor{mygray}
tcpdump   & 13926 & 1924 & 13.8\% \\
tiff2ps   & 1768 & 57 & 3.2\% \\
\rowcolor{mygray}
readelf  & 2476 & 99  & 4.0\%  \\
xmllint  & 5258 & 195  & 3.7\% \\
\rowcolor{mygray}
djpeg  & 9391 & 573  & 6.1\% \\
tiff2pdf   & 3126  & 80 & 2.6\% \\
\rowcolor{mygray}
jasper & 3838  & 228  & 5.9\% \\
objdump & 9025 & 346   & 3.8\% \\
\hline {\bf Average} & {\bf 6106} & {\bf 438} & {\bf 5.36\%} \\

\bottomrule[0.5pt]
\bottomrule[0.5pt]
\end{tabular}
\caption{Number of UBSan labels removed in our benchmark programs. On average, 5.36\% of the labels are reduced.}
\label{tab:labelremove}
\vspace{-2ex}
\end{table}

\subsection{LAVA-M Evaluation}

In the evaluation with LAVA-M, \bugguidedsearch helps identify 
a group of LAVA bugs that are not listed. Table~\ref{tab:new-laval-bug-search} 
shows the IDs of these LAVA bugs.

\begin{table}[h!]

\centering
\scriptsize
\begin{tabular}{l|c}
\toprule[0.5pt]
\toprule[0.5pt]

{\bf{\emph{Program}}} & \bf{\emph{Bugs unlisted by LAVA-M but exposed by \bugguidedsearch}}  \\ \hline

\rowcolor{mygray}
base64 & 274, 521, 526, 527 \\

uniq &  227 \\

\rowcolor{mygray}
md5sum & 281, 287 \\

who &\makecell{1007, 1026, 1034, 1038, 1049, 1054, 1071, 1072, 117, 12\\
125, 1329, 1334, 1339, 1345, 1350, 1355, 1361, 1377, 1382\\
1388, 1393, 1397, 1403, 1408, 1415, 1420, 1429, 1436, 1445\\
1450, 1456, 1461, 16, 165, 169, 1718, 1727, 1728, 173\\
1735, 1736, 1737, 1738, 1747, 1748, 1755, 1756, 177, 181\\
185, 189, 1891, 1892, 1893, 1894, 1903, 1904, 1911, 1912\\
1921, 1925, 193, 1935, 1936, 1943, 1944, 1949, 1953, 197\\
1993, 1995, 1996, 2, 20, 2000, 2004, 2008, 2012, 2014\\
2019, 2023, 2027, 2031, 2034, 2035, 2039, 2043, 2047, 2051\\
2055, 2061, 2065, 2069, 2073, 2077, 2079, 2081, 2083, 210\\
214, 2147, 218, 2181, 2189, 2194, 2198, 2219, 222, 2221\\
2222, 2223, 2225, 2229, 2231, 2235, 2236, 2240, 2244, 2246\\
2247, 2249, 2253, 2255, 2258, 226, 2262, 2266, 2268, 2269\\
2271, 2275, 2282, 2286, 2291, 2295, 2302, 2304, 24, 2462\\
2463, 2464, 2465, 2466, 2467, 2468, 2469, 2499, 2500, 2507\\
2508, 2521, 2522, 2529, 2681, 2682, 2703, 2704, 2723, 2724\\
2742, 2796, 2804, 2806, 2814, 2818, 2823, 2827, 2834, 2838\\
2843, 2847, 2854, 2856, 2919, 2920, 2921, 2922, 294, 2974\\
2975, 298, 2982, 2983, 2994, 2995, 3002, 3003, 3013, 3021\\
303, 307, 3082, 3083, 3099, 312, 316, 3189, 3190, 3191\\
3192, 3198, 3202, 3209, 321, 3213, 3218, 3222, 3237, 3238\\
3239, 3242, 3245, 3247, 3249, 325, 3252, 3256, 3257, 3260\\
3264, 3265, 3267, 3269, 327, 334, 336, 338, 3389, 3439\\
346, 3466, 3468, 3469, 3470, 3471, 3487, 3488, 3495, 3496\\
350, 3509, 3510, 3517, 3518, 3523, 3527, 355, 359, 3939\\
4, 4024, 4025, 4026, 4027, 4222, 4223, 4224, 4225, 4287\\
4295, 450, 454, 459, 463, 468, 472, 477, 481, 483\\
488, 492, 497, 501, 504, 506, 512, 514, 522, 526\\
531, 535, 55, 57, 59, 6, 61, 63, 73, 77\\
8, 81, 85, 89, 974, 975, 994, 995, 996}\\

\bottomrule[0.5pt]
\end{tabular}
\caption{IDs of unlisted bugs in LAVA-M that are triggered with \bugguidedsearch.}
\label{tab:new-laval-bug-search}
\vspace{-2ex}
\end{table}

\subsection{Real World Benchmark Evaluation}
\label{subsec:appendix:bugdiff}
For a better reference of  our evaluation with real-world programs, 
we summarize the number of 
triggered violations at the end of 24 hours in Table~\ref{tab:ubsanviolations}.

In addition, we also compare the UBSan violations triggered by \savior and 
the other 5 fuzzers. The results are summarized in Table~\ref{tab:bugdiff}.
In general, these fuzzers are exploring a similar group of UBSan violations. 
More importantly, for most of the cases, 
\savior triggers a super-set of the violations 
that are made by the other fuzzers (in particular \afl and \aflgo). 
This indicates that \savior has a better thoroughness in 
vulnerability finding.

\begin{table}[t!]
\scriptsize
\centering
\begin{tabular}{l|p{8mm}p{8mm}p{8mm}p{8mm}p{8mm}p{8mm}}
\toprule[0.5pt]
\toprule[0.5pt]

\multirow{2}{*}{\bf{\emph{Prog.}}} & \multicolumn{6}{c}{\bf{\emph{Difference of triggered UBSan violations}}}

 \\ \cline{2-7}
 & \afl  & \aflgo & \angora & \driller  & \qsym & \savior \\\hline
\rowcolor{mygray}
tcpdump & +5/-43  & +0/-61 & +0/-76 & +7/-30  & +15/-28 & +0/-0 \\
tiff2ps & +0/-13  & +0/-6 & +0/-9 & +0/-8  & +0/-8 & +0/-0 \\
\rowcolor{mygray}
readelf & +0/-7  & +1/-7 & +4/-13 & +2/-7  & +2/-7 & +0/-0 \\
xmllint & +0/-6  & +0/-6 & +0/-15 & +0/-6  & +0/-6 & +0/-6\\
\rowcolor{mygray}
djpeg & +0/-0  & +0/-7 & +0/-7 & +0/-0  & +0/-0 & +0/-0\\
tiff2pdf & +0/-7  & +0/-4 & +5/-13 & +0/-6  & +0/-9 & +0/-0\\
\rowcolor{mygray}
jasper & +2/-13  & +0/-13 & +1/-22 & +0/-18  & +0/-8 & +0/-0\\
objdump & +14/-18  & +10/-18 & +16/-20 & +10/-18  & +12/-17 & +0/-0\\

\bottomrule[0.5pt]
\bottomrule[0.5pt]
\end{tabular}
\caption{Difference between violations triggered by \savior and other fuzzers. 
(+X/-Y) means X violations are triggered by the fuzzer but not by \savior and
Y violations are triggered by \savior but not by that fuzzer.}
\label{tab:bugdiff}
\vspace{-2ex}
\end{table}

\begin{table}[t!]
\scriptsize
\centering
\begin{tabular}{l|p{8mm}p{8mm}p{8mm}p{8mm}p{8mm}p{8mm}}
\toprule[0.5pt]
\toprule[0.5pt]

\multirow{2}{*}{\bf{\emph{Prog.}}} & \multicolumn{6}{c}{\bf{\emph{Number of triggered UBSan violations}}}

 \\ \cline{2-7}
 & \afl  & \aflgo & \angora & \driller  & \qsym & \savior \\\hline
\rowcolor{mygray}
tcpdump & 87  & 59 & 43 & 102  & 113 & 128 \\
tiff2ps & 3  & 10 & 7 & 8  & 8 & 16 \\
\rowcolor{mygray}
readelf & 14  & 16 & 14 & 15  & 16 & 22 \\
xmllint & 12  & 12 & 3 & 12  & 12 & 18\\
\rowcolor{mygray}
djpeg & 141  & 134 & 134 & 141  & 141 & 141\\
tiff2pdf & 13  & 13 & 9 & 13  & 10 & 17\\
\rowcolor{mygray}
jasper & 33  & 31 & 23 & 26  & 26 & 44\\
objdump & 64  & 60 & 64 & 60  & 63 & 79\\
\hline {\bf Total} & {\bf 367} & {\bf 335} & {\bf 297} & {\bf 377}  & {\bf 389} & {\bf 465}\\

\bottomrule[0.5pt]
\bottomrule[0.5pt]
\end{tabular}
\caption{Number of unique UBSan violations triggered by different fuzzers in 24 hours. In particular, \MOREBUGDRILLER and \MOREBUGQSYM more violations than \driller and \qsym, respectively.}
\label{tab:ubsanviolations}
\vspace{-2ex}
\end{table}

\input{discuss}

\end{appendices}

%% file: discuss.tex
\section{Technical Discussion and Future Work}
\label{sec:discuss}
In this section, we discuss the limitations of our current
design, insights we learned and possible future directions.

\point{Over-approximation in Vulnerability Labeling} 
As explained in Section~\ref{sec:design}, \savior leverages sound algorithms 
to label vulnerabilities where the over-approximation may introduce many false-positive labels. This imprecision can consequently weaken the performance
of \savior's prioritization. A straightforward reaction to this issue is to 
eliminate as many dummy labels as possible. In our design, we utilize a 
rule-based scheme to filter those false-positive labels in Section~\ref{subsec:system}. 
In the future, we plan to include more precise static analysis for 
finer-grained label pruning. For instance, the \textsc{Stack} system 
developed by Wang et. al~\cite{WangCCJZK12,WangZKS13} and the approach proposed by 
Hathhorn et. al~\cite{HathhornER15} can be incorporated into \savior, which 
are complementary to UBSan in identifying code snippets that may 
lead to undefined behavior.


%

\point{Prediction in Vulnerability Detection}
Once reaching a potentially vulnerable program location in concolic execution, 
\savior extracts the guarding predicates of the vulnerability label. However, 
these predicates may contradict the current path condition. In case of such 
contradiction, \savior terminates the exploration of the labeling site immediately, 
since continuing the analysis cannot contribute to any valuable test input. 

Moreover, in many cases, we can predict whether an execution path can trigger a 
vulnerability or not by studying the runtime information of previous executions. 
Also, more importantly, before that execution arrives the vulnerability site. 
To achieve this goal, we need a method to backwardly summarize path constraints
from the labeled site to its predecessors in the explored paths. The core technique
of this summary is the weakest precondition~\cite{harel2001dynamic} (derived from the
Hoare Logic) which has been applied to both sequential and concurrent program analysis 
domains~\cite{afifi2000method,GuoKWYG15,YiYGWLZ18}.